\documentclass[twoside]{article}

\usepackage{amsfonts}
\usepackage{amsmath}
\usepackage{amssymb}
\usepackage{graphicx}
\usepackage{subfigure}

\def\diag{\hbox{\bf{diag}}}

\newcommand{\TNorm }[1]{\mbox{}\left\|#1\right\|_2  }
\newcommand{\TNormS}[1]{\mbox{}\left\|#1\right\|_2^2}
\newcommand{\FNorm }[1]{\mbox{}\left\|#1\right\|_F  }
\newcommand{\FNormS}[1]{\mbox{}\left\|#1\right\|_F^2}
\newcommand{\argmin}{\text{argmin}}

\newtheorem{definition}{Definition}

\renewcommand{\sectionmark}[1]{\markboth{\textsc{M. W. Mahoney}}
{\textsc{Randomized algorithms for matrices and data}}}
\renewcommand{\subsectionmark}[1]{\markboth{\textsc{M. W. Mahoney}}
{\textsc{Randomized algorithms for matrices and data}}}

\newlength{\figurewidth}
\newlength{\defbaselineskip}
\begin{document}

\title{
Randomized algorithms for matrices and data%
\footnote{Version appearing as a monograph in Now Publishers' ``Foundations and Trends in Machine Learning'' series.}
}

\author{
Michael W. Mahoney%
\thanks{
Department of Mathematics, 
Stanford University, 
Stanford, CA 94305. 
Email: mmahoney@cs.stanford.edu
}
}

\date{}
\maketitle

\begin{center}
{\bf
}
\end{center}
\vspace{-0.25in}

\begin{abstract}
Randomized algorithms for very large matrix problems have received a great 
deal of attention in recent years.
Much of this work was motivated by problems in large-scale data analysis, 
largely since matrices are popular structures with which to model data drawn 
from a wide range of application domains, and this work was performed by 
individuals from many different research communities.
While the most obvious benefit of randomization is that it can lead to 
faster algorithms, either in worst-case asymptotic theory and/or numerical 
implementation, there are numerous other benefits that are at least as 
important.
For example, the use of randomization can lead to simpler algorithms that 
are easier to analyze or reason about when applied in counterintuitive 
settings; it can lead to algorithms with more interpretable output, which is 
of interest in applications where analyst time rather than just 
computational time is of interest; it can lead implicitly to regularization 
and more robust output; and randomized algorithms can often be organized to 
exploit modern computational architectures better than classical numerical 
methods.

This monograph will provide a detailed overview of recent work on the theory 
of randomized matrix algorithms as well as the application of those 
ideas to the solution of practical problems in large-scale data analysis.
Throughout this review, an emphasis will be placed on a few simple core 
ideas that underlie not only recent theoretical advances but also the 
usefulness of these tools in large-scale data applications.
Crucial in this context is the connection with the concept of statistical 
leverage.
This concept has long been used in statistical regression diagnostics to 
identify outliers; and it has recently proved crucial in the development 
of improved worst-case matrix algorithms that are also amenable to 
high-quality numerical implementation and that are useful to domain 
scientists.
This connection arises naturally when one explicitly decouples the effect 
of randomization in these matrix algorithms from the underlying linear
algebraic structure.  This decoupling also permits much finer control in the 
application of randomization, as well as the easier exploitation of 
domain~knowledge.

Most of the review will focus on random sampling algorithms and random 
projection algorithms for versions of the linear least-squares problem and 
the low-rank matrix approximation problem.
These two problems are fundamental in theory and ubiquitous in practice.
Randomized methods solve these problems by constructing and operating on a 
randomized sketch of the input matrix $A$---for random sampling methods, the 
sketch consists of a small number of carefully-sampled and rescaled 
columns/rows of $A$, while for random projection methods, the sketch 
consists of a small number of linear combinations of the columns/rows of $A$.
Depending on the specifics of the situation, when compared with the best
previously-existing deterministic algorithms, the resulting randomized 
algorithms have worst-case running time that is asymptotically faster; their
numerical implementations are faster in terms of clock-time; or they can be 
implemented in parallel computing environments where existing numerical 
algorithms fail to run at~all.
Numerous examples illustrating these observations will be described in~detail.
\end{abstract}

\newpage
\rule{6in}{0.1mm}
\tableofcontents
\rule{6in}{0.1mm}
\newpage

\newpage

\section{Introduction}

This monograph will provide a detailed overview of recent work on the theory 
of \emph{randomized matrix algorithms} as well as the application of those 
ideas to the solution of practical problems in large-scale data analysis.
By ``randomized matrix algorithms,'' we refer to a class of 
recently-developed random sampling and random projection algorithms for 
ubiquitous linear algebra problems such as least-squares regression and 
low-rank matrix approximation.
These and related problems are ubiquitous since matrices are fundamental 
mathematical structures for representing data drawn from a wide range of 
application domains.
Moreover, the widespread interest in randomized algorithms for these 
problems arose due to the need for principled algorithms to deal with the 
increasing size and complexity of data that are being generated in many of 
these application~areas.

Not surprisingly, algorithmic procedures for working with matrix-based data 
have been developed from a range of diverse perspectives by researchers 
from a wide range of areas---including, \emph{e.g.}, researchers from 
theoretical computer science (TCS), numerical linear algebra (NLA), 
statistics, applied mathematics, data analysis, and machine learning, as 
well as domain scientists in physical and biological sciences---and in many 
of these cases they have drawn strength from their domain-specific insight.
Although this has been great for the development of the area, and for the 
``technology transfer'' of theoretical ideas to practical applications, the 
technical aspects of dealing with any one of those areas has obscured for many the 
simplicity and generality of some of the underlying ideas; thus leading 
researchers to fail to appreciate the underlying connections and the 
significance of contributions by researchers outside their own area. 
Thus, rather than focusing on the technical details of proving worst-case 
bounds or of providing high-quality numerical implementations or of relating 
to traditional machine learning tools or of using these algorithms in a 
particular physical or biological domain, in this review we will focus on 
highlighting for a broad audience the simplicity and generality of some core 
ideas---ideas that are often obscured but that are fruitful for using these 
randomized algorithms in large-scale data applications.
To do so, we will focus on two fundamental and ubiquitous matrix 
problems---least-squares approximation and low-rank matrix 
approximation---that have been at the center of these recent developments.

The work we will review here had its origins within TCS.
In this area, one typically considers a particular well-defined problem, and 
the goal is to prove bounds on the running time and quality-of-approximation 
guarantees for algorithms for that particular problem that hold for 
``worst-case'' input.
That is, the bounds should hold for \emph{any} input matrix, independent of 
any ``niceness'' assumptions such as, \emph{e.g.}, that the elements of the 
matrix satisfy some smoothness or normalization condition or that the 
spectrum of the matrix satisfies some decay condition.
Clearly, the generality of this approach means that the bounds will be 
suboptimal---and thus can be improved---in any particular application where 
stronger assumptions can be made about the input.
Importantly, though, it also means that the underlying algorithms and 
techniques will be broadly applicable even in situations where such 
assumptions do not~apply.

An important feature in the use of randomized algorithms in TCS more 
generally is that one must identify and then algorithmically deal with
relevant ``non-uniformity structure'' in the data.%
\footnote{For example, for those readers familiar with Markov chain-based 
Monte Carlo algorithms as used in statistical physics, this non-uniformity 
structure is given by the Boltzmann distribution, in which case the 
algorithmic question is how to sample efficiently with respect to it as an 
importance sampling distribution without computing the intractable partition 
function.  
Of course, if the data are sufficiently nice (or if they have been 
sufficiently preprocessed, or if sufficiently strong assumptions are made 
about them, etc.), then that non-uniformity structure might be uniform, in 
which case simple methods like uniform sampling might be appropriate---but 
this is far from true in general, either in worst-case theory or in practical 
applications.}
For the randomized matrix algorithms to be reviewed here and that have proven 
useful recently in NLA and large-scale data analysis 
applications, the relevant non-uniformity structure
is defined by the so-called \emph{statistical leverage 
scores}.
Defined more precisely below, these leverage scores are basically the 
diagonal elements of the projection matrix onto the dominant part of the 
spectrum of the input matrix.
As such, they have a long history in statistical data analysis, where they 
have been used for outlier detection in regression diagnostics.
More generally, and very importantly for practical large-scale data 
applications of recently-developed randomized matrix algorithms, these 
scores often have a very natural interpretation in terms of the data and 
processes generating the data.
For example, they can be interpreted in terms of the leverage or influence 
that a given data point has on, say, the best low-rank matrix approximation; 
and this often has an interpretation in terms of high-degree nodes in data 
graphs, very small clusters in noisy data, coherence of information, 
articulation points between clusters, etc.

Historically, although the first generation of randomized matrix algorithms 
(to be described in Section~\ref{sxn:background2})
achieved what is known as additive-error bounds and were extremely fast, 
requiring just a few passes over the data from external storage, these 
algorithms did \emph{not} gain a foothold in NLA and only heuristic variants 
of them were used in machine learning and data analysis applications.
In order to ``bridge the gap'' between NLA, TCS, and data applications, 
much finer control over the random sampling process was needed.
Thus, in the second generation of randomized matrix algorithms 
(to be described in Sections~\ref{sxn:least-squares} and~\ref{sxn:low-rank})
that \emph{has} led to high-quality numerical implementations and useful 
machine learning and data analysis applications, two key developments were 
crucial.
\begin{itemize}
\item
\textbf{Decoupling the randomization from the linear algebra.}
This was originally implicit within the analysis of the second generation 
of randomized matrix algorithms, and then it was made explicit.
By making this decoupling explicit, not only were improved quality-of-approximation 
bounds achieved, but also \emph{much} finer control was achieved in the 
application of randomization.
For example, it permitted easier exploitation of domain expertise, in both 
numerical analysis and data analysis applications.
\item
\textbf{Importance of statistical leverage scores.}
Although these scores have been used historically for outlier detection in 
statistical regression diagnostics, they have also been crucial in the 
recent development of randomized matrix algorithms.
Roughly, the best random sampling algorithms use these scores to construct 
an importance sampling distribution to sample with respect to; and 
the best random projection algorithms rotate to a basis where these scores 
are approximately uniform and thus in which uniform sampling is 
appropriate.
\end{itemize}
As will become clear, these two developments are very related.
For example,
once the randomization was decoupled from the linear algebra, it became 
nearly obvious that the ``right'' importance sampling probabilities to use 
in random sampling algorithms are those given by the statistical leverage 
scores, and it became clear how to improve the analysis and numerical implementation of random 
projection algorithms.
It is remarkable, though, that statistical leverage scores define the
non-uniformity structure that is relevant not only to obtain the 
strongest worst-case bounds, but also to lead to high-quality numerical 
implementations (by numerical analysts) as well as algorithms that are 
useful in downstream scientific applications (by machine learners and data 
analysts).

Most of this review will focus on random sampling algorithms and random 
projection algorithms for versions of the linear least-squares problem and 
the low-rank matrix approximation problem.
Here is a brief summary of some of the highlights of what follows.
\begin{itemize}
\item
\textbf{Least-squares approximation.}
Given an $m \times n$ matrix $A$, with $m \gg n$, and an $m$-dimensional 
vector $b$, the overconstrained least-squares approximation problem looks 
for the vector $x_{opt} = \mbox{argmin}_x ||Ax-b||_2  $.
This problem typically arises in statistical models where the rows of $A$ 
and elements of $b$ correspond to constraints and the columns of $A$ and 
elements of $x$ correspond to variables.
Classical methods, including the Cholesky decomposition, versions of the QR 
decomposition, and the Singular Value Decomposition, compute a solution
in $O(mn^2)$ time.
Randomized methods solve this problem by constructing a randomized sketch 
of the matrix $A$---for random sampling methods, the sketch consists of a 
small number of carefully-sampled and rescaled rows of $A$ (and the 
corresponding elements of $b$), while for random projection methods, the 
sketch consists of a small number of linear combinations of the rows of $A$ and elements of $b$.
If one then solves the (still overconstrained) subproblem induced on the 
sketch, then very fine relative-error approximations to the solution of the 
original problem are obtained.
In addition, for a wide range of values of $m$ and $n$, the running time is 
$o(mn^2)$---for random sampling algorithms, the computational bottleneck is 
computing appropriate importance sampling probabilities, while for random 
projection algorithms, the computational bottleneck is implementing the 
random projection operation.
Alternatively, if one uses the sketch to compute a preconditioner for the
original problem, then very high-precision approximations can be obtained by
then calling classical numerical iterative algorithms.
Depending on the specifics of the situation, these numerical implementations 
run in $o(mn^2)$ time; they are faster in terms of clock-time than the best 
previously-existing deterministic numerical implementations; or they can be implemented in 
parallel computing environments where existing numerical algorithms fail to 
run at~all.
\item
\textbf{Low-rank matrix approximation.}
Given an $m \times n$ matrix $A$ and a rank parameter $k$, the low-rank 
matrix approximation problem is to find a good approximation to $A$ of 
rank $k \ll \min\{m,n\}$.
The Singular Value Decomposition provides the best rank-$k$ approximation 
to $A$, in the sense that by projecting $A$ onto its top $k$ left or right 
singular vectors, then one obtains the best approximation to $A$ with 
respect to the spectral and Frobenius norms.
The running time for classical low-rank matrix approximation algorithms 
depends strongly on the specifics of the situation---for dense matrices, the 
running time is typically $O(mnk)$; while for sparse matrices, classical 
Krylov subspace methods are used.
As with the least-squares problem, randomized methods for the low-rank matrix
approximation problem construct a randomized sketch---consisting of a small 
number of either actual columns or linear combinations of columns---of the 
input $A$, and then this sketch is manipulated depending on the specifics of 
the situation.
For example, random sampling methods can use the sketch directly to 
construct relative-error low-rank approximations such as CUR decompositions 
that approximate $A$ based on a small number of actual columns of the input 
matrix.
Alternatively, random projection methods can improve the running time for 
dense problems to $O(mn \log k)$; and while they only match the running time 
for classical methods on sparse matrices, they lead to more robust 
algorithms that can be reorganized to exploit parallel computing 
architectures.
\end{itemize}
These two problems are the main focus of this review since they are both 
fundamental in theory and ubiquitous in practice and since in both cases
novel theoretical ideas have already yielded practical results.
Although not the main focus of this review, other related matrix-based 
problems to which randomized methods have been applied will be referenced 
at appropriate points.

Clearly, when a very new paradigm is compared with very well-established 
methods, a na\"{i}ve implementation of the new ideas will perform poorly by 
traditional metrics.
Thus, in both data analysis and numerical analysis applications of this 
randomized matrix algorithm paradigm, the best results have been achieved 
when coupling closely with more traditional methods.
For example, in data analysis applications, this has meant working closely 
with geneticists and other domain experts to understand how the 
non-uniformity structure in the data is useful for their downstream 
applications.  
Similarly, in scientific computation applications, this has meant coupling
with traditional numerical methods for improving quantities like condition
numbers and convergence rates.
When coupling in this manner, however, qualitatively improved results have 
\emph{already} been achieved.
For example, in their empirical evaluation of the random projection algorithm 
for the least-squares approximation problem, to be described in 
Sections~\ref{sxn:least-squares:faster-th}
and~\ref{sxn:least-squares:faster-pr} below, Avron, Maymounkov, and 
Toledo~\cite{AMT10} began by observing that ``Randomization is arguably the 
most exciting and innovative idea to have hit linear algebra in a long 
time;'' and since their implementation ``beats \textsc{Lapack}'s%
\footnote{\textsc{Lapack} (short for Linear Algebra PACKage) is a 
high-quality and widely-used software library of numerical routines for 
solving a wide range of numerical linear algebra problems.} 
direct 
dense least-squares solver by a large margin on essentially any dense tall 
matrix,'' they concluded that their empirical results ``show the potential 
of random sampling algorithms and suggest that random projection algorithms 
should be incorporated into future versions of \textsc{Lapack}.''

The remainder of this review will cover these topics in greater detail.
To do so, we will start in Section~\ref{sxn:background1} with a few 
motivating applications from one scientific domain where these randomized 
matrix algorithms have already found application, and we will describe in
Section~\ref{sxn:background2} general background on randomized matrix 
algorithms, including precursors to those that are the main subject of this 
review.
Then, in the next two sections, we will describe randomized matrix algorithms
for two fundamental matrix problems: Section~\ref{sxn:least-squares} will 
be devoted to describing several related algorithms for the least-squares 
approximation problem; and Section~\ref{sxn:low-rank} will be devoted to 
describing several related algorithms for the problem of low-rank matrix 
approximation.
Then, Section~\ref{sxn:empirical} will describe in more detail some of these 
issues from an empirical perspective, with an emphasis on the ways that 
statistical leverage scores have been used more generally in large-scale 
data analysis; 
Section~\ref{sxn:thoughts} will provide some more general thought on this
successful technology transfer experience; and
Section~\ref{sxn:conclusion} will provide a brief conclusion.

\section{Matrices in large-scale scientific data analysis}
\label{sxn:background1}

In this section, we will provide a brief overview of examples of 
applications of randomized matrix algorithms in large-scale scientific data 
analysis.
Although far from exhaustive, these examples should serve as a motivation to 
illustrate several complementary perspectives that one can adopt on these 
techniques.

\subsection{A brief background}
\label{sxn:background1-brief}

Matrices arise in machine learning and modern massive data set (MMDS) 
analysis in many guises.
One broad class of matrices to which randomized algorithms have been applied 
is \emph{object-feature~matrices}. 
\begin{itemize}
\item
\textbf{Matrices from object-feature data.}
An $m \times n$ real-valued matrix $A$ provides a natural structure for 
encoding information about $m$ objects, each of which is described by $n$ 
features.
In astronomy, for example, very small angular regions of the sky imaged at 
a range of electromagnetic frequency bands can be represented as a matrix---in 
that case, an object is a region and the features are the elements of the 
frequency bands.
Similarly, in genetics, DNA Single Nucleotide Polymorphism or DNA 
microarray expression data can be represented in such a framework, with 
$A_{ij}$ representing the expression level of the $i$-th gene or SNP in 
the $j$-th experimental condition or individual. 
Similarly, term-document matrices can be constructed in many Internet 
applications, with $A_{ij}$ indicating the frequency of the $j$-th term in 
the $i$-th document. 
\end{itemize}
Matrices arise in many other contexts---\emph{e.g.}, they arise when solving
partial differential equations in scientific computation as discretizations
of continuum operators; and they arise as so-called kernels when describing 
pairwise relationships between data points in machine learning.
In many of these cases, certain conditions---\emph{e.g.}, that the spectrum
decays fairly quickly or that the matrix is structured such that it can be 
applied quickly to arbitrary vectors or that the elements of the matrix 
satisfy some smoothness conditions---are known or are thought to hold.

A fundamental property of matrices that is of broad applicability in both 
data analysis and scientific computing is the Singular Value 
Decomposition~(SVD).
If $A \in \mathbb{R}^{m \times n}$, then there exist orthogonal matrices
$ U=[u^{1} u^{2} \ldots u^{m}]\in \mathbb{R}^{m \times m} $ and 
$ V=[v^{1} v^{2} \ldots v^{n}]\in \mathbb{R}^{n \times n} $
such that
$
U^TAV=\Sigma=\diag(\sigma_1,\ldots,\sigma_\nu)    ,
$
where $\Sigma \in \mathbb{R}^{m \times n}$, $\nu=\min\{m,n\}$ and 
$\sigma_1 \geq \sigma_2 \geq \ldots \geq \sigma_\nu \geq 0$.  
The $\sigma_i$ are the singular values of $A$, the column vectors $u^{i}$, $v^{i}$ 
are the $i$-th left and the $i$-th right singular vectors of $A$, 
respectively.
If $k \leq r = \mbox{rank}(A)$, then the SVD of $A$ may be written as
\begin{equation} 
\label{eqn:svdA}
A = U \Sigma V^T 
  = \left[\begin{array}{cc} 
          U_{k} & U_{k,\perp}
    \end{array}\right]
    \left[\begin{array}{cc}
          \Sigma_{k} & \bf{0}        \\
          \bf{0} & \Sigma_{k,\perp}
    \end{array}\right]
    \left[\begin{array}{c}
           V_{k}^T \\
           V_{k,\perp}^T
    \end{array}\right]     
  = U_k \Sigma_k V_k^T + U_{k,\perp} \Sigma_{k,\perp} V_{k,\perp}^T     .
\end{equation}
Here, $\Sigma_k$ is the $k \times k$ diagonal matrix containing the top $k$ 
singular values of $A$, and
$\Sigma_{k,\perp}$ is the $\left(r-k\right) \times \left(r - k\right)$ 
diagonal matrix containing the bottom $r-k$ nonzero singular values of $A$, 
$V_k^T$ is the $k \times n$ matrix consisting of the corresponding top $k$
right singular vectors,%
\footnote{In the text, we will sometimes overload notation and use $V_k^T$ 
to refer to \emph{any} $k \times n$ orthonormal matrix spanning the space 
spanned by the top-$k$ right singular vectors (and similarly for $U_k$ and 
the left singular vectors).  The reason is that this basis is used only to 
compute the importance sampling probabilities---since those probabilities 
are proportional to the diagonal elements of the projection matrix onto the 
span of this basis, the particular basis does not matter.}
etc.
By keeping just the top $k$ singular vectors, the matrix 
$A_k = U_k \Sigma_k V_k^T$ is the best rank-$k$ approximation to $A$, when 
measured with respect to the spectral and Frobenius norm.
Let
$ ||A||_F^2 = \sum_{i=1}^m \sum_{j=1}^n A_{ij}^2 $ denote the square of 
the
Frobenius norm; let 
$||A||_2 = \sup_{x\in \mathbb{R}^n,\ x\neq 0} ||Ax||_2/||x||_2$
denote 
the
spectral norm;%
\footnote{Since the spectral norm is the largest singular value of the 
matrix, it is an ``extremal'' norm in that it measures the 
worst-case stretch of the matrix, while the Frobenius norm is more of an 
``averaging'' norm, since it involves a sum over every singular direction.
The former is of greater interest in scientific computing and NLA, where one 
is interested in actual columns for the subspaces they define and for their 
good numerical properties, while the latter is of greater interest in data 
analysis and machine learning, where one is more interested in actual 
columns for the features they define.  Both are of interest in this review.}
and,
for any matrix $A \in \mathbb{R}^{m \times n}$, let $A_{(i)}, i \in [m]$ 
denote the $i$-th \emph{row} of $A$ as a row vector, and let $A^{(j)}, j \in [n]$
denote the $j$-th \emph{column} of $A$ as a column vector.

Finally, since they will play an important role in later developments, the 
\emph{statistical leverage scores} of an $m \times n$ matrix, with $m > n$, 
are defined here.
\begin{definition}
\label{def:lev-scores}
Given an arbitrary $m \times n$ matrix $A$, with $m > n$, let $U$ denote
the $m \times n$ matrix consisting of the $n$ left singular vectors of $A$,
and let $U_{(i)}$ denote the $i$-th row of the matrix $U$ as a row vector.
Then, the quantities 
$$
\ell_i = ||U_{(i)}||_2^2  ,
\quad
\mbox{for }
i\in\{1,\ldots,m\}  ,
$$
are the \emph{statistical leverage scores} of the rows of $A$.
\end{definition}

\noindent 
Several things are worth noting about this definition.
First, although we have defined these quantities in terms of a particular 
basis, they clearly do not depend on that particular basis, but instead 
only on the space spanned by that basis. 
To see this, let $P_A$ denote the projection matrix onto the span of the 
columns of $A$; then,
$\ell_i = ||U_{(i)}||_2^2 = \left(UU^T\right)_{ii} = \left(P_A\right)_{ii}$.
That is, the statistical leverage scores of a matrix $A$ are equal to the
diagonal elements of the projection matrix onto the span of its columns.
Second, if $m > n$, then $O(mn^2)$ time suffices to compute all the 
statistical leverage scores exactly: simply perform the SVD or compute a QR 
decomposition of $A$ in order to obtain \emph{any} orthogonal basis for the 
range of $A$, and then compute the Euclidean norm of the rows of the 
resulting matrix.
Third, one could also define leverage scores for the columns of such a 
matrix $A$, but clearly those are all equal to one unless $m<n$ or $A$ is
rank-deficient. 
Fourth, and more generally, given a rank parameter $k$, one can define the
\emph{statistical leverage scores relative to the best rank-$k$
approximation to $A$} to be the $m$ diagonal
elements of the projection matrix onto the span of the columns of $A_k$, the 
best rank-$k$ approximation to~$A$.
Finally, the \emph{coherence} $\gamma$ of the rows of $A$ is
$\gamma = \max_{i\in\{1,\ldots,m\}} \ell_i $, \emph{i.e.}, it is the largest 
statistical leverage score of~$A$.

\subsection{Motivating scientific applications}
\label{sxn:background1:genetics}

To illustrate a few examples where randomized matrix algorithms have already
been applied in scientific data analysis,
recall that ``the human genome'' consists of a sequence of roughly 
$3$ billion base pairs on $23$ pairs of chromosomes, roughly $1.5\%$ of 
which codes for approximately $20,000$ -- $25,000$ proteins.
A DNA microarray is a device that can be used to measure simultaneously the 
genome-wide response of the protein product of each of these genes for an 
individual or group of individuals in numerous different environmental 
conditions or disease states.
This very coarse measure can, of course, hide the individual differences or 
polymorphic variations. 
There are numerous types of polymorphic variation, but the most amenable to 
large-scale applications is the analysis of Single Nucleotide Polymorphisms 
(SNPs), which are known locations in the human genome where two alternate 
nucleotide bases (or alleles, out of $A$, $C$, $G$, and $T$) are observed in 
a non-negligible fraction of the population.
These SNPs occur quite frequently, roughly $1$ base pair per thousand (depending on the minor allele frequency), and thus they
are effective genomic markers for the tracking of disease genes (\emph{i.e.}, 
they can be used to perform classification into sick and not sick) as well as
population histories (\emph{i.e}, they can be used to infer properties about 
population genetics and human evolutionary history).

In both cases, $m \times n$ matrices $A$ naturally arise, either as a
people-by-gene matrix, in which $A_{ij}$ encodes information about the 
response of the $j^{th}$ gene in the $i^{th}$ individual/condition, or as 
people-by-SNP matrices, in which $A_{ij}$ encodes information about the 
value of the $j^{th}$ SNP in the $i^{th}$ individual.
Thus, matrix computations have received attention in these genetics
applications~\cite{Alter_SVD_00,KPS02,Meng03,Horne04,LA04,PPR06}.
To give a rough sense of the sizes involved, if the matrix is constructed in 
the na\"{i}ve way based on data from the International HapMap 
Project~\cite{IHMC03,IHMC05}, then it is of size roughly $400$ people by 
$10^{6}$ SNPs, although more recent technological developments have 
increased the number of SNPs to well into the millions and the number of people 
into the thousands and tens-of-thousands.
Depending on the size of the data and the genetics problem under 
consideration, randomized algorithms can be useful in one or more of 
several~ways.

For example, a common genetics challenge is to determine whether there is 
any evidence that the samples in the data are from a population that is 
structured, \emph{i.e.}, are the individuals from a homogeneous population 
or from a population containing genetically distinct subgroups?
In medical genetics, this arises in case-control studies, where uncorrected 
population structure can induce false positives; and in population genetics, 
it arises where understanding the structure is important for uncovering the
demographic history of the population under study.
To address this question, it is common to perform a procedure such as the 
following.
Given an appropriately-normalized (where, of course, the normalization depends crucially on domain-specific considerations) $m \times n$ matrix $A$:
\begin{itemize}
\item
Compute a full or partial SVD or perform a QR decomposition, thereby 
computing the eigenvectors and eigenvalues of the correlation matrix $AA^T$.
\item
Appeal to a statistical model selection criterion%
\footnote{For example, the model selection rule could compare the top part 
of the spectrum of the data matrix to that of a random matrix of the same 
size~\cite{Paschou07b,FK81}; 
or it could use the full spectrum to compute a test statistic to determine
whether there is more structure in the data matrix than would be present in 
a random matrix of the same size~\cite{PPR06,Joh01}.}
to determine either the number $k$ of principal components to keep in order 
to project the data onto or whether to keep an additional principal 
component as significant.
\end{itemize}

Although this procedure could be applied to any data set $A$, to obtain 
meaningful genetic conclusions one must deal with numerous issues.%
\footnote{For example, how to normalize the data, how to deal with missing 
data, how to correct for linkage disequilibrium (or correlational) structure 
in the genome, how to correct for closely-related individuals within the 
sample,~etc.}
In any case, however, the computational bottleneck is typically computing 
the SVD or a QR decomposition.
For small to medium-sized data, this is not a problem---simply call
\textsc{Matlab} or call appropriate routines from \textsc{Lapack} directly.
The computation of the full eigendecomposition takes 
$O(\min\{mn^2,m^2n\})$ time, and if only $k$ components of the 
eigendecomposition are needed then the running time is typically $O(mnk)$ time.
(This ``typically'' is awkward from the perspective of worst-case
analysis, but it is not usually a problem in practice.  Of course, one could
compute the full SVD in $O(\min\{mn^2,m^2n\})$ time and truncate to obtain 
the partial SVD.  Alternatively, one could use a Krylov subspace method to 
compute the partial SVD in $O(mnk)$ time, but these methods can be less 
robust.  Alternatively, one could perform a rank-revealing QR factorization
such as that of Gu and Eisenstat~\cite{GE96} and then post-process the 
factors to obtain a partial SVD.  The cost of computing the QR decomposition 
is typically $O(mnk)$ time, although these methods can require slightly 
longer time in rare cases~\cite{GE96}.  See~\cite{HMT09_SIREV} for a discussion of 
these topics.)

Thus, these traditional methods can be quite fast even for very large data 
if one of the dimensions is small, \emph{e.g.}, $10^2$ individuals typed at 
$10^7$ SNPs.
On the other hand, if both $m$ and $n$ are large, \emph{e.g.}, $10^3$ 
individuals at $10^6$ SNPs, or $10^4$ individuals at $10^5$ SNPs, then, for
interesting values of the rank parameter $k$, the $O(mnk)$ running time of 
even the QR decomposition can be prohibitive.
As we will see below, however, by exploiting randomness inside the 
algorithm, one can obtain an $O(mn \log k)$ running time.
(All of this assumes that the data matrix is dense and fits in 
memory, as is typical in SNP applications.  More generally, randomized 
matrix algorithms to be reviewed below also help in other computational 
environments, \emph{e.g.}, for sparse input matrices, for matrices too large 
to fit into main memory, when one wants to reorganize the steps of the 
computations to exploit modern multi-processor architectures, etc.  
See~\cite{HMT09_SIREV} for a discussion of these topics.)
Since interesting values for $k$ are often in the hundreds, this improvement
from $O(k)$ to $O(\log k)$ can be quite significant in practice;
and thus one can apply the above procedure for identifying structure in 
DNA SNP data on much larger data sets than would have been possible with 
traditional deterministic methods~\cite{Sayan11-unpub}.

More generally, a common \emph{modus operandi} in applying NLA and matrix 
techniques such as PCA and the SVD to DNA microarray, DNA SNPs, and other 
data problems~is:
\begin{itemize}
\item
Model the people-by-gene or people-by-SNP data as an $m \times n$ matrix $A$.
\item
Perform the SVD (or related eigen-methods such as PCA or recently-popular 
manifold-based methods~\cite{TSL00,RS00,SWHSL06} that boil down to the SVD, 
in that they perform the SVD or an eigendecomposition on nontrivial matrices 
constructed from the data) 
to compute a small number of eigengenes or eigenSNPs or eigenpeople that 
capture most of the information in the data matrix.
\item 
Interpret the top eigenvectors as meaningful in terms of underlying 
biological processes; or apply a heuristic to obtain actual genes or actual 
SNPs from the corresponding eigenvectors in order to obtain such an 
interpretation.
\end{itemize}
In certain cases, such reification may lead to insight and such heuristics 
may be justified. 
For instance, if the data happen to be drawn from a Guassian distribution, 
as in Figure~1A,
then the eigendirections tend to correspond to the axes of the corresponding
ellipsoid, and there are many vectors that, up to noise, point along those
directions.
In most cases, however, \emph{e.g.}, when the data are drawn from the union 
of two normals (or mixture of two Gaussians), as in Figure~1B, such reification is not valid.
In general, the justification for interpretation comes from domain 
knowledge and not the mathematics~\cite{Gould96,KPS02,MPC78,CUR_PNAS}.
The reason is that the eigenvectors themselves, being mathematically defined
abstractions, can be calculated for any data matrix and thus are not easily 
understandable in terms of processes generating the data:
eigenSNPs (being linear combinations of SNPs) cannot be assayed;
nor can eigengenes (being linear combinations of genes) be isolated and purified;
nor is one typically interested in how eigenpatients (being linear combinations of patients) respond to treatment 
when one visits a~physician.

For this and other reasons, a common task in genetics and other areas of data 
analysis is the following: given an input data matrix $A$ and a parameter 
$k$, find the best subset of exactly $k$ \emph{actual} DNA SNPs or 
\emph{actual} genes, \emph{i.e.}, \emph{actual} columns or rows from $A$, to 
use to cluster individuals, reconstruct biochemical pathways, reconstruct 
signal, perform classification or inference, etc.
Unfortunately, common formalizations of this algorithmic problem---including 
looking for the $k$ actual columns that capture the largest amount of 
information or variance in the data or that are maximally 
uncorrelated---lead to intractable optimization 
problems~\cite{CM08,CM09a}.
For example, consider the so-called Column Subset Selection Problem~\cite{BMD09_CSSP_SODA}:
given as input an arbitrary $m \times n$ matrix $A$ and a rank parameter 
$k$, choose the set of exactly $k$ columns of $A$ s.t. the $m \times k$ 
matrix $C$ minimizes (over all ${n \choose k}$ sets of such columns) the error:
\begin{equation}
\min ||A-P_CA||_{\nu} = \min ||A-CC^+A||_{\nu}  
\label{eqn:error-measure}
\end{equation}
where  $\nu\in\{2,F\}$ represents the spectral or Frobenius norm
of $A$, $C^+$ is the Moore-Penrose pseudoinverse of $C$, and $P_C=CC^+$ is the projection onto the subspace spanned by 
the columns of~$C$.
As we will see below, however, by exploiting randomness inside the 
algorithm, one can find a small set of actual columns that is provably 
nearly optimal.
Moreover, this algorithm and obvious heuristics motivated by it have already 
been applied successfully to problems of interest to geneticists such as 
genotype reconstruction in unassayed populations, identifying substructure 
in heterogeneous populations, and inference of individual 
ancestry~\cite{CUR_PNAS,Paschou07a,Paschou07b,Paschou08a,Paschou10a,Paschou10b,Paschou11a}.

In order to understand better the reification issues in scientific data 
analysis, consider a synthetic data set---it was originally introduced 
in~\cite{WRR03} to model oscillatory and exponentially decaying patterns of 
gene expression from~\cite{Cho_cellcycle_98}, although it could just as 
easily be used to describe oscillatory and exponentially decaying patterns 
in stellar spectra, etc.
The data matrix consists of $14$ expression level assays (columns of $A$) and
$2000$ genes (rows of $A$), corresponding to a $2000 \times 14$ matrix $A$.
Genes have one of three types of transcriptional response:
noise ($1600$ genes); noisy sine pattern ($200$ genes); and noisy exponential 
pattern ($200$ genes).
Figures~1C and~1D present the ``biological'' data, \emph{i.e.}, overlays of 
five noisy sine wave genes and five noisy exponential genes, respectively;
Figure~1E presents the first and second singular vectors of the data matrix, 
along with the original sine pattern and exponential pattern that generated 
the data; and 
Figure~1F shows that the data cluster well in the space spanned by the top 
two singular vectors, which in this case account for $64\%$ of the variance 
in the data. 
Note, though, that the top two singular vectors both display a linear combination of oscillatory and 
decaying properties; and thus they are not easily interpretable as ``latent
factors'' or ``fundamental modes'' of the original (sinusoid and exponential) 
``biological'' processes generating the data.
This is problematic more generally when one is interested in extracting 
insight or ``discovering knowledge'' from the output of data analysis 
algorithms~\cite{CUR_PNAS}.%
\footnote{Indeed, after describing the many uses of the vectors provided by 
the SVD and PCA in DNA microarray analysis, Kuruvilla 
\emph{et al.}~\cite{KPS02} bluntly conclude that 
``While very efficient basis vectors, the (singular) vectors 
themselves are completely artificial and do not correspond to actual (DNA
expression) profiles. ... Thus, it would be interesting to try to find basis 
vectors for all experiment vectors, using actual experiment vectors and not 
artificial bases that offer little insight.''}

\begin{figure}
\begin{center}
\includegraphics[width=7in]{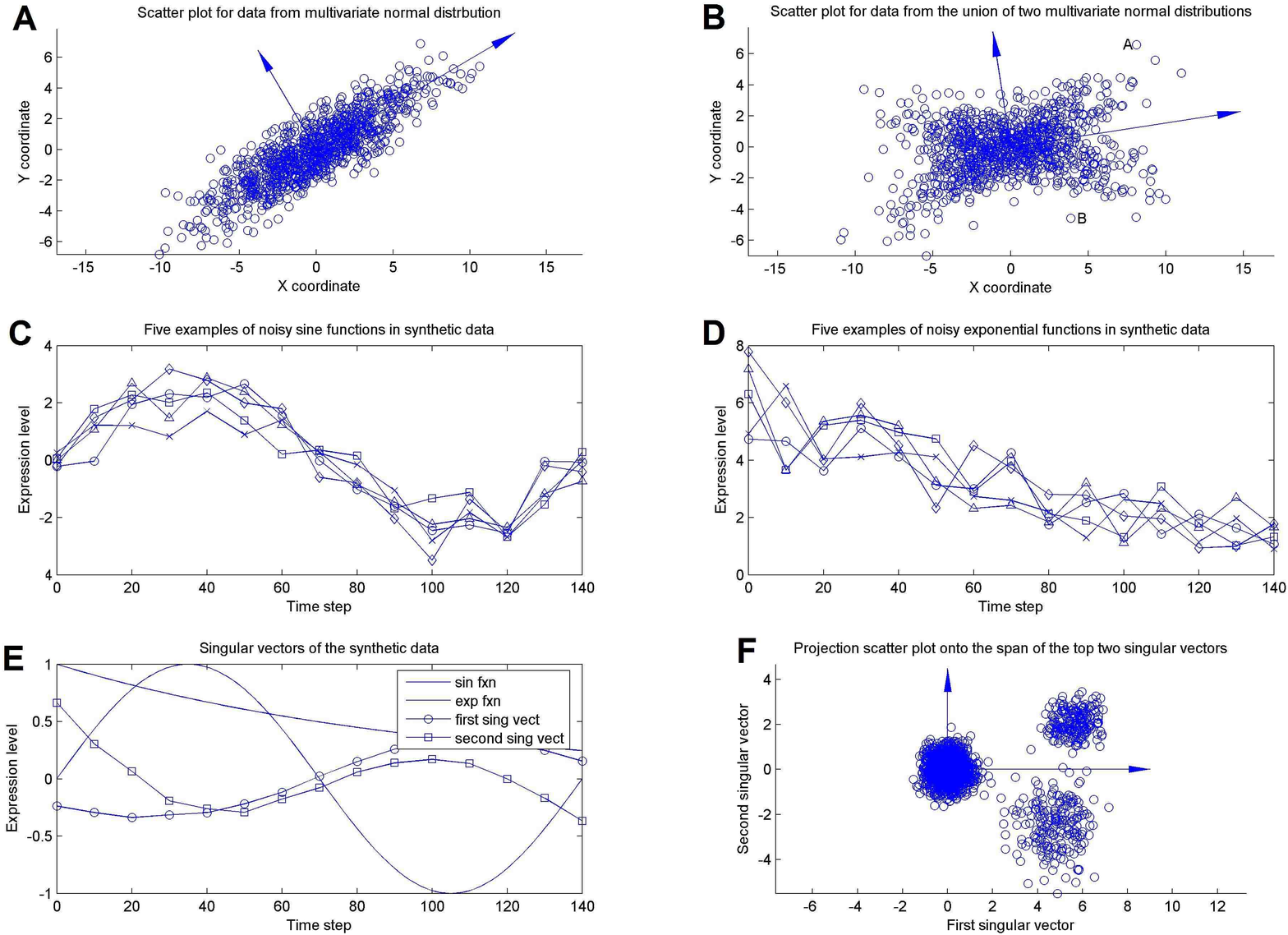}
\caption{Applying the SVD to data matrices $A$.
(A) $1000$ points on the plane, corresponding to a $1000 \times 2$ matrix $A$, 
(and the two principal components) drawn from a multivariate normal 
distribution.
(B) $1000$ points on the plane (and the two principal components) drawn from a 
more complex distribution, in this case the union of two multivariate normal 
distributions.
(C-F) A synthetic data set considered in~\cite{WRR03} to model 
oscillatory and exponentially decaying patterns of gene expression 
from~\cite{Cho_cellcycle_98}, as described in the text.
(C) Overlays of five noisy sine wave genes.
(D) Overlays of five noisy exponential genes.
(E) The first and second singular vectors of the data matrix (which account for 
$64\%$ of the variance in the data), along with the original sine pattern and 
exponential pattern that generated the data.
(F) Projection of the synthetic data on its top two singular vectors.
Although the data cluster well in the low-dimensional space, the top two 
singular vectors are completely artificial and do not offer insight into the 
oscillatory and exponentially decaying patterns that generated the data.}
\label{fig:hardcorepicture}
\end{center}
\end{figure}

Broadly similar issues arise in many other MMDS (modern massive data sets) application areas.  
In astronomy, for example, PCA and the SVD have been used directly for 
spectral classification~\cite{CSBKC95,CS99,MLTT01,YCSetc04}, 
to predict morphological types using galaxy spectra~\cite{FLM96}, 
to select quasar candidates from sky surveys~\cite{YCVetc04}, 
etc.~\cite{BLFetc04,BWSDY09,MKI10,BL10}.
Size is an issue, but so too is understanding the data~\cite{BDPS02,BB10}; 
and many of these studies have found that principal components of galaxy 
spectra (and their elements) correlate with various physical processes such 
as star formation (via absorption and emission line strengths of, 
\emph{e.g.}, the so-called H$\alpha$ spectral line) as well as with galaxy color and morphology.
In addition, there are many applications in scientific computing where 
low-rank matrices appear, \emph{e.g.}, fast numerical algorithms for solving
partial differential equations and evaluating potential fields rely on 
low-rank approximation of continuum operators~\cite{GR97,GH03}, and 
techniques for model reduction or coarse graining of multiscale physical 
models that involve rapidly oscillating coefficients often employ low-rank 
linear mappings~\cite{Engquist01}.
Recent work that has already used randomized low-rank matrix approximations
based on those reviewed here 
include~\cite{CDY07_JRNL,EY07,Mar08_TR,LLY11,LYMLYE11,CB10_DRAFT}.
More generally, many of the machine learning and data analysis applications 
cited below use these algorithms and/or greedy or heuristic variants of 
these algorithms for problems in diagnostic data analysis and for 
unsupervised feature selection for classification and clustering problems.

\subsection{Randomization as a resource}
\label{sxn:background1-resource}

The examples described in the previous subsection illustrate two common reasons 
for using randomization in the design of matrix algorithms for large-scale
data problems:
\begin{itemize}
\item
\textbf{Faster Algorithms.}
In some computation-bound applications, one simply wants \emph{faster 
algorithms} that return more-or-less the exact%
\footnote{Say, for example, that a numerically-stable answer that is precise to, say, 
$10$ digits of significance is more-or-less exact.  Exact answers are often 
impossible to compute numerically, in particular for continuous problems, as 
anyone who has studied numerical analysis knows.  Although they will not be 
the main focus of this review, such issues need to be addressed to provide 
high-quality numerical implementations of the randomized algorithms 
discussed here.}
answer.
In many of these applications, one thinks of the rank parameter $k$ as the
numerical rank of the matrix,%
\footnote{Think of the numerical rank of a matrix as being its ``true'' 
rank, up to issues associated with machine precision and roundoff error.  
Depending on the application, it can be defined in one of several related 
ways.  For example, if $\nu=\min\{m,n\}$, then, given a tolerance parameter 
$\varepsilon$, one way to define it is the largest $k$ such that 
$\sigma_{\nu-k+1} > \varepsilon \cdot \sigma_{\nu}$~\cite{GVL96}.}
and thus one wants to choose the error parameter $\epsilon$ such 
that the approximation is precise on the order of machine~precision.
\item
\textbf{Interpretable Algorithms.}
In other analyst-bound applications, one wants \emph{simpler 
algorithms or more-interpretable output} in order to obtain qualitative 
insight in order to pass to a downstream analyst.%
\footnote{The tension between providing more interpretable decompositions 
versus optimizing any single criterion---say, obtaining slightly better 
running time (in scientific computing) or slightly better prediction 
accuracy (in machine learning)---is well-known~\cite{CUR_PNAS}.  
It was illustrated most prominently recently by the Netflix Prize 
competition---whereas a half dozen or so base models captured the main 
ideas, the winning model was an ensemble of over $700$ base 
models~\cite{KBV09}.}
In these cases, $k$ is determined according to some domain-determined model 
selection criterion, in which case the difference between $\sigma_k$ and 
$\sigma_{k+1}$ may be small or it may be that $\sigma_{k+1} \gg 0$.%
\footnote{Recall that $\sigma_i$ is the $i^{th}$ singular value of the data matrix.}
Thus, it is acceptable (or even desirable since there is substantial noise in the data) 
if $\epsilon$ is chosen such that the approximation is much less precise.  
\end{itemize}
Thus, randomization can be viewed as a computational resource to be 
exploited in order to lead to ``better'' algorithms.
Perhaps the most obvious sense of better is faster running time, either
in worst-case asymptotic theory and/or numerical implementation---we will see below 
that both of these can be achieved.
But another sense of better is that the algorithm is more useful or easier 
to use---\emph{e.g.}, it may lead to more interpretable output, which is of 
interest in many data analysis applications where analyst time rather than 
just computational time is of interest.
Of course, there are other senses of better---\emph{e.g.}, the use of 
randomization and approximate computation can lead implicitly to 
regularization and more robust output;
randomized algorithms can be organized to exploit modern computational 
architectures better than classical numerical methods; and
the use of randomization can lead to simpler algorithms that are easier to 
analyze or reason about when applied in counterintuitive settings%
\footnote{Randomization can also be useful in less obvious 
ways---\emph{e.g.}, to deal with pivot rule issues in 
communication-constrained linear algebra~\cite{BDT08_TR}, or to achieve 
improved condition number properties in optimization 
applications~\cite{Asp08}.}%
---but these will not be the main focus of this review.

\section{Randomization applied to matrix problems}
\label{sxn:background2}

Before describing recently-developed randomized algorithms for 
least-squares  approximation and low-rank matrix approximation that 
underlie applications such as those described in 
Section~\ref{sxn:background1}, in this section we will provide a brief 
overview of the immediate precursors%
\footnote{Although this ``first-generation'' of randomized matrix 
algorithms was extremely fast and came with provable 
quality-of-approximation guarantees, most of these algorithms did 
\emph{not} gain a foothold in NLA and only heuristic variants of them were 
used in machine learning and data analysis applications.
Understanding them, though, was important in the development of a 
``second-generation'' of randomized matrix algorithms that were embraced by 
those communities.
For example, while in some cases these first-generation algorithms yield 
to a more sophisticated analysis and thus can be implemented directly, more 
often these first-generation algorithms represent a set of primitives that 
are more powerful once the randomness is decoupled from the linear algebraic 
structure.}
of that work.

\subsection{Random sampling and random projections}
\label{sxn:background2:misc-stuff}

Given an $m \times n$ matrix $A$, it is often of interest to sample 
randomly a small number of actual columns 
from that matrix.%
\footnote{Alternatively, one might be interested in sampling other things 
like elements~\cite{AM07_JRNL} or submatrices~\cite{FK99}.  Like the 
algorithms described in this section, these other sampling algorithms also 
achieve additive-error bounds.  We will not describe them in this review 
since, although of interest in TCS, they have not (yet?) gained traction in 
either NLA or in machine learning and data analysis applications.}
(To understand why sampling columns (or rows) from a matrix is of interest, 
recall that matrices are ``about'' their columns and 
rows~\cite{Strang88}---that is, linear combinations 
are taken with respect to them; one all but understands a given matrix if one 
understands its column space, row space, and null spaces; and understanding 
the subspace structure of a matrix sheds a great deal of light on the linear 
transformation that the matrix represents.)
A na\"{i}ve way to perform this random sampling would be to select 
those columns uniformly at random in i.i.d. trials.
A more sophisticated and much more powerful way to do this would be to 
construct an \emph{importance sampling distribution} $\{p_i\}_{i=1}^{n}$, 
and then perform the random sample according to~it.

To illustrate this importance sampling approach in a simple setting, 
consider the problem of approximating the product of two matrices.
Given as input any arbitrary $m \times n$ matrix $A$ and any arbitrary 
$n \times p$ matrix $B$:
\begin{itemize}
\item
Compute the importance sampling probabilities $\{p_i\}_{i=1}^{n}$, where 
\begin{equation}
p_i = \frac{||A^{(i)}||_2||B_{(i)}||_2 }{\sum_{i^\prime=1}^{n} ||A^{(i^\prime)}||_2||B_{(i^\prime)}||_2 }   .
\label{eqn:optimal_probs}
\end{equation}
\item
Randomly select (and rescale appropriately---if the $j^{th}$ column of $A$
is chosen, then scale it by $1/\sqrt{cp_{j}}$; see~\cite{dkm_matrix1} for 
details) 
$c$ 
columns of $A$ and the corresponding rows of $B$ (again rescaling in the 
same manner), thereby forming $m \times c$ and $c \times p$ matrices $C$ 
and $R$, respectively.
\end{itemize}
Two quick points are in order regarding the sampling process in this and 
other randomized algorithms to be described below.
First, the sampling here is with replacement.
Thus, in particular, if $c=n$ one does not necessarily recover the ``exact'' 
answer, but of course one should think of this algorithm as being most 
appropriate when $c \ll n$.
Second, if a given column-row pair is sampled, then it must be rescaled by a 
factor depending on the total number of samples to be drawn and the 
probability that that given column-row pair was chosen.
The particular form of $1/cp_{j}$ ensures that appropriate estimators are 
unbiased; see \cite{dkm_matrix1} for details.

This algorithm (as well as other algorithms that sample based on 
the Euclidean norms of the input matrices) requires just two passes over the 
data from external storage.  Thus, it can be implemented in 
pass-efficient~\cite{dkm_matrix1} or streaming~\cite{Muthu05} models of 
computation.
This algorithm is described in more detail in~\cite{dkm_matrix1},
where it is shown that Frobenius norm bounds of the~form
\begin{equation}
||AB-CR||_F \le \frac{O(1)}{\sqrt{c}} ||A||_F||B||_F  ,
\label{eqn:mm-bound-1}
\end{equation}
where $O(1)$ refers to some constant,
hold both in expectation and with high probability.
(Issues associated with potential failure probabilities, big-O notation, 
etc. for this pedagogical example are addressed in~\cite{dkm_matrix1}---%
these issues will be addressed in more detail for the algorithms of the 
subsequent sections.)
Moreover, 
if, instead of using 
importance sampling probabilities of the form~(\ref{eqn:optimal_probs}) 
that depend on both $A$ and $B$, one uses probabilities of the form 
\begin{equation}
p_i = ||A^{(i)}||_2^2/||A||_F^2
\label{eqn:optimal_probs_symmetric}
\end{equation}
that depend on only $A$ (or alternatively ones that depend only on
$B$), then (slightly weaker) bounds of the form~(\ref{eqn:mm-bound-1}) 
still~hold~\cite{dkm_matrix1}.
As we will see, this algorithm (or variants of it, as well as their associated 
bounds) is a primitive that underlies many of the randomized matrix 
algorithms that have been developed in recent years; for very recent
examples of this, see~\cite{MZ11,ESSWAI11}.

To gain insight into ``why'' this algorithm works, recall that the product 
$AB$ may be written as the outer product or sum of $n$ rank one matrices
$
AB = \sum_{t=1}^{n} A^{(t)} B_{(t)}   .
$
When matrix multiplication is formulated in this manner, a simple randomized 
algorithm to approximate the product matrix $AB$ suggests itself: 
randomly sample with replacement from the terms in the summation $c$ times, 
rescale each term appropriately, and output the sum of the scaled terms.
If $m=p=1$ then $A^{(t)},B_{(t)} \in \mathbb{R}$ and it is straightforward 
to show that this sampling procedure produces an unbiased estimator for the 
sum.
When the terms in the sum are rank one matrices, similar results hold.  
In either case, using importance sampling probabilities to exploit 
non-uniformity structure in the data---\emph{e.g.}, to bias the sample
toward ``larger'' terms in the sum, as~(\ref{eqn:optimal_probs}) 
does---produces an estimate with \emph{much} better variance properties.
For example, importance sampling probabilities of the 
form~(\ref{eqn:optimal_probs}) are optimal with respect to minimizing the
expectation of $||AB-CR||_F$.

The analysis of the Frobenius norm bound~(\ref{eqn:mm-bound-1}) is quite 
simple~\cite{dkm_matrix1}, using very simple linear algebra and only 
elementary probability, and it can be improved.
Most relevant for the randomized matrix algorithms of this review is the 
bound of~\cite{Rud99,RV07}, where much more sophisticated methods were used 
to shown that if $B=A^T$ is an $n \times k$ orthogonal matrix $Q$ 
(\emph{i.e.}, its $k$ columns consist of $k$ orthonormal vectors in 
$\mathbb{R}^{n}$),%
\footnote{In this case, $Q^TQ=I_k$, $\TNorm{Q}=1$, and $\FNormS{Q}=k$.  
Thus, the right hand side of~(\ref{eqn:mm-bound-1}) would be 
$O(1)\sqrt{k^2/c}$.
The tighter spectral norm bounds of the form~(\ref{eqn:mm-bound-2}) on the 
approximate product of two orthogonal matrices can be used to show that all 
the singular values of $Q^TS$ are nonzero and thus that rank 
is not lost---a crucial step in relative-error and high-precision 
randomized matrix algorithms.}
then, under certain assumptions satisfied by orthogonal matrices, spectral 
norm bounds of the~form 
\begin{equation}
\TNorm{I - CC^T} 
   = \TNorm{Q^TQ - Q^TSS^TQ} 
   \leq O(1)\sqrt{\frac{k \log c}{c}} 
\label{eqn:mm-bound-2}
\end{equation}
hold both in expectation and with high probability.
In this and other cases below, one can represent the random sampling operation 
with a \emph{random sampling matrix} $S$---\emph{e.g.}, if the random 
sampling is implemented by choosing $c$ columns, one in each of $c$ i.i.d. trials, then the $n \times c$ matrix 
$S$ has entries $S_{ij}=1/\sqrt{cp_i}$ if the $i^{th}$
column is picked in the $j^{th}$ independent trial, and $S_{ij}=0$ 
otherwise---in which case $C=AS$.

Alternatively, given an $m \times n$ matrix $A$, one might be interested in 
performing a random projection by post-multiplying $A$ by an 
$n \times \ell$ \emph{random projection matrix} $\Omega$, thereby selecting 
$\ell$ linear combinations of the columns of $A$.
There are several ways to construct such a matrix.
\begin{itemize}
\item
Johnson and Lindenstrauss consider an orthogonal projection onto a random 
$\ell$-dimensional space~\cite{JL84}, where $\ell=O(\log m)$, 
and~\cite{FH87} considers a projection onto $\ell$ random orthogonal vectors.
(In both cases, as well as below, the obvious scaling factor of 
$\sqrt{n/\ell}$ is needed.)
\item
\cite{IM98} and~\cite{DG02} 
choose
the entries of $\Omega$ as independent, 
spherically-symmetric random vectors, the coordinates of which are $\ell$ 
i.i.d. Gaussian $N(0,1)$ random variables. 
\item
\cite{Ach03_JRNL} 
chooses
the entries of $n \times \ell$ matrix $\Omega$ as $\{-1,+1\}$ random 
variables and also shows that a constant factor---up to $2/3$---of the 
entries of $\Omega$ can be set to $0$.
\item
\cite{AC06,AC06-JRNL09,Matousek08_RSA} choose $\Omega = D H P$, where $D$ 
is a $n \times n$ diagonal matrix, where each $D_{ii}$ is drawn 
independently from $\{-1,+1\}$ with probability $1/2$; $H$ is an 
$n \times n$ normalized Hadamard transform matrix, defined below; and 
$P$ is an $n \times \ell$ random matrix constructed as follows:
$P_{ij}=0$ with probability $1-q$, where $q=O(\log^2(m)/n)$;
and otherwise either $P_{ij}$ is drawn from a Gaussian distribution with 
an appropriate variance, or $P_{ij}$ is drawn independently from 
$\left\{-\sqrt{1/\ell q},+\sqrt{1/\ell q}\right\}$, each with probability $q/2$.
\end{itemize}
As with random sampling matrices, post-multiplication by the $n \times \ell$ 
random projection matrix $\Omega$ amounts to operating on the columns---in 
this case, choosing linear combinations of columns; and thus pre-multiplying 
by $\Omega^T$ amounts to choosing a small number of linear combinations of 
rows.
Note that, aside from the original constructions, these randomized linear 
mappings are \emph{not} random projections in the usual linear algebraic 
sense; but instead they satisfy certain approximate metric preserving 
properties satisfied by ``true'' random projections, and they are useful 
much more generally in algorithm design.
Vanilla application of such random projections has been used in data 
analysis and machine learning applications for clustering and classification 
of data~\cite{Kas98,BM01,FM03,FB03,GBN05,VW11}.

An important technical point is that the last Hadamard-based construction
is of particular importance for fast implementations (both in theory and in 
practice).  
Recall that the (non-normalized) $n \times n$ matrix of the Hadamard 
transform $H_n$ may be defined recursively as 
$$ H_n = \left[
\begin{array}{cc}
  H_{n/2} & H_{n/2} \\
  H_{n/2} & -H_{n/2}
\end{array}\right]   ,
\qquad \mbox{with} \qquad
H_2 = \left[
\begin{array}{cc}
  +1 & +1 \\
  +1 & -1
\end{array}\right] ,
$$
in which case the $n \times n$ normalized matrix of the Hadamard transform,
to be denoted by $H$ hereafter, is equal to $\frac{1}{\sqrt{n}}H_n$.
(For readers not familiar with the Hadamard transform, note that it 
is an orthogonal transformation, and one should think of it as a 
real-valued version of the complex Fourier transform.  Also, as defined 
here, $n$ is a power of $2$, but variants of this construction exist for 
other values of $n$.)
Importantly, applying the \emph{randomized Hadamard transform}, \emph{i.e.},
computing the product $xDH$ for any vector $x \in \mathbb{R}^n$ takes 
$O(n\log n)$ time (or even $O(n \log r )$ time if only $r$ elements in the 
transformed vector need to be accessed).
Applying such a \emph{structured random projection} was first proposed 
in~\cite{AC06,AC06-JRNL09}, it was first applied in the context of 
randomized matrix algorithms in~\cite{Sarlos06,DMMS07_FastL2_NM10}, and 
there has been a great deal of research in recent years on variants of this 
basic structured random projection that are better in theory or in 
practice~\cite{Matousek08_RSA,DMMS07_FastL2_NM10,LWFMRT07,RT08,AL08,LAS08,AMT10,AC10,DKT10,KN10_TR,KN10b_TR,AL11}.
For example, one could choose $\Omega = DHS$, where $S$ is a random sampling
matrix, as defined above, that represents the operation of uniformly 
sampling a small number of columns from the randomized Hadamard~transform.

Random projection matrices constructed with any of these methods exhibit a 
number of similarities, and the choice of which is appropriate depends on 
the application---\emph{e.g.}, a random unitary matrix or a matrix with 
i.i.d.  Gaussian entries may be the simplest conceptually or provide the 
strongest bounds; for TCS algorithmic applications, one may prefer a 
construction with i.i.d. Gaussian, $\{-1,+1\}$, etc. entries, or randomized 
Hadamard methods that are theoretically efficient to implement; for 
numerical implementations, one may prefer i.i.d. Gaussians if one is working 
with structured matrices that can be applied rapidly to arbitrary vectors 
and/or if one wants to be very aggressive in minimizing the oversampling 
factor needed by the algorithm, while one may prefer fast-Fourier-based 
methods that are better by constant factors than simple Hadamard-based 
constructions when working with arbitrary dense~matrices.

Intuitively, these random projection algorithms work since, if 
$\Omega^{(j)}$ is the $j^{th}$ column of $\Omega$, then $A\Omega^{(j)}$ is a 
random vector in the range of $A$.
Thus if one generates several such vectors, they will be 
linearly-independent (with very high probability, but perhaps poorly conditioned), and so one might hope 
to get a good approximation to the best rank-$k$ approximation to $A$ by 
choosing $k$ or slightly more than $k$ such vectors.
Somewhat more technically, one can prove that these random projection 
algorithms work by establishing variants of the basic 
\emph{Johnson-Lindenstrauss (JL) lemma}, which states:
\begin{itemize}
\item
Any set of $n$ points in a high-dimensional Euclidean space can be
embedded (via the constructed random projection) into an $\ell$-dimensional 
Euclidean space, where $\ell$ is logarithmic in $n$ and independent of the 
ambient dimension, such that all the pairwise distances are preserved to 
within an arbitrarily-small 
multiplicative
(or $1\pm\epsilon$)
factor~\cite{JL84,FH87,IM98,DG02,Ach03_JRNL,AC06,AC06-JRNL09,Matousek08_RSA}.
\end{itemize}
This result can then be applied to ${n \choose 2}$ vectors associated with 
the columns of $A$.
The most obvious (but not necessarily the best) such set of vectors is the 
rows of the original matrix $A$, in which case one shows that the random 
variable $||A_{(i)}\Omega-A_{(i')}\Omega||_2^2$ equals 
$||A_{(i)}-A_{(i')}||_2^2$ in expectation (which is usually easy to show) and that 
the variance is sufficiently small (which is usually harder to show).

By now, the relationship between sampling algorithms and projection 
algorithms should be clear.
Random sampling algorithms identify a coordinate-based non-uniformity 
structure, and they use it to construct an importance sampling distribution.
For these algorithms, the ``bad'' case is when that distribution is 
extremely nonuniform, \emph{i.e.}, when most of the probability mass is 
localized on a small number of columns.
This is the bad case for sampling algorithms in the sense that a na\"{i}ve
method like uniform sampling will perform poorly, while using an 
importance sampling distribution that provides a bias toward these columns 
will perform much better (at preserving distances, angles, subspaces, and
other quantities of interest).
On the other hand, random projections and randomized Hadamard transforms
destroy or ``wash out'' or uniformize that coordinate-based non-uniformity 
structure by rotating to a basis where the importance sampling distribution 
is very delocalized and thus where uniform sampling is nearly optimal (but 
by satisfying the above JL lemma they too preserve metric properties of 
interest).
For readers more familiar with Dirac $\delta$ functions and sinusoidal 
functions, recall that a similar situation holds---$\delta$ functions are 
extremely localized, but when they are multiplied by a Fourier transform, 
they are converted into delocalized sinusoids.
As we will see, making such structure explicit has numerous~benefits.

\subsection{Randomization for large-scale matrix problems}
\label{sxn:background2:previous}

Consider the following random projection algorithm that was introduced in 
the context of understanding the success of latent semantic 
analysis~~\cite{PRTV00}.
Given an $m \times n$ matrix $A$ and a rank parameter~$k$:
\begin{itemize}
\item
Construct an $n \times \ell$, with $\ell \geq \alpha\log m/\epsilon^2$ for 
some constant $\alpha$, random projection matrix $\Omega$, as in the previous 
subsection.
\item
Return $B=A\Omega$.
\end{itemize}
This algorithm, which amounts to choosing uniformly a small number $\ell$ 
of columns in a randomly rotated basis, was introduced in~\cite{PRTV00}, 
where it is proven that
\begin{equation}
||A-P_{B_{2k}}A||_F \le ||A-P_{U_k}A ||_F + \epsilon ||A||_F
\label{eqn:project-add-err}
\end{equation}
holds with high probability.
(Here, $B_{2k}$ is the best rank-$2k$ approximation to the matrix $B$; 
$P_{B_{2k}}$ is the projection matrix onto this $2k$-dimensional space; and
$P_{U_k}$ is the projection matrix onto $U_k$, the top $k$ left singular 
vectors of $A$.)
The analysis of this algorithm boils down to the JL ideas of the previous 
subsection applied to the rows of the input matrix $A$.
That is, the error $||A-P_{B_{2k}}A||_F$ boils down to the error incurred by
the best rank-$2k$ approximation plus an additional error term.
By applying the relative-error JL lemma to the rows of the matrix $A$, the 
additional error can be shown to be no greater than $\epsilon\FNorm{A}$.

Next, consider the following random sampling algorithm that was 
introduced in the context of clustering large data sets~\cite{DFKVV04_JRNL}.
Given an $m \times n$ matrix $A$ and a rank parameter $k$:
\begin{itemize}
\item
Compute the importance sampling probabilities $\{p_i\}_{i=1}^{n}$, where 
$p_i=||A^{(i)}||_2^2/||A||_F^2$.
\item
Randomly select and rescale $c = O(k \log k /\epsilon^2)$ columns of $A$
according to these probabilities to form the matrix $C$.
\end{itemize}
This algorithm was introduced in~\cite{DFKVV04_JRNL}, although a more 
complex variant of it appeared in~\cite{FKV04}.
The original analysis was extended and simplified in~\cite{dkm_matrix2}, 
where it is proven that
\begin{eqnarray}
\label{eqn:sample-add-err-2}
\TNorm{A-P_{C_k}A} &\le& \TNorm{A-P_{U_k}A } + \epsilon \FNorm{A}  \quad\mbox{and} \\
\FNorm{A-P_{C_k}A} &\le& \FNorm{A-P_{U_k}A } + \epsilon \FNorm{A}
\label{eqn:sample-add-err-F}
\end{eqnarray}
hold with high probability.
(Here, $C_k$ is the best rank-$k$ approximation to the matrix $C$, and 
$P_{C_k}$ is the projection matrix onto this $k$-dimensional space.)
This additive-error column-based matrix decomposition, as well as heuristic 
variants of it, has been applied in a range of data analysis 
applications~\cite{MMD06,Paschou07a,SXZF07_cmdmatrix,TPSYF08,PZW08}. 

Note that, in a theoretical sense, this and related random sampling 
algorithms that sample with respect to the Euclidean norms of the input 
columns are particularly appropriate for very large-scale settings.  
The reason is that these algorithms can be implemented efficiently in the
pass-efficient or streaming models of computation, in which 
the scarce computational resources are the number of passes over the data, 
the additional RAM space required, and the additional time required. 
See~\cite{dkm_matrix1,dkm_matrix2} for details about this.

The analysis of this random sampling algorithm boils down to an approximate matrix 
multiplication result, in a sense that will be constructive to consider 
in some~detail.
As an intermediate step in the proof of the previous results, that was 
made explicit in~\cite{dkm_matrix2}, it was shown that
\begin{eqnarray*}
\TNormS{A-P_{C_k}A} &\le& ||A-P_{U_k}A ||_2^2 + 2        \TNorm{AA^T-CC^T} \quad\mbox{and} \\
\FNormS{A-P_{C_k}A} &\le& ||A-P_{U_k}A ||_F^2 + 2\sqrt{k}\FNorm{AA^T-CC^T} .
\end{eqnarray*}
These bounds decouple the linear 
algebra from the randomization in the following sense: they hold for 
\emph{any} set of columns, \emph{i.e.}, for any matrix $C$, and the effect of the randomization enters only 
through the ``additional error'' term.
By using $p_i=||A^{(i)}||_2^2/||A||_F^2$ as the importance sampling 
probabilities, this algorithm is effectively saying that the relevant 
non-uniformity structure in the data is defined by the Euclidean norms of 
the original matrix.
(This may be thought to be an appropriate non-uniformity structure to 
identify since, \emph{e.g.}, it is one that can be identified and extracted
in two passes over the data from external storage.)
In doing so, this algorithm can take advantage of~(\ref{eqn:mm-bound-1}) to 
provide additive-error bounds.
A similar thing was seen in the analysis of the random projection 
algorithm---since the JL lemma was applied directly to the columns of $A$, 
additive-error bounds of the form~(\ref{eqn:project-add-err}) were 
obtained.

This is an appropriate point to pause to describe different notions of 
approximation that a matrix algorithm might provide.
In the theory of algorithms, bounds of the form provided by 
(\ref{eqn:sample-add-err-2}) and (\ref{eqn:sample-add-err-F}) are known as 
\emph{additive-error bounds}, the reason being that the ``additional'' 
error (above and beyond that incurred by the SVD) is an additive factor of 
the form $\epsilon$ times the scale $\FNorm{A}$.
Bounds of this form are very different and in general weaker than when the 
additional error enters as a multiplicative factor, such as when the error 
bounds are of the form 
$ ||A-P_{C_k}A|| \le f(m,n,k,\eta) ||A-P_{U_k}A ||$, where $f(\cdot)$ is 
some function and $\eta$ represents other parameters of the problem. 
Bounds of this type are of greatest interest when $f(\cdot)$ does not 
depend on $m$ or $n$, in which case they are known as a \emph{constant-factor 
bounds}, or when they depend on $m$ and $n$ only weakly.
The strongest bounds are when $f = 1+\epsilon$, for an error parameter 
$\epsilon$, \emph{i.e.}, when the bounds are of the form
$ ||A-P_{C_k}A|| \le (1+\epsilon) ||A-P_{U_k}A ||$.
These \emph{relative-error bounds} are the gold standard, and they provide a 
\emph{much} stronger notion of approximation than additive-error or 
weaker multiplicative-error bounds.
We will see bounds of all of these forms below.  

One application of these random sampling ideas that deserves special mention 
is when the input matrix $A$ is symmetric positive semi-definite. 
Such matrices are common in kernel-based machine learning, and sampling 
columns in this context often goes by the name \emph{the Nystr\"{o}m 
method}.
Originating in integral equation theory, the Nystr\"{o}m method was 
introduced into machine learning in~\cite{WS01} and it was analyzed and
discussed in detail in~\cite{dm_kernel_JRNL}.
Applications to large-scale machine learning problems 
include~\cite{TKR08,KMT09,KMT09c} and~\cite{ZTK08,LKL10,ZK10}, and
applications in statistics and signal processing 
include~\cite{PWT05,BW07_WKSHP,BW07_TR,SW08,BW08,BW09_PNAS,BW09_JRNL}.
As an example, the Nystr\"{o}m method can be used to provide an 
approximation to a matrix without even looking at the entire matrix---under 
assumptions on the input matrix, of course, such as that the leverage scores 
are approximately uniform.

\subsection{A retrospective and a prospective}

Much of the early work in TCS focused on randomly sampling columns according 
to an importance sampling distribution that depended on the Euclidean norm 
of those 
columns~\cite{FKV04,DFKVV04_JRNL,dkm_matrix1,dkm_matrix2,dkm_matrix3,RV07}.
This had the advantage of being ``fast,'' in the sense that it could be
performed in a small number of ``passes'' over that data from external 
storage, and also that additive-error quality-of-approximation bounds could 
be proved.
On the other hand, this had the disadvantage of being less 
immediately-applicable to scientific computing and large-scale data analysis 
applications.
At root, the reason is that these algorithms didn't highlight ``interesting'' or 
``relevant'' non-uniformity structure, which then led to bounds that were 
rather coarse.
For example, columns are easy to normalize and are 
often normalized during data preprocessing.
Even when not normalized, column norms can still be uninformative, as in 
heavy-tailed graph data,%
\footnote{By \emph{heavy-tailed graph}, consider a graph---or equivalently 
the adjacency matrix of such a graph---in which quantities such as the 
degree distribution or eigenvalue distribution decay in a heavy-tailed or 
power law manner.}
where they often correlate strongly with simpler statistics such as node 
degree.

Relatedly, bounds of the form~(\ref{eqn:mm-bound-1}) do not exploit the 
underlying vector space structure.
This is analogous to how the JL lemma was applied in the analysis of the 
random projection algorithm---by applying the JL lemma to the actual rows 
of $A$, as opposed to some other more refined vectors associated with the 
rows of $A$, the underlying vector space structure was ignored and only 
coarse additive-error bounds were obtained.
To obtain improvements and to bridge the gap between TCS, NLA, and data
applications, much finer bounds that take into account the vector space 
structure in a more refined way were needed.
To do so, it helped to identify more refined structural properties that 
decoupled the random matrix ideas from the underlying linear algebraic 
ideas---understanding this will be central to the next two sections.

Although these newer algorithms identified more refined structural 
properties, they have the same general structure as the original 
randomized matrix algorithms.
Recall that the general structure of the algorithms just reviewed is 
the~following.
\begin{itemize}
\item
Preprocess the input by: defining a non-uniformity structure over the 
columns of the input matrix; or performing a random projection/rotation to 
uniformize that structure.
\item
Draw a random sample of columns from the input matrix, either using the
non-uniformity structure as an importance sampling distribution to select 
actual columns, or selecting columns uniformly at random in the rotated 
basis.
\item
Postprocess the sample with a traditional deterministic NLA method.
\end{itemize}
In the above algorithms, the preprocessing was very fast, in that the 
importance sampling distribution could be computed by simply passing over 
the data a few times from external storage; and the postprocessing consists 
of just computing the best rank-$k$ or best rank-$2k$ approximation to the 
sample.
As will become clear below, by making one or both of these steps more 
sophisticated, very substantially improved results can be obtained, both in 
theory and in practice.
This can be accomplished, \emph{e.g.}, by using more sophisticated sampling 
probabilities or coupling the randomness in more sophisticated ways with 
traditional NLA methods, which in some cases will require additional 
computation.

\section{Randomized algorithms for least-squares approximation}
\label{sxn:least-squares}

In this section and the next, we will describe randomized matrix algorithms
for the least-squares approximation and low-rank approximation problems.
The analysis of low-rank matrix approximation algorithms described in 
Section~\ref{sxn:low-rank} boils down to a randomized approximation 
algorithm for the least-squares approximation 
problem~\cite{DMM08_CURtheory_JRNL,CUR_PNAS,BMD09_CSSP_SODA,BMD08_CSSP_TRv2}.
For this reason, for pedagogical reasons, and due to the fundamental 
importance of the least-squares problem more generally, randomized algorithms for the least-squares 
problem will be the topic of this section.

\subsection{Different perspectives on least-squares approximation}
\label{sxn:least-squares:perspectives}

Consider the problem of finding a vector $x$ such that $Ax \approx b$, 
where the rows of $A$ and elements of $b$ correspond to constraints and the 
columns of $A$ and elements of $x$ correspond to variables.
In the very \emph{overconstrained least-squares approximation problem}, 
where the $m \times n$ matrix $A$ has $m \gg n$,%
\footnote{In this section only, we will assume that $m \gg n$.}
there is in general no vector $x$ such that $Ax=b$, and it is 
common to quantify ``best'' by looking for a vector $x_{opt}$ such that the 
Euclidean norm of the residual error is small, \emph{i.e.}, to solve 
the least-squares (LS) approximation problem
\begin{eqnarray}
\label{eqn:orig_ls_prob}
x_{opt} &=& \mbox{argmin}_x ||Ax-b||_2  .
\end{eqnarray}
This problem is ubiquitous in applications, where it often arises from 
fitting the parameters of a model to experimental data, and it is central 
to theory.
Moreover, it has a natural statistical interpretation as providing the best 
estimator within a natural class of estimators; and it has a natural 
geometric interpretation as fitting the part of the vector $b$ that resides 
in the column space of $A$.
From the viewpoint of low-rank matrix approximation, this LS problem arises 
since measuring the error with a Frobenius or spectral norm, as 
in~(\ref{eqn:error-measure}), amounts to choosing columns that are ``good'' 
in a least squares sense.%
\footnote{Intuitively, these low-rank approximation algorithms find columns 
that provide a space that is good in a least-squares sense, when compared 
to the best rank-$k$ space, at reconstructing every row of the input 
matrix.  Thus, the reason for the connection is that the merit function 
that describes the quality of those algorithms is typically reconstruction 
error with respect to the spectral or Frobenius norm.} 

There are a number of different perspectives one can adopt on this LS 
problem.
Two major perspectives of central importance in this review are the 
following.
\begin{itemize}
\item
\textbf{Algorithmic perspective.}
From an algorithmic perspective, the relevant question is: how long does it 
take to compute $x_{opt}$?
The answer to this question is that is takes $O(mn^2)$ time~\cite{GVL96}.
This can be accomplished with one of several algorithms---with the Cholesky 
decomposition (which is good if $A$ has full column rank and is very 
well-conditioned); or with a variant of the QR decomposition (which is 
somewhat slower, but more numerically stable); or by computing the full 
SVD $A=U \Sigma V^T$ (which is often, but certainly not always, overkill, 
but which can be easier to explain%
\footnote{The SVD has been described as the ``Swiss Army Knife'' of 
NLA~\cite{MMDS06summary}.  That is, given it, one can do nearly anything one
wants, but it is almost always overkill, as one rarely if ever needs its full
functionality.  Nevertheless, for pedagogical reasons, since other 
decompositions are typically better in terms of running time by only constant 
factors, and since numerically-stable algorithms for these latter 
decompositions can be quite complex, it is convenient to formulate results in
terms of the SVD and the best rank-$k$ approximation to the SVD.}), 
and letting $x_{opt} = V \Sigma^{+} U^T b$.
Although these methods differ a great deal in practice and in terms of 
numerical implementation, asymptotically each of these methods takes a 
constant times $mn^2$ time to compute a vector $x_{opt}$.
Thus, from an algorithmic perspective, a natural next question to ask is:
can the general LS problem be solved, \emph{either exactly or approximately}, 
in $o(mn^2)$ time,%
\footnote{Formally, $f(n) = o(g(n))$ as $n \rightarrow \infty$ means that for 
every positive constant $\varepsilon$ there exists a constant $N$ such that
$|f(n)| \le \varepsilon|g(n)|$, for all $n \ge N$.  Informally, it means that 
$f(n)$ grows more slowly than $g(n)$.  Thus, if the running time of an 
algorithm is $o(mn^2)$ time, then it is asymptotically faster than any 
(arbitrarily small) constant times $mn^2$.}
with no assumptions at all on the input data?
\item
\textbf{Statistical perspective.}
From a statistical perspective, the relevant question is: when is computing 
the $x_{opt}$ the right thing to do?
The answer to this question is that this LS optimization is the right problem to
solve when the relationship between the ``outcomes'' and ``predictors'' is 
roughly linear and when the error processes generating the data are ``nice'' 
(in the sense that they have mean zero, constant variance, are uncorrelated, 
and are normally distributed; or when we have adequate sample size to rely 
on large sample theory)~\cite{ChatterjeeHadi88}.
Thus, from a statistical perspective, a natural next question to ask is:
what should one do when the assumptions underlying the use of LS methods 
are not satisfied or are only imperfectly~satisfied?
\end{itemize}
Of course, there are also other perspectives that one can adopt.
For example, from a numerical perspective, whether the algorithm is 
numerically stable, issues of forward versus backward stability, condition
number issues, and whether the algorithm takes time that is a large or small 
constant multiplied by $\min\{mn^2,m^2n\}$ are of paramount importance.

When adopting the statistical perspective, it is common to check the extent 
to which the assumptions underlying the use of LS have been satisfied.
To do so, it is common to assume that $b=Ax+\varepsilon$, where $b$ is the 
response, the columns $A^{(i)}$ are the carriers, and $\varepsilon$ is a 
``nice'' error process.%
\footnote{This is typically done by assuming that the error process 
$\varepsilon$ consists of i.i.d. Gaussian entries.  As with the 
construction of random projections in 
Section~\ref{sxn:background2:misc-stuff}, numerous other constructions are 
possible and will yield similar results.  
Basically, it is important that no one or small number of data points has a
particularly large influence on the LS fit, in order to ensure that 
techniques from large-sample theory like measure concentration apply.}
Then $x_{opt}=(A^TA)^{-1}A^Tb$, and thus $\hat{b}=Hb$, where the projection 
matrix onto the column space of $A$, 
$$
H=A(A^TA)^{-1}A^T  ,
$$ is the so-called \emph{hat matrix}.
It is known that $H_{ij}$ measures the influence or statistical leverage 
exerted on the prediction $\hat{b}_i$ by the observation 
$b_j$~\cite{HW78,ChatterjeeHadi88,CH86,VW81,ChatterjeeHadiPrice00}.
Relatedly, if the $i^{th}$ diagonal element of $H$ is particularly large 
then the $i^{th}$ data point is particularly sensitive or influential in 
determining the best LS fit, thus justifying the interpretation of the 
elements $H_{ii}$ as \emph{statistical leverage scores}~\cite{CUR_PNAS}.
These leverage scores have been used extensively in classical 
regression diagnostics to identify potential outliers by, \emph{e.g.}, 
flagging data points with leverage score greater than $2$ or $3$ times the 
average value in order to be investigated as errors or potential 
outliers~\cite{ChatterjeeHadi88}.
Moreover, in the context of recent graph theory applications, this concept 
has proven useful under the name of graph 
\emph{resistance}~\cite{SS08a_STOC}; and, for the matrix problems considered 
here, some researchers have used the term \emph{coherence} to measure the 
degree of non-uniformity of these statistical leverage 
scores~\cite{CR09,TalRos10,MTJ11_TR}.

In order to compute these quantities \emph{exactly}, recall that if $U$ is 
\emph{any} orthogonal matrix spanning the column space of $A$, then 
$H=P_U=UU^T$ and thus 
$$
H_{ii}=||U_{(i)}||_2^2  , 
$$
\emph{i.e.}, the statistical leverage scores equal the Euclidean norm of 
the \emph{rows} of any such matrix $U$~\cite{DMM08_CURtheory_JRNL,CUR_PNAS}.
Recall Definition~\ref{def:lev-scores} from 
Section~\ref{sxn:background1-brief}.
(Clearly, the columns of such a matrix $U$ are orthonormal, but the rows of 
$U$ in general are not---they can be uniform if, \emph{e.g.}, $U$ consists 
of columns from a truncated Hadamard matrix; or extremely nonuniform if, 
\emph{e.g.}, the columns of $U$ come from a truncated identity matrix; or 
anything in between.)
More generally, and of interest for the low-rank matrix approximation 
algorithms in Section~\ref{sxn:low-rank}, the \emph{statistical leverage 
scores relative to the best rank-$k$ approximation to $A$} are the diagonal 
elements of the projection matrix onto the best rank-$k$ approximation 
to~$A$. 
Thus, they can be computed from 
$$
(P_{U_k})_{ii} = ||U_{k,(i)}||_2^2  , 
$$
where $U_{k,(i)}$ is the $i^{th}$ row of any
matrix spanning the space spanned by the top $k$ left singular vectors 
of~$A$ (and similarly for the right singular subspace if columns rather than 
rows are of interest).

In many diagnostic applications, \emph{e.g.}, when one is interested in 
exploring and understanding the data to determine what would be the 
appropriate computations to perform, the time to compute or approximate 
$(P_{U})_{ii}$ or $(P_{U_k})_{ii}$ is not the bottleneck.
On the other hand, in cases where this time is the bottleneck, an algorithm 
we will describe in Section~\ref{sxn:least-squares:faster-th:rand-samp}
will provide very fine approximations to all these 
leverage scores in time that is qualitatively faster than that required to 
compute them exactly.

\subsection{A simple algorithm for approximating least-squares approximation}
\label{sxn:least-squares:algorithm}

Returning to the algorithmic perspective, consider the following random 
sampling algorithm for the LS approximation 
problem~\cite{DMM06,DMM08_CURtheory_JRNL}.
Given a very overconstrained LS problem, where the input matrix
$A$ and vector $b$ are \emph{arbitrary}, but $m \gg n$:
\begin{itemize}
\item
Compute the normalized statistical leverage scores $\{p_i\}_{i=1}^{m}$, 
\emph{i.e.}, compute $p_i = ||U_{(i)}||_2^2/n$, where $U$ is the 
$m \times n$ matrix consisting of the left singular vectors of $A$.%
\footnote{Stating this in terms of the singular vectors is a convenience, 
but it can create confusion.  In particular, although computing the SVD 
is sufficient, it is by no means necessary---here, $U$ can be \emph{any} 
orthogonal matrix spanning the column space of $A$~\cite{CUR_PNAS}.  
Moreover, these probabilities are robust, in that any probabilities that 
are close to the leverage scores will suffice; see~\cite{dkm_matrix1} for 
a discussion of approximately-optimal sampling probabilities.
Finally, as we will describe below, these probabilities can be approximated
quickly, \emph{i.e.}, more rapidly than the time needed to compute a basis
exactly, or the matrix can be preprocessed quickly to make them nearly 
uniform.}
\item
Randomly sample and rescale%
\footnote{Recall from the discussion in 
Section~\ref{sxn:background2:misc-stuff} that each sampled row should be 
rescaled by a factor of $1/r p_i$.  Thus, it is these sampled-and-rescaled 
rows that enter into the subproblem that this algorithm constructs and solves.}
$r=O(n \log n /\epsilon^2 )$ constraints, 
\emph{i.e.}, rows of $A$ and the corresponding elements of $b$, using these
scores as an importance sampling distribution.
\item
Solve the induced subproblem
$ \tilde{x}_{opt} = \mbox{argmin}_x || S Ax - S b||_2  $, where the 
$r \times m$ matrix $S$ represents the sampling-and-rescaling operation.
\end{itemize}

\noindent
The induced subproblem can be solved using any appropriate direct or 
iterative LS solver as a black box.
For example, one could compute the solution directly via the generalized 
inverse as $ \tilde{x}_{opt} = \left(SA\right)^{\dagger}Sb $, which would 
take $O(rn^2)$ time~\cite{GVL96}.
Alternatively, one could use iterative methods such as the Conjugate 
Gradient Normal Residual method, which can produce an 
$\epsilon$-approximation to the optimal solution of the sampled problem in 
$O(\kappa(SA) rn \log(1/\epsilon) )$ time, where $\kappa(SA)$ is the 
condition number of $SA$~\cite{GVL96}.
As stated, this algorithm will compute all the statistical leverage scores
exactly, and thus it will not be faster than the traditional algorithm for
the LS problem---importantly, we will see below how to get around 
this~problem.

Since this overconstrained%
\footnote{Not surprisingly, similar ideas apply to underconstrained LS 
problems, where $m \ll n$, and where the goal is to compute the 
minimum-length solution.  In this case, one randomly samples columns, and 
one uses a somewhat more complicated procedure to construct an approximate 
solution, for which relative-error bounds also hold.  In fact, as we will 
describe in Section~\ref{sxn:least-squares:faster-th:rand-samp},
the quantities $\{p_i\}_{i=1}^{m}$ can be approximated in $o(mn^2)$ 
time by relating them to an underconstrained LS approximation problem and 
running such a procedure.}
LS algorithm samples constraints and not variables, the 
dimensionality of the vector $\tilde{x}_{opt}$ that solves the subproblem is 
the same as that of the vector $x_{opt}$ that solves the original problem.
The algorithm just presented is described in more detail 
in~\cite{DMM06,DMM08_CURtheory_JRNL,DMMS07_FastL2_NM10},
where it is shown that relative-error bounds of the form 
\begin{eqnarray}
\label{eqn:ls-bound-eq1}
||b-A\tilde{x}_{opt}||_2 
   &\leq& (1+\epsilon) ||b-Ax_{opt}||_2  \hspace{2mm} \mbox{ and } \\
\label{eqn:ls-bound-eq2}
||x_{opt} - \tilde{x}_{opt}||_2
   &\leq& \sqrt{\epsilon} \left( \kappa(A)\sqrt{\gamma^{-2}-1} \right) ||x_{opt}||_2  
\end{eqnarray}
hold, where $\kappa(A)$ is the condition number of $A$ and where 
$\gamma = ||UU^Tb||_2/||b||_2$ is a parameter defining the amount of the 
mass of $b$ inside the column space of $A$.%
\footnote{We should reemphasize that such relative-error bounds, either on the 
optimum value of the objective function, as in~(\ref{eqn:ls-bound-eq1}) and 
as is more typical in TCS, or on the vector or ``certificate'' achieving 
that optimum, as in~(\ref{eqn:ls-bound-eq2}) and as is of greater interest 
in NLA, provide an \emph{extremely} strong notion of approximation.}
Of course, there is randomization inside this algorithm, and it is possible
to flip a fair coin ``heads'' $100$ times in a row.
Thus, as stated, with $r=O(n \log n/\epsilon^2)$, the randomized 
least-squares algorithm just described might fail with a probability 
$\delta$ that is no greater than a constant (say $1/2$ or $1/10$ or $1/100$, 
depending on the (modest) constant hidden in the $O(\cdot)$ notation) that 
is independent of $m$ and~$n$.
Of course, using standard methods~\cite{MotwaniRaghavan95}, this failure 
probability can easily be improved to be an arbitrarily small $\delta$ 
failure probability.
For example, this holds if $r = O(n \log(n)\log(1/\delta)/\epsilon^2)$ in the above 
algorithm;
alternatively, it holds if one repeats the above algorithm $O(\log(1/\delta))$ times 
and keeps the best of the results. 

As an aside, it is one thing for TCS researchers, well-known to be cavalier 
with respect to constants and even polynomial factors, to make such 
observations about big-O notation and the failure probability, but by now 
these facts are acknowledged more generally.
For example, a recent review of coupling randomized low-rank matrix 
approximation algorithms with traditional NLA methods~\cite{HMT09_SIREV} starts 
with the following observation.
``Our experience suggests that many practitioners of scientific computing
view randomized algorithms as a desperate and final resort. Let us address
this concern immediately.  Classical Monte Carlo methods are highly 
sensitive to the random number generator and typically produce output with 
low and uncertain accuracy.  In contrast, the algorithms discussed herein 
are relatively insensitive to the quality of randomness and produce highly 
accurate results. The probability of failure is a user-specified parameter 
that can be rendered negligible (say, less than $10^{-15}$) with a nominal 
impact on the computational resources~required.''

Finally, it should be emphasized that modulo this failure probability 
$\delta$ that can be made arbitrarily small without adverse effect and an 
error $\epsilon$ that can also be made arbitrarily small, the above 
algorithm (as well as the basic low-rank matrix approximation algorithms 
of Section~\ref{sxn:low-rank} that boil down to this randomized approximate 
LS algorithm) returns an answer $\tilde{x}_{opt}$ that satisfies bounds of 
the form~(\ref{eqn:ls-bound-eq1}) and~(\ref{eqn:ls-bound-eq2}), independent 
of \emph{any} assumptions at all on the input matrices $A$ and~$b$.

\subsection{A basic structural result}
\label{sxn:least-squares:structural}

What the above random sampling algorithm highlights is that the 
``relevant non-uniformity structure'' that needs to be dealt with in order 
to solve the LS approximation problem is defined by the statistical leverage 
scores.
(Moreover, since the randomized algorithms for low-rank matrix 
approximation to be described in Section~\ref{sxn:low-rank} boil down to a 
least-squares problem, an analogous statement is true for these low-rank 
approximation algorithms.)
To see ``why'' this works, it is helpful to identify a deterministic 
structural condition sufficient for relative-error approximation---doing so
decouples the linear algebraic part of the analysis from the randomized 
matrix part.
This condition, implicit in the analysis 
of~\cite{DMM06,DMM08_CURtheory_JRNL}, was made explicit 
in~\cite{DMMS07_FastL2_NM10}.

Consider preconditioning or premultiplying the input matrix $A$ and the 
target vector $b$ with some \emph{arbitrary} matrix $Z$,
and consider the solution to the LS approximation problem
\begin{eqnarray}
\label{eqn:orig_ls_prob_Xrotated}
\tilde{x}_{opt} &=& \mbox{argmin}_x ||Z(Ax-b)||_2   .
\end{eqnarray}
Thus, for the random sampling algorithm described in 
Section~\ref{sxn:least-squares:algorithm}, the matrix $Z$ is a 
carefully-constructed data-dependent random sampling matrix, and for the 
random projection algorithm below it is a data-independent random 
projection, but more generally it could be \emph{any} arbitrary 
matrix~$Z$.
Recall that the SVD of $A$ is $A=U_A\Sigma_AV_A^T$; 
and, for notational simplicity, let 
$b^{\perp} = U_A^{\perp}{U_A^{\perp}}^{T}b$ denote the  part of the right 
hand side vector $b$ lying outside of the column space of $A$.
Then, the following structural condition holds.
\begin{itemize}
\item
\textbf{Structural condition underlying the randomized least-squares 
algorithm.}
Under the assumption that $Z$ satisfies the following two conditions:
\begin{eqnarray}
\label{eqn:lemma1_ass1}
& & \sigma_{min}^2 \left( ZU_A \right) \ge 1/\sqrt{2} \mbox{; and}  \\
\label{eqn:lemma1_ass2} & &
||U_A^TZ^TZb^{\perp}||_2^2
      \le \frac{\epsilon}{2} ||Ax_{opt}-b||_2^2  ,
\end{eqnarray}
for some $\epsilon \in (0,1)$,
the solution vector $\tilde{x}_{opt}$ to the LS approximation 
problem~(\ref{eqn:orig_ls_prob_Xrotated})
satisfies relative-error bounds of the form~(\ref{eqn:ls-bound-eq1})
and~(\ref{eqn:ls-bound-eq2}).
\end{itemize}
In this condition, the particular constants $1/\sqrt{2}$ and $1/2$ clearly 
don't matter---they have been chosen for ease of comparison 
with~\cite{DMMS07_FastL2_NM10}.  
Also, recall that $||b^{\perp}||_2=||Ax_{opt}-b||_2$.
Several things should be noted about these two structural conditions:
\begin{itemize}
\item
First, since $\sigma_i(U_A)=1$, for all $i \in [n]$, 
Condition~(\ref{eqn:lemma1_ass1}) indicates that the rank of $ZU_A$ is the
same as that of $U_A$.
Note that although Condition~(\ref{eqn:lemma1_ass1}) only states that 
$\sigma_i^2(ZU_A)\geq 1/\sqrt{2}$, for all $i \in [n]$, for the randomized 
algorithm of Section~\ref{sxn:least-squares:algorithm}, it will follow that 
$\left|1-\sigma_i^2(ZU_A)\right| \le 1-2^{-1/2}$, for all $i \in [n]$.
Thus, one should think of Condition~(\ref{eqn:lemma1_ass1}) as stating 
that $ZU_A$ is an approximate~isometry. 
Thus, this expression can be bounded with the approximate matrix multiplication 
spectral norm bound of~(\ref{eqn:mm-bound-2}).
\item
Second, since before preprocessing by $Z$, 
$b^{\perp}=U_A^{\perp}{U_A^{\perp}}^{T}b$ is clearly orthogonal to $U_A$, 
Condition~(\ref{eqn:lemma1_ass2}) simply states that after preprocessing
$Zb^{\perp}$ remains approximately orthogonal
to $ZU_A$. 
Although Condition~(\ref{eqn:lemma1_ass2}) depends on the right hand 
side vector $b$, the randomized algorithm of 
Section~\ref{sxn:least-squares:algorithm} satisfies it without using any 
information from $b$.
The reason for not needing information from $b$ is that the left hand side 
of Condition~(\ref{eqn:lemma1_ass2}) is of the form of an approximate 
product of two different matrices---where for the randomized algorithm of
Section~\ref{sxn:least-squares:algorithm} the importance sampling 
probabilities depend only on one of the two matrices---and thus one can apply an
approximate matrix multiplication bound of the form~(\ref{eqn:mm-bound-1}).
\item
Third, as the previous two points indicate, 
Condition~(\ref{eqn:lemma1_ass1}) and~(\ref{eqn:lemma1_ass2}) 
both boil down to the problem of approximating the product of two matrices,
and thus the algorithmic primitives on approximate matrix multiplication 
from Section~\ref{sxn:background2:misc-stuff} will be useful, either 
explicitly or within the analysis.
\end{itemize}
It should be emphasized that there is no randomization in these two 
structural conditions---they are 
deterministic statements about an arbitrary matrix $Z$ that represent a 
structural condition sufficient for relative-error approximation.
Of course, if $Z$ happens to be a random matrix, \emph{e.g.}, representing 
a random projection or a random sampling process, then 
Conditions~(\ref{eqn:lemma1_ass1}) or~(\ref{eqn:lemma1_ass2}) may fail to 
be satisfied---but, conditioned on their being satisfied, the 
relative-error bounds of the form~(\ref{eqn:ls-bound-eq1}) 
and~(\ref{eqn:ls-bound-eq2}) follow.
Thus, the effect of randomization enters only via $Z$, and it is decoupled 
from the linear algebraic structure.

\subsection{Making this algorithm fast---in theory}
\label{sxn:least-squares:faster-th}

Given this structural insight, what one does with it depends on the 
application.
In some cases, as in certain genetics 
applications~\cite{Paschou07b,Paschou08a,CUR_PNAS} or when solving the Column Subset Selection Problem as
described in Section~\ref{sxn:low-rank:exactlyk}, the time for the 
computation of the leverage scores is not the bottleneck, and thus they may 
be computed with the traditional procedure. 
In other cases, as when dealing with not-extremely-sparse random matrices 
or other situations where it is expected or hoped that no single data point 
is particularly influential, it is assumed that the scores are exactly or 
approximately uniform, in which case uniform sampling is appropriate.
In general, however, the simple random sampling algorithm of 
Section~\ref{sxn:least-squares:algorithm} requires the computation of the
normalized statistical leverage scores, and thus it runs in $O(mn^2)$ time.

In this subsection, we will describe \emph{two} ways to speed up the random 
sampling algorithm of Section~\ref{sxn:least-squares:algorithm} so that it 
runs in $o(mn^2)$ time for arbitrary~input.
The first way involves preprocessing the input with a randomized Hadamard 
transform and then calling the algorithm of 
Section~\ref{sxn:least-squares:algorithm} with uniform sampling probabilities.
The second way involves computing a quick approximation to the statistical 
leverage scores and then using those approximations as the importance sampling 
probabilities in the algorithm of Section~\ref{sxn:least-squares:algorithm}. 
Both of these methods provide fast and accurate algorithms in 
theory, and straightforward extensions of them provide very fast and very 
accurate algorithms in practice.

\subsubsection{A fast random projection algorithm for the LS problem}
\label{sxn:least-squares:faster-th:rand-proj}

Consider the following structured random projection algorithm for 
approximating the solution to the LS approximation problem.
\begin{itemize}
\item
Premultiply $A$ and $b$ with an $n \times n$ randomized Hadamard 
transform~$HD$.
\item
Uniformly sample roughly 
$r=O\left(n(\log n)(\log m)+\frac{n (\log m)}{\epsilon}\right)$ rows from 
$HDA$ and the corresponding elements from $HDb$.
\item
Solve the induced subproblem
$ \tilde{x}_{opt} = \mbox{argmin}_x || SHD Ax - SHD b||_2  $, where the 
$r \times m$ matrix $S$ represents the sampling operation.
\end{itemize}
This algorithm, which first preprocesses the input with a structured random 
projection and then solves the induced subproblem, as well as a variant 
of it that uses the original ``fast'' 
Johnson-Lindenstrauss transform~\cite{AC06,AC06-JRNL09,Matousek08_RSA}, was 
presented in~\cite{Sarlos06,DMMS07_FastL2_NM10}, (where a precise 
statement of $r$ is given and) where it is shown that 
relative-error bounds of the form~(\ref{eqn:ls-bound-eq1}) 
and~(\ref{eqn:ls-bound-eq2}) hold.

To understand this result, recall that premultiplication by a randomized 
Hadamard transform is a unitary operation and thus does not change the 
solution; and that from the SVD of $A$ and of $HDA$ it follows that 
$U_{HDA}=HDU_A$.
Thus, the ``right'' importance sampling distribution for the preprocessed 
problem is defined by the diagonal elements of the projection matrix onto 
the span of $HDU_A$.
Importantly, application of such a Hadamard transform tends to 
``uniformize'' the leverage scores,%
\footnote{As noted above, this is for very much the same reason that a 
Fourier matrix delocalizes a localized $\delta$-function; and it also holds 
for the application of an unstructured random orthogonal matrix or random 
projection.}
in the sense that all the leverage scores associated with $U_{HDA}$ are (up 
to logarithmic fluctuations) uniform~\cite{AC06,DMMS07_FastL2_NM10}.
Thus, uniform sampling probabilities are optimal, up to a logarithmic factor 
which can be absorbed into the sampling complexity.
Overall, this relative-error approximation algorithm for the LS problem 
run in $o(mn^2)$ time~\cite{Sarlos06,DMMS07_FastL2_NM10}---essentially 
$O\left( mn \log(n/\epsilon) + \frac{n^3 \log^2 m}{\epsilon} \right)$ time, 
which is much less than $O(mn^2)$ when $m \gg n$.
Although the ideas underlying the Fast Fourier Transform have been around 
and used in many applications for a long time~\cite{CT65,GR87}, they were 
first applied in the context of randomized matrix algorithms only 
recently~\cite{AC06,Sarlos06,DMMS07_FastL2_NM10}.

The $o(mn^2)$ running time is most interesting when the input matrix to 
the overconstrained LS problem is dense; if the input matrix is sparse, then 
a more appropriate comparison might have to do with the number of nonzero 
entries.
In general, however, random Gaussian projections and randomized
Hadamard-based transforms tend not to respect sparsity.
In some applications, \emph{e.g.}, the algorithm of~\cite{MSM11_TR} that 
is described in Section~\ref{sxn:least-squares:faster-pr} and that 
automatically speeds up on sparse matrices and with fast linear operators, 
as well as several of the randomized algorithms for low-rank approximation 
to be described in Section~\ref{sxn:low-rank:proj}, this can be worked around.
In general, however, the question of using sparse random projections and 
sparsity-preserving random projections is an active topic of 
research~\cite{DKT10,KN10_TR,KN10b_TR,GI10}.

\subsubsection{A fast random sampling algorithm for the LS problem}
\label{sxn:least-squares:faster-th:rand-samp}

Next, consider the following algorithm which takes as input an arbitrary 
$m \times n$ matrix $A$, with $m \gg n$, and which returns as output
approximations to all $m$ of the statistical leverage scores of $A$.
\begin{itemize}
\item
Premultiply $A$ by a structured random projection, \emph{e.g.}, 
$\Omega_1 = S H D$ from Section~\ref{sxn:background2:misc-stuff}, which 
represents uniformly sampling roughly $r_1=O(m \log n /\epsilon)$ rows from 
a randomized Hadamard transform.
\item
Compute the $m \times r_2$ matrix $X = A (\Omega_1 A )^{\dagger} \Omega_2 $, 
where $\Omega_2$ is an $r_1 \times r_2$ unstructured random projection matrix 
and where the dagger represents the Moore-Penrose pseudoinverse.
\item
For each $i = 1,\ldots,m$,
compute and return $\tilde{\ell}_i = ||\tilde{X}_{(i)}||_2^2$.
\end{itemize}
This algorithm was introduced in~\cite{DMMW11_TR}, based 
on ideas in~\cite{Malik10_TR}.
In~\cite{DMMW11_TR}, it is proven that 
$$
|\ell_i- \tilde{\ell}_i | \le \epsilon \ell_i 
\quad 
\mbox{for all } 
\quad 
i = 1,\ldots,m   ,
$$
where $\ell_i$
are the statistical leverage scores of Definition~\ref{def:lev-scores}.
That is, this algorithm returns a relative-error 
approximation to \emph{every} one of the $m$ statistical leverage scores.
Moreover, in~\cite{DMMW11_TR} it is also proven that this algorithm runs in 
$o(mn^2)$ time---due to the structured random projection in the first step, 
the running time is basically the same time as that of the fast random projection 
algorithm described in the previous subsection.%
\footnote{Recall that since the coherence of a matrix is equal 
to the largest leverage score, this algorithm also computes a relative-error
approximation to the coherence of the matrix in $o(mn^2)$ time---which 
is qualitatively faster than the time needed to compute an orthogonal basis
spanning the original matrix.} 
In addition, given an arbitrary rank parameter $k$ and an arbitrary-sized 
$m \times n$ matrix $A$, this algorithm can be extended to approximate the 
leverage scores relative to the best rank-$k$ approximation to $A$ in 
roughly $O(mn k)$ time.
See~\cite{DMMW11_TR} for a discussion of the technical issues associated 
with this.  In particular, note that the problem of asking for approximations 
to the leverage scores relative to the best rank-$k$ space of a matrix is 
ill-posed; and thus one must replace it by computing approximations to the 
leverage scores for some space that is a good approximation to the best 
rank-$k$~space.

Within the analysis, this algorithm for computing rapid approximations to 
the statistical leverage scores of an arbitrary matrix basically boils down 
to an \emph{under}constrained LS problem, in which a structured random 
projection is carefully applied, in a manner somewhat analogous to the fast 
\emph{over}constrained LS random projection algorithm in the previous 
subsection.
In particular, let $A$ be an $m \times n$ matrix, with $m \ll n$, and 
consider the problem of finding the minimum-length solution to
$ x_{opt} = \argmin_{x}||Ax-b||_2 = A^+b$.
Sampling variables or columns from $A$ can be represented by postmultiplying
$A$ by a $n \times c$ (with $c>m$) column-sampling matrix $S$ to
construct the (still underconstrained) least-squares problem:
$ \tilde{x}_{opt} = \argmin_{x}||ASS^Tx-b||_2 = A^T(AS)^{T+}(AS)^{+}b $.
The second equality follows by inserting $P_{A^T}=A^TA^{T+}$ to obtain
$ASS^TA^TA^{T+}x-b$ inside the $||\cdot||_2$ and recalling that
$A^+=A^TA^{T+}A^+$ for the Moore-Penrose pseudoinverse.
If one constructs $S$ by randomly sampling 
$c=O((n/\epsilon^2) \log(n /\epsilon))$ columns
according to ``column-leverage-score'' probabilities, \emph{i.e.}, exact or approximate
diagonal elements of the projection matrix onto the row space of $A$, then it can
be proven that $||x_{opt}-\tilde{x}_{opt}||_2 \le \epsilon||x_{opt}||_2$
holds, with high probability.
Alternatively, this underconstrained LS problem problem can also be solved 
with a random projection.
By using ideas from the previous subsection, one can show that if $S$ 
instead represents a random projection matrix, then by projecting to a 
low-dimensional space (which takes $o(m^2n)$ time with  a structured random 
projection matrix), then relative-error approximation guarantees also hold.

Thus, one can run the following algorithm for approximating the solution to 
the general overconstrained LS approximation problem.
\begin{itemize}
\item
Run the algorithm of this subsection to obtain numbers
$\tilde{\ell}_i $, for each $i = 1,\ldots,m$, rescaling them to form a 
probability distribution over $\{1,\ldots,m\}$.
\item
Call the randomized LS algorithm that is described in 
Section~\ref{sxn:least-squares:algorithm}, except using these numbers 
$\tilde{\ell}_i $
(rather than the exact statistical leverage scores $\ell_i $) to construct the importance 
sampling distribution over the rows of $A$.
\end{itemize}
That is: approximate the normalized statistical leverage scores in $o(mn^2)$ 
time; and then call the random sampling algorithm of 
Section~\ref{sxn:least-squares:algorithm} using those approximate scores 
as the importance sampling distribution.  
Clearly, this combined algorithm provides relative-error guarantees of the 
form~(\ref{eqn:ls-bound-eq1}) and~(\ref{eqn:ls-bound-eq2}), and overall it 
runs in $o(mn^2)$ time.

\subsubsection{Some additional thoughts}
\label{sxn:least-squares:faster-th:addl}

The previous two subsections have shown that the random sampling algorithm of 
Section~\ref{sxn:least-squares:algorithm}, which na\"{i}vely needs $O(mn^2)$ 
to compute the importance sampling distribution, can be sped up in two 
different ways.
One can spend $o(mn^2)$ to uniformize the sampling probabilities and then 
sample uniformly; or one can spend $o(mn^2)$ time to compute approximately 
the sampling probabilities and then use those approximations as an 
importance sampling distribution. 
Both procedures take basically the same time, which should not be surprising 
since the approximation of the statistical leverage scores boils down to an 
underconstrained LS problem; and which procedure is preferable in any 
given situation depends on 
the downstream~application.

Finally, to understand better the relationship between random sampling and 
random projection algorithms for the least squares problem, consider the
following vanilla random projection algorithm.
Given a matrix $A$ and vector $b$, representing a very overconstrained LS 
problem, one~can:
\begin{itemize}
\item
Construct an $O(n \log n/\epsilon^2) \times m$ random projection matrix 
$\Omega$, where the entries consist (up to scaling) of i.i.d. Gaussians 
or~$\{-1,+1\}$.
\item
Solve the induced subproblem
$ \tilde{x}_{opt} = \mbox{argmin}_x ||\Omega Ax - \Omega b||_2  $.
\end{itemize}
It is relatively-easy to show that this algorithm also satisfies relative-error bounds of the 
form~(\ref{eqn:ls-bound-eq1}) and~(\ref{eqn:ls-bound-eq2}).
Importantly, though, this random projection algorithm 
requires $O(mn^2)$ time to implement.
For the random sampling algorithm of 
Section~\ref{sxn:least-squares:algorithm}, the $O(mn^2)$ computational 
bottleneck is computing the importance sampling probabilities; while for 
this random projection algorithm the $O(mn^2)$ computational bottleneck is 
actually performing the matrix-matrix multiplication representing the random 
projection.
Thus, one can view the $o(mn^2)$ random projection algorithm that uses a 
small number of rows from a randomized Hadamard transform in one of two 
ways: either as a structured approximation to a usual random projection 
matrix with i.i.d. Gaussian or~$\{-1,+1\}$ entries that can be rapidly 
applied to arbitrary vectors; or as preprocessing the input with an 
orthogonal transformation to uniformize the leverage scores so that uniform 
sampling is~appropriate.

\subsection{Making this algorithm fast---in practice}
\label{sxn:least-squares:faster-pr}

Several ``rubber-hits-the-road'' issues need to be dealt with in order for 
the algorithms of the previous subsection to yield to high-precision 
numerical implementations that beat traditional numerical code, either in 
specialized scientific applications or when compared with popular numerical 
libraries.
The following issues are most significant.
\begin{itemize}
\item
\textbf{Awkward $\epsilon$ dependence.}
The sampling complexity, \emph{i.e.}, the number of columns or rows needed 
by the algorithm, scales as $1/\epsilon^2$ or $1/\epsilon$, which is the 
usual poor asymptotic complexity for Monte Carlo methods.
Even though there exist lower bounds for these problems for certain models
of data access~\cite{CW09}, this is problematic if one wants to choose 
$\epsilon$ to be on the order of machine precision.
\item
\textbf{Numerical conditioning and preconditioning.}
Thus far, nothing has been said about numerical conditioning issues, although
it is well-known that these issues are crucial when matrix algorithms are 
implemented numerically.
\item
\textbf{Forward error versus backward error.}
The bounds above, \emph{e.g.},~(\ref{eqn:ls-bound-eq1}) 
and~(\ref{eqn:ls-bound-eq2}), provide so-called forward error bounds.
The standard stability analysis in NLA is in terms of the backward error, 
where the approximate solution $\tilde{x}_{opt}$ is shown to be the exact 
solution of some slightly perturbed problem 
$x_{opt} = \mbox{argmin}_x ||(A+ \delta A)x-b||_2 $, where 
$||\delta A|| \le \tilde{\epsilon}||A||$ for some small $\tilde{\epsilon}$.
\end{itemize}
All three of these issues can be dealt with by coupling the randomized 
algorithms of the previous subsection with traditional tools from 
iterative NLA algorithms (as opposed to applying a traditional NLA algorithm 
as a black box on the random sample or random projection, as the above 
algorithms do).
This was first done by~\cite{RT08}, and these issues were addressed in much 
greater detail by~\cite{AMT10} and then by~\cite{CRT11} and~\cite{MSM11_TR}.

Both of the implementations of~\cite{RT08,AMT10} take the following~form.
\begin{itemize}
\item
Premultiply $A$ by a structured random projection, \emph{e.g.}, 
$\Omega = S H D$ from Section~\ref{sxn:background2:misc-stuff}, which 
represents uniformly sampling a few rows from a randomized Hadamard 
transform.
\item
Perform a QR decomposition on $\Omega A$, so that $\Omega A = QR$.
\item
Use the $R$ from the QR decomposition as a preconditioner for an iterative
Krylov-subspace~\cite{GVL96} method.
\end{itemize}
In general, iterative algorithms compute an $\epsilon$-approximate solution 
to a LS problem like~(\ref{eqn:orig_ls_prob}) by performing 
$O(\kappa(A)\log(1/\epsilon))$ iterations, where 
$\kappa(A)=\frac{\sigma_{max}(A)}{\sigma_{min}(A)}$ is the 
condition number of the input matrix (which could be quite large, thereby 
leading to slow convergence of the iterative algorithm).%
\footnote{These iterative algorithms replace the solution 
of~(\ref{eqn:orig_ls_prob}) with two problems: first solve 
$x_{opt} = \mbox{argmin}_x ||A \Pi^{-1}y-b||_2 $ iteratively, 
where $\Pi$ is the preconditioner; and then solve $\Pi x = y$.
Thus, a matrix $\Pi$ is a good preconditioner if $\kappa(A\Pi^{-1})$ 
is small and if $\Pi x = y$ can be solved quickly.}
In this case, by choosing the dimension of the random projection 
appropriately, \emph{e.g.}, as discussed in the previous subsections, one 
can show that $\kappa(AR^{-1})$ is bounded above by a small data-independent 
constant.
That is, by using the $R$ matrix from a QR decomposition of $\Omega A$, one 
obtains a good preconditioner for the original 
problem~(\ref{eqn:orig_ls_prob}), independent of course of any assumptions 
on the original matrix $A$.
Overall, 
applying the structured random projection in the first step takes $o(mn^2)$ 
time; 
performing a QR decomposition of $\Omega A$ is fast since $\Omega A$ is much 
smaller than $A$; 
and one needs to perform only $O(\log(1/\epsilon))$ iterations, each of 
which needs $O(mn)$ time, to compute the approximation.

The algorithm of~\cite{RT08} used CGLS (Conjugate Gradient Least Squares) as the iterative Krylov-subspace 
method, while the algorithm of~\cite{AMT10} used LSQR~\cite{PS82}; and both demonstrate 
that randomized algorithms can outperform traditional deterministic NLA 
algorithms in terms of clock-time (for particular implementations or 
compared with \textsc{Lapack}) for computing high-precision solutions 
for LS systems with as few as thousands of constraints and hundreds of 
variables.
The algorithm of~\cite{AMT10} 
considered five different classes of structured random projections 
(\emph{i.e.}, Discrete Fourier Transform, Discrete Cosine Transform, 
Discrete Hartely Transform, Walsh-Hadamard Transform, and a Kac random 
walk), explicitly addressed conditioning and backward stability issues, and
compared their implementation with \textsc{Lapack} on a wide range of 
matrices with uniform, moderately nonuniform, and very nonuniform leverage
score importance sampling distributions.
Similar ideas have also been applied to other common NLA tasks; for 
example,~\cite{CRT11} shows that these ideas can be used to develop fast 
randomized algorithms for computing projections onto the null space and row
space of $A$, for structured matrices $A$ such that both $A$ and $A^T$ can 
be rapidly applied to arbitrary~vectors.

The implementations of~\cite{CRT11,MSM11_TR} are similar, except that the 
random projection matrix in the first step of the above procedure is a 
traditional Gaussian random projection matrix.
While this does \emph{not} lead to a $o(mn^2)$ running time, it can be 
appropriate in certain situations: for example, if both $A$ and its adjoint
$A^T$ are structured such that they can be applied rapidly to arbitrary 
vectors~\cite{CRT11}; or for solving large-scale problems on distributed 
clusters with high communication cost~\cite{MSM11_TR}.
For example, due to the Gaussian random projection, the preconditioning 
phase of the algorithm of~\cite{MSM11_TR} is very well-conditioned, which 
implies that the number of iterations is fully predictable when LSQR or the 
Chebyshev semi-iterative method is applied to the preconditioned system.
The latter method is more appropriate for parallel computing environments, 
where communication is a major issue,
and thus~\cite{MSM11_TR} illustrates the empirical behavior of the algorithm 
on Amazon Elastic Compute Cloud (EC2)~clusters.

\section{Randomized algorithms for low-rank matrix approximation}
\label{sxn:low-rank}

In this section, we will describe several related randomized algorithms 
for low-rank matrix approximation that underlie applications such as those 
described in Section~\ref{sxn:background1}.
These algorithms build on the algorithms of 
Section~\ref{sxn:background2:previous}; and they achieve much-improved
worst-case bounds and are more useful in both numerical analysis and data 
analysis applications (when compared with the algorithms of Section~\ref{sxn:background2:previous}).
First, the algorithm of Section~\ref{sxn:low-rank:relerr} is a random 
sampling algorithm that improves the additive-error bounds to much finer 
relative-error bounds---the analysis of this algorithm boils down to an 
immediate application of the least-squares approximation algorithm of 
Section~\ref{sxn:least-squares}.
Next, the algorithm of Section~\ref{sxn:low-rank:exactlyk} is a random 
sampling algorithm that returns \emph{exactly} $k$, rather than 
$O(k \log k/\epsilon^2)$, columns, for an input rank parameter $k$---the 
proof of this result involves a structural property that decouples the 
randomization from the linear algebra in a somewhat more refined way than 
we saw in Section~\ref{sxn:least-squares}.
Finally, the algorithms of Section~\ref{sxn:low-rank:proj} are random 
projection algorithms that take advantage of this more refined structural 
property to couple these randomized algorithms with traditional methods 
from NLA and scientific computing.%
\footnote{Most of the results in this section will be formulated in terms of 
the amount of spectral or Frobenius norm that is captured by the (full or 
rank-$k$ approximation to the) random sample or random projection.
Given a basis for this sample or projection, it is straightforward to 
compute other common decompositions such as the pivoted QR factorization, 
the eigenvalue decomposition, the partial SVD, etc. using traditional NLA
methods; see~\cite{HMT09_SIREV} for a good discussion of~this.}


\subsection{A basic random sampling algorithm}
\label{sxn:low-rank:relerr}

Additive-error bounds (of the form proved for the low-rank algorithms of 
Section~\ref{sxn:background2:previous}) are rather coarse, and the gold 
standard in TCS is to establish much finer relative-error bounds of the 
form provided in~(\ref{eqn:rel-err}) below.
To motivate the importance sampling probabilities used to achieve such 
relative-error guarantees, recall that if one is considering a matrix with 
$k-1$ large singular values and one much smaller singular value, then the 
directional information of the $k^{th}$ singular direction will be hidden 
from the Euclidean norms of the input matrix.
The reason is that, since $A_k=U_k \Sigma_k V_k^T$, the 
Euclidean norms of the columns of $A$ are convolutions of ``subspace 
information'' (encoded in $U_k$ and $V_k^T$) and ``size-of-$A$ information'' 
(encoded in $\Sigma_k$).
This \emph{suggests} deconvoluting subspace information and size-of-$A$ 
information by choosing importance sampling probabilities that depend on the
Euclidean norms of the columns of $V_k^T$.
This importance sampling distribution defines a non-uniformity 
structure over $\mathbb{R}^{n}$ that indicates \emph{where} in the 
$n$-dimensional space the information in $A$ is being sent, independent of 
\emph{what} that (singular value) information is.
More formally, these quantities are proportional to the diagonal elements 
of the projection matrix onto the span of $V_k^T$,%
\footnote{Here, we are sampling columns and not rows, as in the algorithms 
of Section~\ref{sxn:least-squares}, and thus we are dealing with the right,
rather than the left, singular subspace; but clearly the ideas 
are analogous.  Thus, in particular, note that the ``span of $V_k^T$'' 
refers to the span of the \emph{rows} of $V_k^T$, whereas the ``span of 
$U_k$,'' as used in previous sections, refers to the span of the 
\emph{columns} of $U_k$.}
and thus they are examples of generalized statistical leverage scores. 

This idea was suggested in~\cite{DMM08_CURtheory_JRNL,CUR_PNAS}, and it 
forms the basis for the algorithm from the TCS literature that achieves the 
strongest Frobenius norm bounds.%
\footnote{Subsequent work in TCS that has not (yet?) found application in 
NLA and data analysis also achieves similar relative-error bounds but with different
methods.  For example, the algorithm of~\cite{HarPeled06_relerr_DRAFT} runs 
in roughly $O(mnk^2 \log k)$ time and uses geometric ideas involving 
sampling and merging approximately optimal $k$-flats.  Similarly, the 
algorithm of~\cite{DV06_relerr_TR} randomly samples in a more complicated 
manner and runs in $O(Mk^2 \log k)$, where $M$ is the number of nonzero 
elements in the matrix; alternatively, it runs in $O(k \log k)$ passes over the data from external 
storage.}
Given an $m \times n$ matrix $A$ and a rank parameter~$k$:
\begin{itemize}
\item
Compute the importance sampling probabilities $\{p_i\}_{i=1}^{n}$, where 
$p_i=\frac{1}{k}||{V_k^T}^{(i)}||_2^2$, where $V_k^T$ is \emph{any} 
$k \times n$ orthogonal matrix spanning the top-$k$ right singular subspace 
of~$A$.
\item
Randomly select and rescale $c = O(k \log k /\epsilon^2)$ columns of $A$
according to these probabilities to form the matrix $C$.
\end{itemize}
A more detailed description of this basic random sampling algorithm may be 
found in~\cite{DMM08_CURtheory_JRNL,CUR_PNAS}, where it is proven that
\begin{equation}
||A-P_{C_k}A||_F \le (1+\epsilon) ||A-P_{U_k}A ||_F  
\label{eqn:rel-err}
\end{equation}
holds.   
(As above, $C_k$ is the best rank-$k$ approximation to the matrix $C$, and 
$P_{C_k}$ is the projection matrix onto this $k$-dimensional space.)
As with the relative-error random sampling algorithm of 
Section~\ref{sxn:least-squares:algorithm}, the dependence of the sampling 
complexity and running time on the failure probability $\delta$ is 
$O(\log(1/\delta))$;
thus, the failure probability for this randomized low-rank approximation, 
as well as the subsequent algorithms of this section, can be made to be 
negligibly-small, both in theory and in practice.
The analysis of this algorithm boils down to choosing a set of columns 
that are relative-error good at capturing the Frobenius norm of $A$, when
compared to the basis provided by the top-$k$ singular vectors.
That is, it boils down to the randomized algorithm for the least-squares 
approximation problem from Section~\ref{sxn:least-squares}; 
see~\cite{DMM08_CURtheory_JRNL,CUR_PNAS} for~details.

This algorithm and related algorithms that randomly sample columns and/or 
rows provide what is known as \emph{CX or CUR matrix 
decompositions}~\cite{dkm_matrix3,DMM08_CURtheory_JRNL,CUR_PNAS}.%
\footnote{Within the NLA community, Stewart developed the quasi-Gram-Schmidt 
method and applied it to a matrix and its transpose to obtain such a CUR 
matrix decomposition~\cite{Ste99,BPS04_TR}; and Goreinov, Tyrtyshnikov, and 
Zamarashkin developed a CUR matrix decomposition (a so-called pseudoskeleton 
approximation) and related the choice of columns and rows to a ``maximum 
uncorrelatedness'' concept~\cite{GTZ97,GT01}.  Note that the Nystr\"{o}m 
method is a type of CUR decomposition and that the pseudoskeleton
approximation is also a generalization of the Nystr\"{o}m method.}
In addition, this relative-error column-based CUR decomposition, as well as 
heuristic variants of it, has been applied in a range of data analysis 
applications, ranging from term-document data to DNA SNP 
data~\cite{CUR_PNAS,Paschou07a,Paschou07b}.
The computational bottleneck for this relative-error approximation algorithm 
is computing the importance sampling distribution $\{p_i\}_{i=1}^{n}$, for 
which it suffices to compute \emph{any} $k \times n$ matrix $V_k^T$ that 
spans the top-$k$ right singular subspace of $A$.
That is, it \emph{suffices} (but is \emph{not} necessary) to compute any 
orthonormal basis spanning $V_k^T$, which typically requires $O(mnk)$ 
running time, and it is \emph{not} necessary to compute the full or partial 
SVD.
Alternatively, the leverage scores can all be 
approximated to within $1\pm\epsilon$ in roughly $O(mn k)$ time using 
the algorithm of~\cite{DMMW11_TR} from 
Section~\ref{sxn:least-squares:faster-th:rand-samp}, in which case these 
approximations can be used as importance sampling probabilities in the 
above random sampling algorithm.

\subsection{A more refined random sampling algorithm}
\label{sxn:low-rank:exactlyk}

The algorithm of the previous subsection randomly samples 
$O(k \log k/\epsilon^2)$ columns, and then in order to compare with the 
bound provided by the SVD, it ``filters'' those columns through an exactly 
rank-$k$ space.
In this subsection, we will describe a randomized algorithm for the Column 
Subset Selection Problem (CSSP): the problem of selecting \emph{exactly} $k$ 
columns from an input matrix.
Clearly, bounds for algorithms for the CSSP will be worse than the very 
strong relative-error bounds, provided by (\ref{eqn:rel-err}), that hold 
when $O(k \log k / \epsilon^2)$ columns are selected.
Importantly, though, this CSSP algorithm extends the ideas of the previous 
relative-error TCS algorithm to obtain bounds of the form used historically 
in NLA.
Moreover, the main structural result underlying the analysis of this 
algorithm permits \emph{much} finer control on the application of 
randomization, \emph{e.g.}, for high-performance numerical implementation of 
both random sampling and random projection algorithms.
We will start with a review of prior approaches to the CSSP;
then describe the algorithm and the  
quality-of-approximation bounds; and finally highlight the main structural 
result underlying its analysis.

\subsubsection{A formalization of and prior approaches to this problem}

Within NLA, a great deal of work has focused on this 
CSSP~\cite{BG65,Fos86,Cha87,CH90,BH91,HP92,CI94,GE96,BQ98a,PT99,Pan00}.
Several general observations about the NLA approach include:
\begin{itemize}
\item
The focus in NLA is on \emph{deterministic algorithms}.
Moreover, these algorithms are greedy, in that at each iterative step, the 
algorithm makes a decision about which columns to keep according to a 
pivot-rule that depends on the columns it currently has, the spectrum of 
those columns, etc.
Differences between different algorithms often boil down to how to deal with 
such pivot rules decisions, and the hope is that more sophisticated 
pivot-rule decisions lead to better algorithms in theory or in practice.
\item
There are deep \emph{connections with QR factorizations} and in particular 
with the so-called Rank Revealing QR factorizations.
Moreover, there is an emphasis on optimal conditioning questions, backward 
error analysis issues, and whether the running time is a large or small 
constant multiplied by $n^2$ or~$n^3$.
\item
Good \emph{spectral norm bounds} are obtained.
A typical spectral norm bound is:
\begin{equation}
||A-P_CA||_2 \le O\left(\sqrt{k(n-k)}\right)||A-P_{U_k}A||_2 ,
\label{eqn:cssp-1}
\end{equation}
and these results are algorithmic, in that the running time is a low-degree 
polynomial in $m$ and $n$~\cite{GE96}.
(The $\sqrt{n}$ multiplicative factor might seem large, but recall that 
the spectral norm is much less ``forgiving'' than the Frobenius norm and 
that it is not even known whether there exists columns that do better.)
On the other hand, the strongest result for the Frobenius norm in this 
literature is
\begin{equation}
||A-P_CA||_F \le \sqrt{(k+1)(n-k)}||A-P_{U_k}A||_2  ,
\label{eqn:cssp-2}
\end{equation}
but it is only an existential result, \emph{i.e.}, the only known algorithm 
essentially involves exhaustive enumeration~\cite{HP92}.
(In these two expressions, $U_k$ is the $m \times k$ matrix consisting of the
top $k$ left singular vectors of $A$, and $P_{U_k}$ is a projection matrix 
onto the span of $U_k$.)
\end{itemize}

Within TCS, a great deal of work has focused on the related problem of 
choosing good columns from a 
matrix~\cite{DFKVV04_JRNL,dkm_matrix1,dkm_matrix2,dkm_matrix3,RV07,DMM08_CURtheory_JRNL}.
Several general observations about the TCS approach~include:
\begin{itemize}
\item
The focus in TCS is on \emph{randomized 
algorithms}.
In particular, with these algorithms, there exists some nonzero 
probability, which can typically be made extremely small, say 
$\delta=10^{-20}$, that the algorithm will return columns that fail to 
satisfy the desired quality-of-approximation bound. 
\item
The algorithms select \emph{more than $k$ columns}, and the best 
rank-$k$ projection onto those columns is considered.
The number of columns is typically a low-degree polynomial in $k$, most often
$O(k \log k)$, where the constants hidden in the big-O notation are quite 
reasonable.
\item
Very good \emph{Frobenius norm bounds} are obtained. 
For example, the algorithm (described above) that provides the strongest 
Frobenius norm bound achieves
(\ref{eqn:rel-err}),
while running in time of the order of computing an exact or approximate 
basis for the top-$k$ right singular subspace~\cite{DMM08_CURtheory_JRNL}.
The TCS literature also demonstrates that there exists a set of $k$ columns 
that achieves a constant-factor approximation:
\begin{equation}
||A-P_{C}A||_F \le \sqrt{k} ||A-P_{U_k}A ||_F  ,
\label{eqn:cssp-3}
\end{equation}
but note that this is an existential result~\cite{DV06}.
\end{itemize}

Note that, prior to the algorithm of the next subsection, it was not 
immediately clear how to combine these two very different approaches.
For example, if one looks at the details of the pivot rules in the 
deterministic NLA methods, it isn't clear that keeping more than exactly 
$k$ columns will help at all in terms of reconstruction error.
Similarly, since there is a version of the so-called Coupon Collector 
Problem at the heart of the usual TCS analysis, keeping fewer than 
$\Omega(k \log k)$ will fail with respect to this worst-case analysis.
Moreover, the obvious hybrid algorithm of first randomly sampling 
$O(k \log k)$ columns and then using a deterministic QR procedure to select 
exactly $k$ of those columns does not seem to perform so well (either in 
theory or in practice).

\subsubsection{A two-stage hybrid algorithm for this problem}

Consider the following more sophisticated version of a two-stage hybrid 
algorithm.
Given an arbitrary $m \times n$ matrix $A$ and rank parameter $k$:
\begin{itemize}
\item
(Randomized phase)
Compute the importance sampling probabilities $\{p_i\}_{i=1}^{n}$, where 
$p_i=\frac{1}{k}||{V_k^T}^{(i)}||_2^2$, where $V_k^T$ is \emph{any} 
$k \times n$ orthogonal matrix spanning the top-$k$ right singular subspace 
of~$A$.
Randomly select and rescale $c = O(k \log k)$ columns of $V_k^T$ according 
to these probabilities.
\item
(Deterministic phase)
Let $\tilde{V}^T$ be the $k \times O(k \log k)$ non-orthogonal matrix 
consisting of the down-sampled and rescaled columns of $V_k^T$.
Run a deterministic QR algorithm on $\tilde{V}^T$ to select exactly $k$ 
columns of $\tilde{V}^T$.
Return the corresponding columns of $A$.
\end{itemize}
A more detailed description of this algorithm may be found 
in~\cite{BMD09_CSSP_SODA,BMD08_CSSP_TRv2}, where it is shown that with 
extremely high probability the following spectral%
\footnote{Note that to establish the spectral norm 
bound,~\cite{BMD09_CSSP_SODA,BMD08_CSSP_TRv2} used slightly more complicated 
(but still depending only on information in $V_k^T$) importance sampling 
probabilities, but this may be an artifact of the analysis.}
and Frobenius norm bounds hold:
\begin{eqnarray}
\label{eqn:cssp-4}
||A-P_CA||_2 &\le& O(k^{3/4} \log^{1/2}(k) n^{1/2}) ||A-P_{U_k}A||_2  \\
\label{eqn:cssp-5}
||A-P_CA||_F &\le& O(k \log^{1/2} k) ||A-P_{U_k}A||_F   .
\end{eqnarray}
Note that both the original choice of columns in the first phase, as well 
as the application of the QR algorithm in the second phase,%
\footnote{Note that QR (as opposed to the SVD) is \emph{not} performed in 
the second phase to speed up the computation of a relatively cheap part of 
the algorithm, but instead it is performed since the goal of the algorithm 
is to return actual columns of the input matrix.}
involve the sampled version of the matrix $V_k^T$, \emph{i.e.}, the matrix defining the relevant 
non-uniformity structure over the columns of $A$ in the relative-error 
algorithm of Section~\ref{sxn:low-rank:relerr}.
In particular, it is critical to the success of this algorithm that the QR 
procedure in the second phase be applied to the randomly-sampled version 
of $V_k^T$, rather than of $A$ itself.
This algorithm may be viewed as post-processing the relative-error 
random sampling algorithm of the previous subsection to remove redundant 
columns; and it has been applied successfully to a range of data analysis 
problems.  See, \emph{e.g.},~\cite{Paschou08a,Paschou10a,Paschou10b}
and~\cite{BMD08_CSSP_KDD,BMD09_kmeans_NIPS,SD11}, as well as~\cite{BBP10}
for a discussion of numerical issues associated with this algorithm.

With respect to running time, the computational bottleneck for this 
algorithm is computing $\{p_i\}_{i=1}^{n}$, for which it suffices to 
compute \emph{any} $k \times n$ matrix $V_k^T$ that spans the top-$k$ right 
singular subspace of $A$.
(In particular, a full or partial SVD computation is \emph{not} necessary.)
Thus, this running time is of the same order as the running time of the QR 
algorithm used in the second phase when applied to the original matrix $A$, 
typically roughly $O(mnk)$ time.
(Not surprisingly, one could also perform a random projection, such as 
those described in Section~\ref{sxn:low-rank:proj} below, to approximate 
this basis, and then use that approximate basis to compute approximate 
importance sampling probabilities, as described in 
Section~\ref{sxn:least-squares:faster-th:rand-samp} above.  In that case, 
similar bounds would hold, but the running time would be improved to 
$O(mn \log k)$ time.)
Moreover, this algorithm scales up to matrices with thousands of rows and 
millions of columns, whereas existing off-the-shelf implementations of 
traditional QR algorithms may fail to run at all.
With respect to the worst-case quality of approximation bounds, this 
algorithm selects columns that are comparable to the state-of-the-art 
algorithms for constant $k$ (\emph{i.e.}, (\ref{eqn:cssp-4}) is only
$O(k^{1/4}\log^{1/2}k)$ worse than (\ref{eqn:cssp-1})) for the spectral norm; 
and (\ref{eqn:cssp-5}) is only a factor of at most $O((k \log k)^{1/2})$ 
worse than (\ref{eqn:cssp-3}), the best previously-known existential result 
for the Frobenius norm.

\subsubsection{A basic structural result}

As with the relative-error LS algorithm of Section~\ref{sxn:least-squares}, 
in order to see ``why'' this algorithm for the CSSP works, it is helpful to 
identify a structural condition that decouples the randomization from the 
linear algebra.
This structural condition was first identified 
in~\cite{BMD09_CSSP_SODA,BMD08_CSSP_TRv2}, and it was subsequently improved 
by~\cite{HMT09_SIREV}.
To identify it, consider preconditioning or postmultiplying the input matrix 
$A$ by some \emph{arbitrary} matrix~$Z$.
Thus, for the above randomized algorithm, the matrix $Z$ is a 
carefully-constructed random sampling matrix, but it could be a random 
projection, or more generally \emph{any} other arbitrary matrix~$Z$.
Recall that if $k \leq r = \mbox{rank}(A)$, then the SVD of $A$ may be 
written as
\begin{equation*} 
A = U_A \Sigma_A V_A^T 
  = U_k \Sigma_k V_k^T + U_{k,\perp} \Sigma_{k,\perp} V_{k,\perp}^T     ,
\end{equation*}
where $U_k$ is the $m \times k$ matrix consisting of the top $k$ singular 
vectors, $U_{k,\perp}$ is the $m \times (r-k)$ matrix consisting of the bottom
$r-k$ singular vectors, etc.
Then, the following structural condition~holds.
\begin{itemize}
\item
\textbf{Structural condition underlying the randomized low-rank algorithm.}
If $V_k^TZ$ has full rank, then for $\nu\in\{2,F\}$, \emph{i.e.}, for both the 
Frobenius and spectral norms,
\begin{equation}
\label{eqn:struct-cond-low-rank}
\left|\left|A-P_{AZ}A\right|\right|_{\nu}^{2} 
   \leq \left|\left|A-A_k\right|\right|_{\nu}^{2} 
      + \left|\left| \Sigma_{k,\perp}\left(V_{k,\perp}^TZ\right)\left(V_{k}^TZ\right)^{\dagger} \right|\right|_{\nu}^{2} 
\end{equation}
holds, where $P_{AZ}$ is a projection onto the span of $AZ$, and where the 
dagger symbol represents the Moore-Penrose pseudoinverse.
\end{itemize}
This structural condition characterizes the manner in which the behavior of 
the low-rank algorithm depends on the interaction between the right singular 
vectors of the input matrix and the matrix $Z$.
(In particular, it depends on the interaction between the subspace 
associated with the top part of the spectrum and the subspace associated 
with the bottom part of the spectrum via the 
$\left(V_{k,\perp}^TZ\right)\left(V_{k}^TZ\right)^{\dagger} $ term.)
Note that the assumption that $V_k^TZ$ does not lose rank is a
generalization of Condition~(\ref{eqn:lemma1_ass2}) of 
Section~\ref{sxn:least-squares}.
Also, note that the form of this structural condition is the same for both 
the spectral and Frobenius~norms.

As with the LS problem, given this structural insight, what one does with 
it depends on the application:
one can compute the basis $V_k^T$ exactly if that is not computationally 
prohibitive and if one is interested in extracting exactly $k$ columns; or
one can perform a random projection and ensure that with high probability 
the structural condition is satisfied.
Moreover, by decoupling the randomization from the linear algebra, it is 
easier to parameterize the problem in terms more familiar to NLA and 
scientific computing: for example, one can consider sampling $\ell > k$ 
columns and projecting onto a rank-$k^\prime$, where $k^\prime > k$, 
approximation to those columns; or one can couple these ideas with 
traditional methods such as the power iteration method.
Several of these extensions will be the topic of the next subsection.

\subsection{Several related random projection algorithms}
\label{sxn:low-rank:proj}

In this subsection, we will describe three random projection algorithms that 
draw on the ideas of the previous subsections in progressively finer ways.

\subsubsection{A basic random projection algorithm}
\label{sxn:low-rank:proj-first}

To start, consider the following basic random projection algorithm.
Given an $m \times n$ matrix $A$ and a rank parameter~$k$:
\begin{itemize}
\item
Construct an $n \times \ell$, with $\ell =O(k/\epsilon)$, structured random 
projection matrix $\Omega$, \emph{e.g.}, $\Omega = DHS $ from 
Section~\ref{sxn:background2:misc-stuff}, which represents uniformly 
sampling a few rows from a randomized Hadamard transform.
\item
Return $B=A\Omega$.
\end{itemize}
This algorithm, which amounts to choosing uniformly a small number $\ell$ 
of columns in a randomly rotated basis, was introduced in~\cite{Sarlos06},
where it is proven that
\begin{equation}
||A-P_{B_k}A||_F \le (1+\epsilon) ||A-P_{U_k}A ||_F
\label{eqn:project-rel-err}
\end{equation}
holds with high probability.
(Recall that $B_k$ is the best rank-$k$ approximation to the matrix $B$, and 
$P_{B_k}$ is the projection matrix onto this $k$-dimensional space.)
This algorithm runs in $O(Mk/\epsilon +(m+n)k^2/\epsilon^2)$ time, where $M$
is the number of nonzero elements in $A$, and it requires $2$ passes over the
data from external storage.

Although this algorithm is very similar to the additive-error random 
projection algorithm of~\cite{PRTV00} that was described in 
Section~\ref{sxn:background2:previous}, this algorithm achieves much 
stronger relative-error bounds by performing a much more refined analysis.
Basically,~\cite{Sarlos06} (and also the improvement~\cite{NDT09}) modifies 
the analysis of the relative-error random sampling 
of~\cite{DMM08_CURtheory_JRNL,CUR_PNAS} that was described in 
Section~\ref{sxn:low-rank:relerr}, which in turn relies on the 
relative-error random sampling algorithm for LS 
approximation~\cite{DMM06,DMM08_CURtheory_JRNL} that was described in 
Section~\ref{sxn:least-squares}.
In the same way that we saw in Section~\ref{sxn:least-squares:faster-th} that
fast structured random projections could be used to uniformize 
coordinate-based non-uniformity structure for the LS problem, after which 
fast uniform sampling was appropriate, here uniform sampling in the 
randomly-rotated basis achieves relative-error bounds.
In showing this,~\cite{Sarlos06} also states a ``subspace'' analogue to the 
JL lemma, in which the geometry of an entire subspace of vectors (rather 
than just $N$ pairs of vectors) is preserved.
Thus, one can view the analysis of~\cite{Sarlos06} as applying JL ideas, 
not to the rows of $A$ itself, as was done by~\cite{PRTV00}, but instead
to vectors defining the subspace structure of $A$.
Thus, with this random projection algorithm, the subspace information and 
size-of-$A$ information are deconvoluted \emph{within the analysis}, whereas 
with the random sampling algorithm of Section~\ref{sxn:low-rank:relerr}, 
this took place \emph{within the algorithm} by modifying the importance 
sampling probabilities.

\subsubsection{An improved random projection algorithm}
\label{sxn:low-rank:proj-second}

As with the randomized algorithms for the LS problem, several 
rubber-hits-the-road issues need to be dealt with in order for randomized 
algorithms for the low-rank matrix approximation problem to yield to 
high-precision numerical implementations that beat traditional deterministic
numerical code.
In addition to the issues described in 
Section~\ref{sxn:least-squares:faster-pr}, the main issue here is the 
following.
\begin{itemize}
\item
\textbf{Minimizing the oversampling factor.}
In practice, choosing even $O(k \log k)$ columns, in either the original or 
a randomly-rotated basis, even when the big-O notation hides only modest 
constants, can make it difficult for these randomized matrix algorithms to 
beat previously-existing high-quality numerical implementations.
Ideally, one could parameterize the problem so as to choose some number 
$\ell=k+p$ columns, where $p$ is a modest additive oversampling factor, 
\emph{e.g.}, $10$ or $20$ or $k$, and where there is no big-O constant.
\end{itemize}
When attempting to be this aggressive at minimizing the size of the sample, 
the choice of the oversampling factor $p$ is more sensitive to the input
than in the algorithms we have reviewed so far.
That is, whereas the previous bounds held for any input, here the proper 
choice for the oversampling factor $p$ can be quite sensitive to the input
matrix.
For example, when parameterized this way, $p$ could depend on the size of 
the matrix dimensions, the decay properties of the spectrum, and the 
particular choice made for the 
random projection matrix~\cite{MRT11,WLRT08,LWFMRT07,RST09,HMT09_SIREV,HMST10_TR}.
Moreover, for worst-case input matrices, such a procedure may fail.
For example, one can very-easily construct matrices such that if one randomly 
samples $o(k \log k)$ columns, in either the original canonical basis or in 
the randomly-rotated basis provided by the structured Hadamard transform, 
then the algorithm will fail.
Basically, one can easily encode the so-called Coupon 
Collector Problem~\cite{MotwaniRaghavan95} in the columns, and it is known 
that $\Theta( k \log k)$ samples are necessary for this problem.

That being said, running the risk of such a failure might be acceptable if
one can efficiently couple to a diagnostic to check for such a 
failure, and if one can then correct for it by choosing more 
samples if necessary.
The best numerical implementations of randomized matrix algorithms for 
low-rank matrix approximation do just this, and the strongest results in 
terms of minimizing $p$ take advantage of 
Condition~(\ref{eqn:struct-cond-low-rank}) in a somewhat different way than 
was originally used in the analysis of the CSSP~\cite{HMT09_SIREV}.
For example, rather than choosing $O(k \log k)$ dimensions and then 
filtering them through \emph{exactly} $k$ dimensions, as the relative-error 
random sampling and relative-error random projection algorithms do, one can 
choose some number $\ell$ of dimensions and project onto a 
$k^{\prime}$-dimensional subspace, where $k < k^{\prime} \le \ell$, while 
exploiting Condition~(\ref{eqn:struct-cond-low-rank}) to bound the error, as 
appropriate for the computational environment at hand~\cite{HMT09_SIREV}.

Next, consider a second random projection algorithm that will address this 
issue.
Given an $m \times n$ matrix $A$, a rank parameter~$k$, and an oversampling
factor~$p$:
\begin{itemize}
\item
Set $\ell = k + p$.
\item
Construct an $n \times \ell$ random projection matrix $\Omega$, either with 
i.i.d.  Gaussian entries or in the form of a structured random projection 
such as $\Omega = DHS$ which represents uniformly sampling a few rows from 
a randomized Hadamard transform.
\item
Return $B = A \Omega$
\end{itemize}
Although this is quite similar to the algorithms of~\cite{PRTV00,Sarlos06},
algorithms parameterized in this form were introduced 
in~\cite{MRT11,WLRT08,LWFMRT07}, where a suite of bounds of the form
$$
||A-Z||_2 \lesssim 10 \sqrt{\ell \min\{m,n\}}||A-A_k||_2
$$
are shown to hold with high probability.
Here, $Z$ is a rank-$k$-or-greater matrix easily-constructed from $B$.
This result can be used to obtain a so-called \emph{interpolative 
decomposition} (a variant of the basic CSSP with explicit numerical 
conditioning properties), and~\cite{MRT11,WLRT08,LWFMRT07} also provide an 
\emph{a posteriori} error estimate (that is useful for situations in which 
one wants to choose the rank parameter $k$ to be the numerical rank, as 
opposed to the \emph{a priori} specification of $k$ as part of the input, 
which is more common in the TCS-style algorithms that preceded 
this~algorithm).

\subsubsection{A third random projection algorithm}
\label{sxn:low-rank:proj-third}

Finally, consider a third random projection algorithm that will address the 
issue that the decay properties of the spectrum can be important when it is 
of interest to minimize the oversampling very aggressively.%
\footnote{Of course, this should not be completely unexpected, given that 
Condition~(\ref{eqn:struct-cond-low-rank}) shows that the behavior of 
algorithms depends on the interaction between different subspaces 
associated with the input matrix $A$.  When stronger assumptions are made
about the data, stronger bounds can often be obtained.} 
Given an $m \times n$ matrix $A$, a rank parameter~$k$, an oversampling
factor~$p$, and an iteration parameter~$q$:
\begin{itemize}
\item
Set $\ell = k + p$.
\item
Construct an $n \times \ell$ random projection matrix $\Omega$, either with 
i.i.d.  Gaussian entries or in the form of a structured random projection 
such as $\Omega = DHS$ which represents uniformly sampling a few rows from 
a randomized Hadamard transform.
\item
Return $B = (AA^T)^{q} A \Omega$
\end{itemize}
This algorithm (as well as a numerically-stable variant of it) was 
introduced in~\cite{RST09}, where it is shown that bounds of the form
$$
||A-Z||_2 \lesssim \left(10 \sqrt{\ell \min\{m,n\}} \right)^{1/(4q+2)} ||A-A_k||_2
$$
hold with high probability.
(This bound should be compared with the bound for the previous algorithm, and 
thus $Z$ is a rank-$k$-or-greater matrix easily-constructed from $B$.)
Basically, this random projection algorithm modifies the previous algorithm 
by coupling a form of the power iteration method within the random 
projection step.
This has the effect of speeding up the decay of the spectrum while leaving 
the singular vectors unchanged, and it is observed in~\cite{RST09,HMT09_SIREV} 
that $q=2$ or $q=4$ is often sufficient for certain data matrices of 
interest.
This algorithm was analyzed in greater detail for the case of Gaussian 
random matrices in~\cite{HMT09_SIREV}, and an out-of-core implementation (meaning, 
appropriate for data sets that are too large to be stored in RAM) of it was 
presented in~\cite{HMST10_TR}.

The running time of these last two random projection algorithms depends on 
the details of the computational environment, \emph{e.g.}, whether the 
matrix is large and dense but fits into RAM or is large and sparse or is too 
large to fit into RAM; how precisely the random projection matrix is 
constructed; whether the random projection is being applied to an arbitrary 
matrix $A$ or to structured input matrices, etc.~\cite{HMT09_SIREV}.
For example, if random projection matrix $\Omega$ is constructed from i.i.d.
Gaussian entries then in general the algorithm requires $O(mnk)$ time to 
implement the random projection, \emph{i.e.}, to perform the matrix-matrix 
multiplication $A\Omega$, which is no faster than traditional deterministic 
methods.
On the other hand, if the projection is to be applied to matrices $A$
such that $A$ and/or $A^T$ can be applied rapidly to arbitrary vectors
(\emph{e.g.}, very sparse matrices, or structured matrices such as those 
arising from Toeplitz operators, or matrices that arise from discretized 
integral operators that can be applied via the fast multipole method), 
then Gaussian random projections may be preferable.
Similarly, in general, if $\Omega$ is structured, \emph{e.g.}, is of the 
form $\Omega = DHS$, then it can be implemented in $O(mn \log k)$ time, and
this can lead to dramatic clock-time speed-up over classical techniques even 
for problems of moderate sizes.
On the other hand, for out-of-core implementations these additional speed-ups 
have a negligible effect, \emph{e.g.}, since matrix-matrix multiplications 
can be faster than a QR factorization, and so using Gaussian projections can 
be preferable.
Working through these issues in theory and practice is still very much an 
active research area.

\section{Empirical observations}
\label{sxn:empirical}

In this section, we will make some empirical observations, with an emphasis
on the role of statistical leverage in these algorithms and in MMDS 
applications more generally.

\subsection{Traditional perspectives on statistical leverage} 
\label{sxn:empirical-traditional}

\begin{figure}
   \begin{center}
      \subfigure[Wood Beam Data]{
         \includegraphics[width=0.45\textwidth]{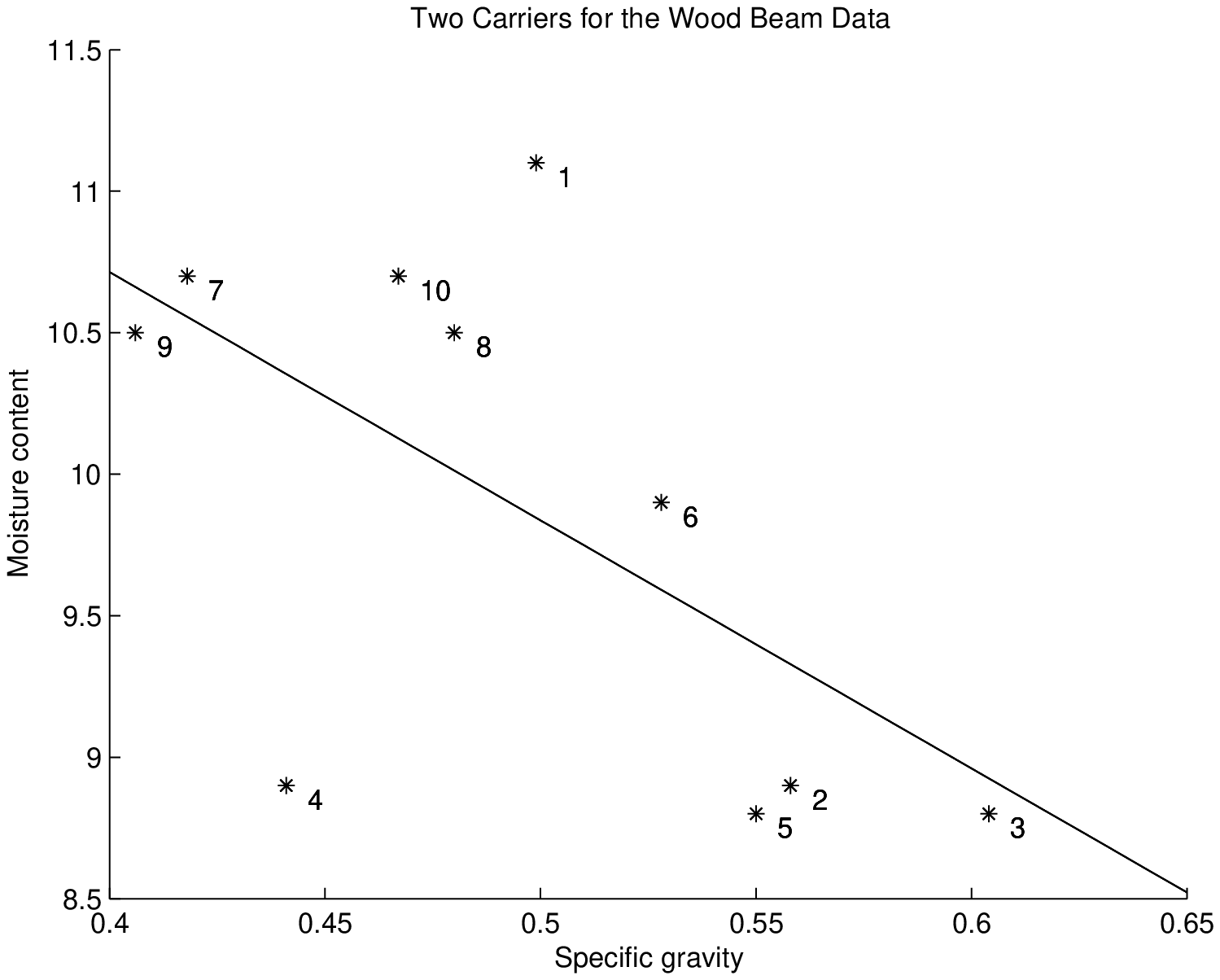}
         \label{fig:leverageI:wooddata}
      } \qquad 
      \subfigure[Corresponding Leverage Scores]{
         \hspace{20mm}
         \includegraphics[width=0.08\textwidth]{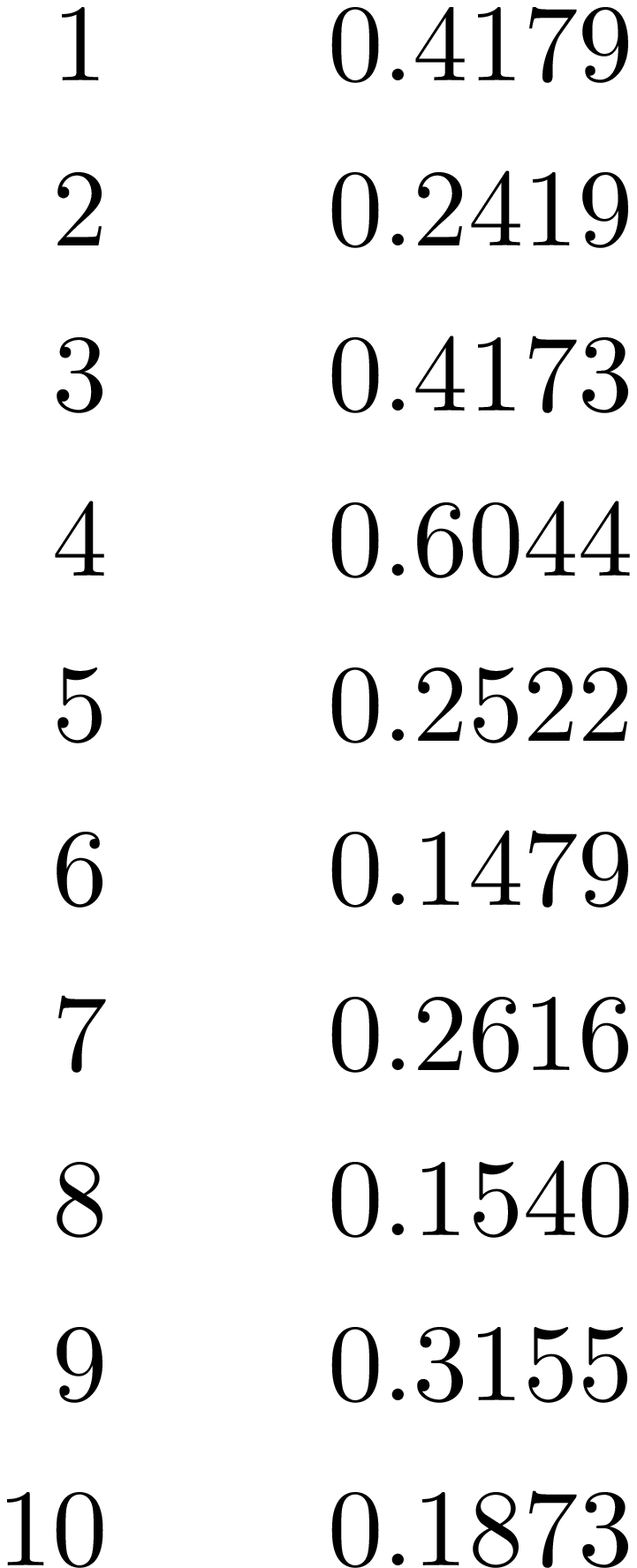}
         \hspace{20mm}
         \label{fig:leverageI:woodscores}
      } 
   \end{center}
\caption{Statistical leverage scores historically in diagnostic regression
analysis.
(\ref{fig:leverageI:wooddata})
The Wood Beam Data described in~\cite{HW78} is an example illustrating the
use of statistical leverage scores in the context of least-squares 
approximation.
Shown are the original data and the best least-squares fit.
(\ref{fig:leverageI:woodscores})
The leverage scores for each of the ten data points in the Wood Beam Data.
Note that the data point marked ``4'' has the largest leverage score, as 
might be expected from visual~inspection.
}
\label{fig:leverageI}
\end{figure}

As mentioned previously, the use of statistical leverage scores has a long
history in statistical and diagnostic regression 
analysis~\cite{HW78,ChatterjeeHadi88,CH86,VW81,ChatterjeeHadiPrice00}. 
To gain insight into these statistical leverage scores, consider the 
so-called ``wood beam data'' example~\cite{DS66,HW78}, which is visually 
presented in Figure~\ref{fig:leverageI:wooddata}, along with the best-fit 
line to that data. 
In Figure~\ref{fig:leverageI:woodscores}, the leverage scores for these ten 
data points are shown.
Intuitively, data points that ``stick out'' have particularly high 
leverage---\emph{e.g.}, the data point that has the most influence or 
leverage on the best-fit line to the wood beam data is the point marked 
``4'', and this is reflected in the relative magnitude of the corresponding 
statistical leverage score.
(Note that the point ``1'' exhibits some similar behavior; and that although 
``3'' and ``9'' don't ``stick out'' in the same sense, they are at the 
``ends'' of the data and possess a relatively-high leverage for that reason.)
Indeed, since $\mbox{Trace}(H)=n$, where $H$ is the hat matrix defined in 
Section~\ref{sxn:least-squares:perspectives}, 
a rule of thumb that has been suggested in
diagnostic regression analysis to identify errors and outliers in a data 
set is to investigate the $i^{th}$ data point if 
$H_{ii} > 2n/m$~\cite{VW81,ChatterjeeHadiPrice00}, 
\emph{i.e.}, if $H_{ii}$ is larger that $2$ or $3$ times the ``average'' 
size.
On the other hand, of course, if it happens to turn out that such a point is 
a legitimate data point, 
then one might expect that such an outlying data point will be a 
particularly important or informative data point.

That leverage scores ``should be'' fairly uniform---indeed, typical 
conditions even in recent work on the so-called coherence of 
matrices~\cite{CR07,CSPW11,AMT10,TalRos10} make just such an 
assumption---is supported by the idea that if they are not then a small 
number of data points might be particularly important, in which case a 
different or more refined statistical model might be appropriate.
Furthermore, they are fairly uniform in various limiting cases where 
measure concentration occurs, \emph{e.g.}, for not-extremely-sparse 
random graphs, and for matrices such as Laplacians associated with 
well-shaped low-dimensional manifolds, basically since eigenfunctions tend 
to be delocalized in those situations.
Of course, their actual behavior in realistic data applications is an 
empirical~question.

\subsection{More recent perspectives on statistical leverage} 
\label{sxn:empirical-morerecent}

\begin{figure}
   \begin{center}
      \subfigure[Zachary Karate Club Data]{
         \includegraphics[width=0.40\textwidth]{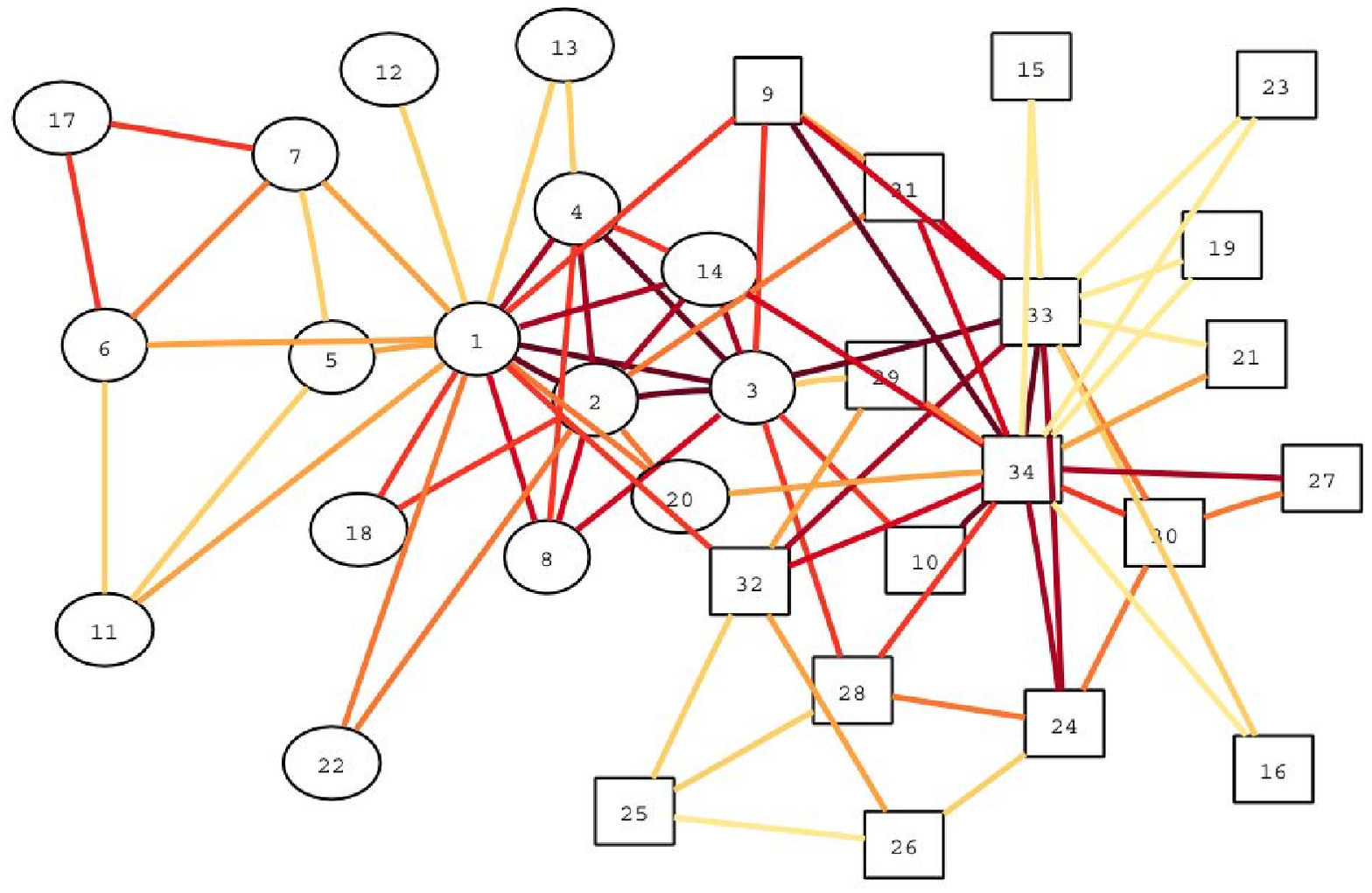}
         \label{fig:leverageII:zachary}
      } \qquad 
      \subfigure[Cumulative Leverage]{
         \includegraphics[width=0.45\textwidth]{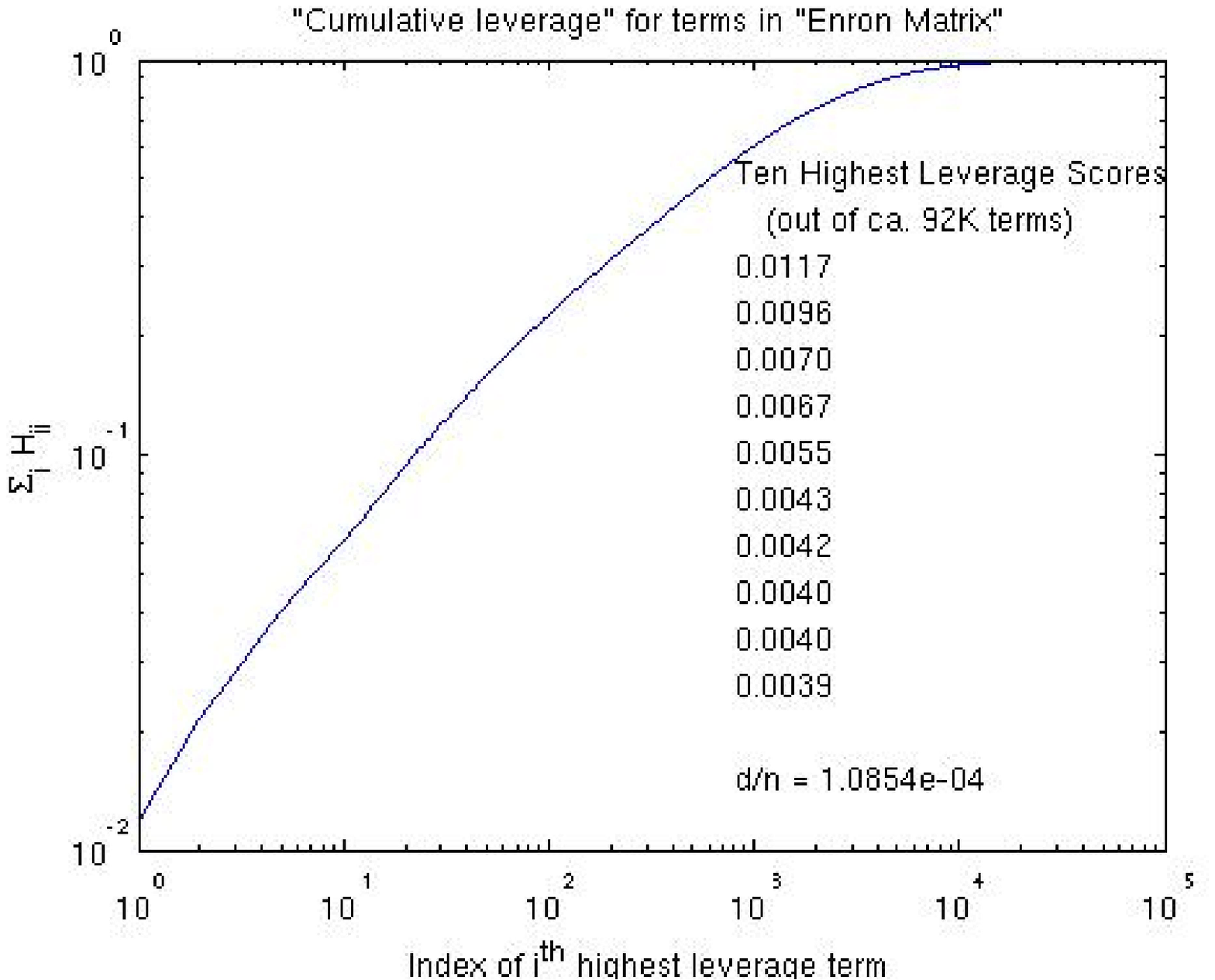}
         \label{fig:leverageII:cumlev}
      } \\
      \subfigure[Leverage Score and Information Gain for DNA Microarray Data]{
         \includegraphics[width=1.00\textwidth]{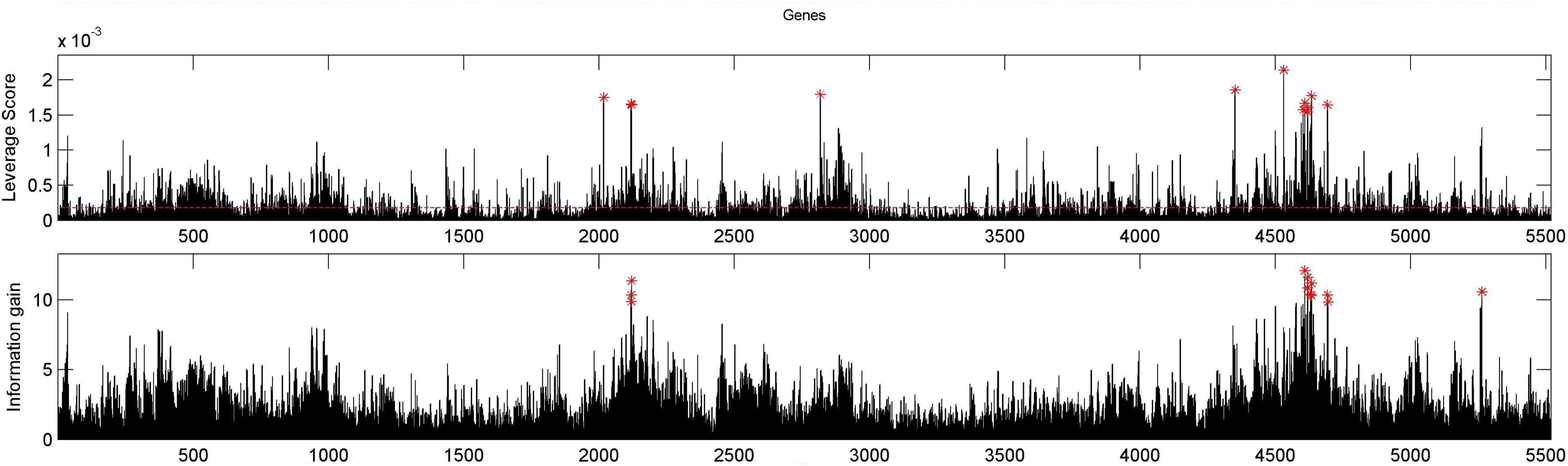}
         \label{fig:leverageII:bio}
      } 
   \end{center}
\caption{Statistical leverage scores in more modern applications.
(\ref{fig:leverageII:zachary})
The so-called Zachary karate club network~\cite{zachary77karate}, with edges
color-coded such that leverage scores for a given edge increase from yellow
to red.
(\ref{fig:leverageII:cumlev})
Cumulative leverage (with $k=10$) for a $65,031 \times 92,133$ term-document 
matrix constructed Enron electronic mail collection, illustrating 
that there are a large number of data points with very high leverage score.
(\ref{fig:leverageII:bio})
The normalized statistical leverage scores and information gain 
score---information gain is a mutual information-based metric popular in 
the application area~\cite{Paschou07b,CUR_PNAS}---for each of the $n = 5520$ 
genes, a situation in which the data cluster well in the low-dimensional 
space defined by the maximum variance axes of the data~\cite{CUR_PNAS}.
Red stars indicate the $12$ genes with the highest leverage scores, and the 
red dashed line indicates the average or uniform leverage scores.
Note the strong correlation between the unsupervised leverage score metric 
and the supervised information gain metric.
}
\label{fig:leverageII}
\end{figure}

To gain intuition for the behavior of the statistical leverage scores in a 
typical application, consider Figure~\ref{fig:leverageII:zachary}, which 
illustrates the so-called Zachary karate club 
network~\cite{zachary77karate}, a small but popular network in the community 
detection literature.
Given such a network $G=(V,E)$, with $n$ nodes, $m$ edges, and corresponding 
edge weights $w_e \ge 0$, define the $n \times n$ Laplacian matrix as 
$L=B^T W B$, where $B$ is the $m \times n$ edge-incidence matrix and $W$ is 
the $m \times m$ diagonal weight matrix.
The effective resistance between two vertices is given by the diagonal 
entries of the matrix $R= B L^{\dagger} B^T$ (where $L^{\dagger}$ denotes the 
Moore-Penrose generalized inverse) and is related to notions of ``network 
betweenness''~\cite{newman05_betweenness}.
For many large graphs, this and related betweenness measures tend to be strongly correlated with
node degree and tend to be large for edges that form articulation points
between clusters and communities, \emph{i.e.}, for edges that ``stick out'' 
a lot.
It can be shown that the effective resistances of the edges of $G$ are 
proportional to the statistical leverage scores of the $m$ rows of the 
$m \times n$ matrix $W^{1/2}B$---consider the $m \times m$ matrix
$$ P = W^{1/2}RW^{1/2} =  \Phi(\Phi^T\Phi)^{\dagger}\Phi^T ,$$
where $ \Phi = W^{1/2}B $, and note that if $U_{\Phi}$ denotes any 
orthogonal matrix spanning the column space of $\Phi$,~then
$$ P_{ii} = (U_{\Phi}U_{\Phi}^T)_{ii} = ||(U_{\Phi})_{(i)}||_2^2 .$$
Figure~\ref{fig:leverageII:zachary} presents a color-coded illustration of 
these scores for Zachary karate club network.
Note that the higher-leverage red edges tend to be those associated with 
higher-degree nodes and those at the articulation point between the two 
clusters.

Next, to gain intuition for the (non-)uniformity properties of statistical 
leverage scores in a typical application, consider a term-document matrix 
derived from the publicly-released 
Enron electronic mail collection~\cite{BB06}, which is typical of the type 
of data set to which SVD-based latent semantic analysis (LSA) 
methods~\cite{DDLFH90} have been applied.
This is a $65,031 \times 92,133$ matrix, as described in~\cite{BB06}, 
and let us choose the rank parameter as $k=10$.
Figure~\ref{fig:leverageII:cumlev} plots the cumulative leverage, \emph{i.e.}, 
the running sum of top $t$ statistical leverage scores, as a function of 
increasing $t$.
Since $\frac{k}{n}=\frac{10}{92,133}\approx1.0854\times10^{-4}$, we see that 
the highest leverage term has a leverage score nearly two orders of 
magnitude larger than this ``average'' size scale, that the second 
highest-leverage score is only marginally less than the first, that the 
third highest score is marginally less than the second, etc.
Thus, by the traditional metrics of diagnostic data 
analysis~\cite{VW81,ChatterjeeHadiPrice00}, which suggests flagging a data 
point if 
$$ (P_{U_k})_{ii} = (H_k)_{ii}> 2k/n ,$$
there are a \emph{huge} number of data points that are \emph{extremely} 
outlying, \emph{i.e.}, that are extreme outliers by the metrics of 
traditional regression diagnostics.
In retrospect, of course, this might not be surprising since the Enron email 
corpus is extremely sparse, with nowhere on the order of $\Omega(n)$ 
nonzeros per row.
Thus, even though LSA methods have been successfully applied, plausible 
generative models associated with these data are clearly not Gaussian, and 
the sparsity structure is such that there is no reason to expect that nice 
phenomena such as measure concentration occur.

Finally, note that DNA microarray and DNA SNP data often exhibit a similar 
degree of non-uniformity, although for somewhat different reasons.
To illustrate, Figure~\ref{fig:leverageII:bio} presents two plots for a data 
matrix, as was described in~\cite{CUR_PNAS}, consisting of $m = 31$ patients 
with $3$ different cancer types with respect to $n = 5520$ genes. 
First, this figure plots the information gain---information gain is a mutual 
information-based metric popular in that application 
area~\cite{Paschou07b,CUR_PNAS}---for each of the $n = 5520$ genes; and
second, it plots the normalized statistical leverage scores for each of 
these genes.
In each case, red dots indicate the genes with the highest values.
A similar plot illustrating the remarkable non-uniformity in statistical 
leverage scores for DNA SNP data was presented in~\cite{Paschou07b}.
Empirical evidence suggests that two phenomena may be responsible for this 
non-uniformity.
First, as with the term-document data, there is no domain-specific reason to 
believe that nice properties like measure concentration occur---on the 
contrary, there are reasons to expect that they do not.
Recall that each DNA SNP corresponds to a single mutational event in human 
history.
Thus, it will ``stick out,'' as its description along its one axis in the 
vector space will likely not be well-expressed in terms of the other axes, 
\emph{i.e.}, in terms of the other SNPs, and by the time it ``works its way 
back'' due to population admixing, etc., other SNPs will have occurred 
elsewhere.
Second, the correlation between statistical leverage and supervised mutual 
information-based metrics is particularly prominent in examples where the 
data cluster well in the low-dimensional space defined by the maximum 
variance axes.
Considering such data sets is, of course, a strong selection bias, but it is 
common in applications.
It would be of interest to develop a model that quantifies the observation
that, conditioned on clustering well in the low-dimensional space, an 
unsupervised measure like leverage scores should be expected to correlate 
well with a supervised measure like informativeness~\cite{Paschou07b} or
information gain~\cite{CUR_PNAS}.

\subsection{Statistical leverage and selecting columns from a matrix} 
\label{sxn:empirical-cssp}

With respect to some of the more technical and implementational issues
underlying the CSSP algorithm of Section~\ref{sxn:low-rank}, recall that an 
important aspect of QR algorithms is how they make so-called pivot rule 
decisions about which columns to keep~\cite{GVL96} and that such decisions 
can be tricky when the columns are not orthogonal or spread out in similarly 
nice~ways.
Several empirical observations~\cite{BMD08_CSSP_KDD,BMD08_CSSP_TRv2} are 
particularly relevant for large-scale data~applications.
\begin{itemize}
\item
We looked at several versions of the QR algorithm, and we compared each 
version of QR to the CSSP using that version of QR in the second phase.
One observation we made was that different QR algorithms behave 
differently---\emph{e.g.}, some versions such as the Low-RRQR algorithm 
of~\cite{CH94} tend to perform much better than other versions such as the 
qrxp algorithm of~\cite{BQ98a,BQ98b}.
Although not surprising to NLA practitioners, this observation indicates 
that some care should be paid to using ``off the shelf'' implementations in 
large-scale applications.
A second less-obvious observation is that preprocessing with the randomized 
first phase tends to improve more poorly-performing variants of QR more than 
better variants.
Part of this is simply that the more poorly-performing variants have more 
room to improve, but part of this is also that more sophisticated versions 
of QR tend to make more sophisticated pivot rule decisions, which are 
relatively less important after the randomized bias toward directions that 
are ``spread out.''
\item
We also looked at selecting columns by applying QR on $V_k^T$ and then 
keeping the corresponding columns of $A$, \emph{i.e.}, just running the 
classical deterministic QR algorithm with no randomized first phase on the 
matrix $V_k^T$.
Interestingly, with this ``preprocessing'' we tended to get better columns 
than if we ran QR on the original matrix $A$.
Again, the interpretation seems to be that, since the norms of the columns of 
$V_k^T$ define the relevant non-uniformity structure with which to sample 
with respect to, working directly with those columns tends make things 
``spread out,'' thereby avoiding (even in traditional deterministic 
settings) situations where pivot rules have problems.
\item
Of course, we also observed that randomization further improves the results, 
assuming that care is taken in choosing the rank parameter $k$ and the 
sampling parameter $c$.
In practice, the choice of $k$ should be viewed as a ``model selection'' 
question.
Then, by choosing $c=k,1.5k,2k,\ldots$, we often observed a 
``sweet spot,'' in bias-variance sense, as a function of increasing $c$.
That is, for a fixed $k$, the behavior of the deterministic QR algorithms 
improves by choosing somewhat more than $k$ columns, but that improvement 
is degraded by choosing too many columns in the randomized~phase.
\end{itemize}
These and related observations~\cite{BMD08_CSSP_KDD,BMD08_CSSP_TRv2} shed 
light on the inner workings of the CSSP algorithm, the effect of providing a 
randomized bias toward high-leverage data points at the two stages of the 
algorithm, and potential directions for the usefulness of this type of 
randomized algorithm in very large-scale data~applications.


\subsection{Statistical leverage in large-scale data analysis}
\label{sxn:empirical-forward}

Returning to the genetics applications where the algorithms described in 
this review have already been applied, we will consider one example each of 
the two common reasons (faster algorithms and more-interpretable algorithms) 
described in Section~\ref{sxn:background1-resource} for using randomization 
in the design of matrix algorithms for large-scale data problems.
To start with the former motivation,~\cite{Sayan11-unpub} applies the 
algorithm of~\cite{RST09} that was described in 
Section~\ref{sxn:low-rank:proj-third} to problems of subspace estimation and 
prediction in case-control studies in genetics.
In particular,~\cite{Sayan11-unpub} extends Principal Component Regression, 
Sliced Inverse Regression, and Localized Sliced Inverse Regression, three 
statistical techniques for matrix-based data that use as a critical step a 
matrix eigendecomposition, to much larger-scale data by using randomization 
to compute an approximately-optimal basis. 
Three goals were of interest:
evaluate the ability of randomized matrix algorithms to provide a good 
approximation to the dimension-reduced subspace; 
evaluate their relative predictive performance, when compared to the exact 
methods; and 
evaluate their ability to provide useful graphical summaries for the domain 
experts.

\begin{figure}
   \begin{center}
         \includegraphics[width=1.00\textwidth]{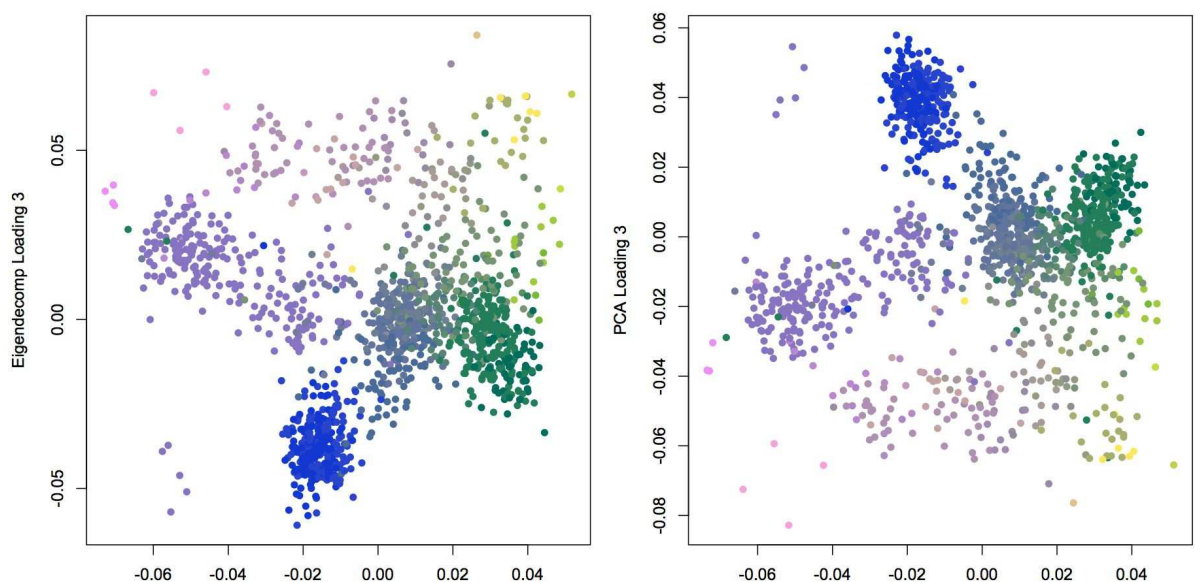}
   \end{center}
\caption{Pictorial illustration of the method of~\cite{Sayan11-unpub} 
recovering a good approximation to the top principal components of the 
DNA SNP data of~\cite{Nov08}.  Left panel is with the randomized algorithm; 
and right panel is with an exact computation.  See the text 
and~\cite{Sayan11-unpub} for details.
}
\label{fig:genetics-fast}
\end{figure}

As an example of their results,
Figure~\ref{fig:genetics-fast} illustrates pictorially the ability of the 
randomized algorithm to recover a good approximation to the top principal 
components of the DNA SNP data of~\cite{Nov08}.
For appropriate parameter choices, the empirical correlations of the exact 
sample principal components with the approximate estimates are (starting 
from second principal component): $0.999$, $-0.996$, $0.930$, $0.523$, 
$0.420$, $0.255$, etc.
Thus, the first few components are extremely-well reproduced; and then it 
appears as if there is some ``component splitting,'' as would be expected 
from standard matrix perturbation theory, as the seventh principal component 
from the randomized algorithm correlates relatively-well with three of the 
exact components.
The performance of the randomized algorithm in a Crohn's disease application 
on a subset of the DNA SNP data from the Wellcome Trust Case Control 
Consortium~\cite{WTC07} illustrates the run-time advantages of exploiting 
randomization in a real case-control application.
For example, when applied to a $4,686 \times 6,041$ matrix of SNPs from one 
chromosome, a single iteration of the randomized algorithm took $6$ seconds, 
versus $37$ minutes for the exact deterministic computation (a call to the 
DGESVD routine in the \textsc{Lapack} package), and it achieved a distance 
of $0.01$ to the exact subspace. 
By iterating just a few additional times, this distance could be decreased 
to less than $10^{-6}$ with a nominal increase in running time relative to 
the exact computation.
Similar results were achieved with matrices of larger sizes, up to a 
$4,686 \times 29,406$ matrix consisting of SNP data from four chromosomes.

\begin{figure}
   \begin{center}
         \includegraphics[width=1.00\textwidth]{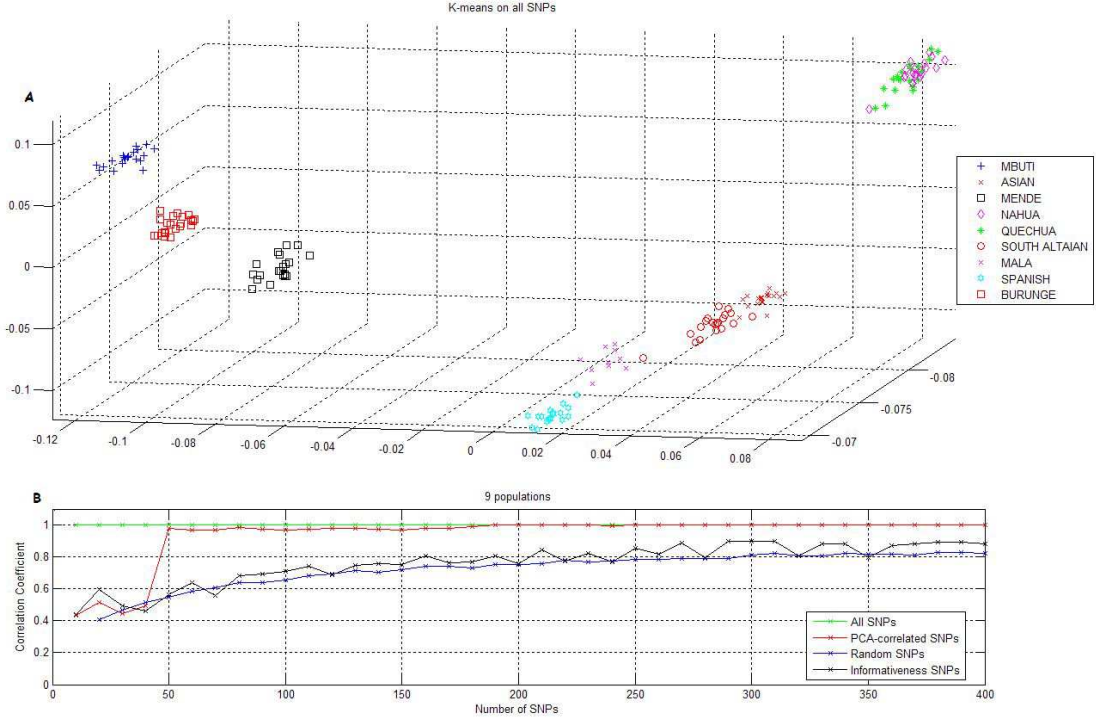}
   \end{center}
\caption{
Pictorial illustration of the clustering of individuals from nine 
populations typed for $9,160$ SNPs, from~\cite{Paschou07b}.
Top panel illustrates $k$-means clustering on the full data set.
Bottom panel plots, as a function of the number of actual SNPs chosen and 
for  three different SNP selection procedures, the correlation coefficient 
between the true and predicted membership of the individuals in the nine 
populations.  See the text and~\cite{Paschou07b} for details.
}
\label{fig:genetics-interp}
\end{figure}

In a somewhat different genetics application,~\cite{Paschou07b} was 
interested in obtaining a small number of actual DNA SNPs that could be 
typed and used for ancestry inference and the study of population structure 
within and across continents around the world.
As an example of their results,
Figure~\ref{fig:genetics-interp} illustrates pictorially the clustering of 
individuals from nine indigenous populations typed for $9,160$ SNPs. 
$k$-means clustering was run on the detected significant eigenvectors, and 
it managed successfully to assign each individual to their country of origin.
In order to determine the possibility of identifying a small set of 
actual SNPs to reproduce this clustering structure,~\cite{Paschou07b} used 
the statistical leverage scores as a ranking function and selected sets of 
$10$ to $400$ actual SNPs and repeated the analysis.
Figure~\ref{fig:genetics-interp} also illustrates the correlation 
coefficient between the true and predicted membership of the individuals in 
the nine populations, when the SNPs are chosen in this manner, as well as 
when the SNPs are chosen with two other procedures (uniformly at random and 
according to a mutual information-based measure popular in the field).
Surprisingly, by using only $50$ such ``PCA-correlated'' SNPs, all 
individuals were correctly assigned to one of the nine studied populations, 
as illustrated in the bottom panel of Figure~\ref{fig:genetics-interp}. 
(Interestingly, that panel also illustrates that, in this case, the mutual 
information-based measure performed much worse at selecting SNPs that could 
then be used to cluster individuals.)

At root, statistical leverage provides one way to quantify the notion of 
``eigenvector localization,'' \emph{i.e.}, the idea that most of the mass of 
an eigenvector is concentrated on a small number of coordinates.
This notion arises (often indirectly) in a wide range of scientific 
computing and data analysis applications; and in \emph{some} of those cases 
the localization can be meaningfully interpreted in terms of the domain from 
which the data are drawn.
To conclude this section, we will briefly discuss these issues more 
generally, with an eye toward forward-looking applications.

When working with networks or graph-based data, the so-called 
\emph{network values} are the eigenvector components associated with the 
largest eigenvalue of the graph adjacency matrix.
In many of these applications, the network values exhibits very high 
variability; and thus they have been used in a number of 
contexts~\cite{CF06_survey}, most notably to measure the value or worth of 
a customer for consumer-based applications such as viral 
marketing~\cite{RD02}.
Relatedly, sociologists have used so-called \emph{centrality} measures to 
measure the importance of individual nodes in a network.
The most relevant centrality notions for us are the Bonacich 
centrality~\cite{Bon87} and the random walk centrality of 
Newman~\cite{newman05_betweenness}, both of which are variants of these 
network values.
In still other network applications, effective resistances (recall the 
connection discussed in Section~\ref{sxn:empirical-morerecent}) have been 
used to characterize distributed control and estimation problems~\cite{BH06} 
as well as problems of asymptotic space localization in sensor 
networks~\cite{JLHB07}.

In many scientific applications, localized eigenvectors have a very natural 
interpretation---after all, physical particles (as well as photons, phonons, 
etc.) themselves are localized eigenstates of appropriate Hamiltonian 
operators.
If the so-called density matrix of a physical system is given by
$\rho(r,r^{'}) = \sum_{i=1}^{N} \psi_i(r)^{T}\psi_i(r^{'})$, then if 
$V$ is the matrix whose column vectors are the normalized eigenvectors 
$\psi_i, i=1,\ldots,s$, for the $s$ occupied states, then $P=VV^{T}$ is a
projection matrix, and the charge density at a point $r_i$ in space is the 
$i^{th}$ diagonal element of $P$.
(Here the transpose actually refers to the Hermitian conjugate since the 
wavefunction $\psi_i(r)$ is a complex quantity.)
Thus, the magnitude of this entry as well as other derived quantities like
the trace of $\rho$ give empirically-measurable quantities; see, 
\emph{e.g.}, Section $6.2$ of~\cite{SCS10}.
More practically, improved methods for estimating the diagonal of a 
projection matrix may have significant implications for leading to 
improvements in large-scale numerical computations in scientific computing 
applications such as the density functional theory of many-atom 
systems~\cite{SCS10,BKS07}.

In other physical applications, localized eigenvectors arise when extreme 
sparsity is coupled with randomness or quasi-randomness.
For example, \cite{BR88,RB88} describe a model for diffusion in a
configuration space that combines features of infinite dimensionality and 
very low connectivity---for readers more familiar with the 
Erd\H{o}s-R\'{e}nyi $G_{np}$ random graph model~\cite{erdos60random} than 
with spin glass theory, the parameter region of interest in these 
applications corresponds to the extremely sparse region 
$1/n \lesssim p \lesssim \log n/n$.
For ensembles of very sparse random matrices, there is a 
localization-delocalization transition which has been studied in 
detail~\cite{FM91,MF91,Eva92,Kuh08}. 
In these applications, as a rule of thumb, eigenvector localization occurs 
when there is some sort of ``structural heterogeneity,'' \emph{e.g.}, 
the degree (or coordination number) of a node is significantly higher or 
lower than~average.

Many large complex networks that have been popular in recent years, 
\emph{e.g.}, social and information networks, networks constructed from 
biological data, networks constructed from financial transactions, etc., 
exhibit similar qualitative properties, largely since these networks are 
often very sparse and relatively-unstructured at large size scales.
See, \emph{e.g.},~\cite{FDBV01,GKK01,DGMS03,MT09} for detailed discussions 
and empirical evaluations.
Depending on whether one is considering the adjacency matrix or the 
Laplacian matrix, localized eigenvectors can correspond to structural 
inhomogeneities such as very high degree nodes or very small cluster-like 
sets of nodes.
In addition, localization is often preserved or modified in characteristic 
ways when a graph is generated by modifying an existing graph in a 
structured manner; and thus it has been used as a diagnostic in certain 
network applications~\cite{BJ09a,BJ08a}.
The implications of the algorithms described in this review remain to be 
explored for these and other applications where eigenvector localization is 
a significant phenomenon.

\section{A few general thoughts, and a few lessons learned}
\label{sxn:thoughts}

\subsection{Thoughts about statistical leverage in MMDS applications}

One high-level question raised by the theoretical and empirical results 
reviewed here is: 
\begin{itemize}
\item
Why are the statistical leverage scores 
so nonuniform in many large-scale data analysis applications?
\end{itemize}
The answer to this seems to be that, intuitively, in many MMDS application 
areas, statistical models are \emph{implicitly} assumed based on 
computational and not statistical considerations.
That is, when computational decisions are made, often with little regard 
for the statistical properties of the data, they carry with them statistical 
consequences, in the sense that the computations are the ``right thing'' or 
the ``wrong thing'' to do for different classes of data.
Thus, in these cases, it is not surprising that some interesting data points 
``stick out'' relative to obviously inappropriate models.
This suggests the use of these importance sampling scores as cheap 
signatures of the ``inappropriateness'' of a statistical model (chosen for 
algorithmic and not statistical reasons) in large-scale exploratory or
diagnostic applications.

A second high-level question raised by the results reviewed here is: 
\begin{itemize}
\item
Why should statistical leverage, a traditional concept 
from regression diagnostics, be useful to obtain improved worst-case 
approximation algorithms for traditional NLA matrix problems?
\end{itemize}
Here, the answer seems to be that, 
intuitively, if a data point has a high 
leverage score and is not an error then it might be a particularly 
important or informative data point.  
Since worst-case analysis takes the input matrix as given, each row/column 
is assumed to be reliable, and so worst-case guarantees are obtained by 
focusing effort on the most informative data points.
It would be interesting to see if this perspective is applicable more 
generally in the design of matrix and graph algorithms for other 
MMDS~applications.

\subsection{Lessons learned about transferring theory to practice}

More generally, it should be emphasized that the randomized matrix algorithms 
reviewed here represent a real success story in bringing novel theoretical 
ideas to the solution of practical problems.
This often-cited (but less frequently-achieved) goal arises in many MMDS applications---in fact, bridging 
the gap between TCS, NLA, and data applications was at the origin of the 
MMDS meetings~\cite{MMDS06summary,MMDS08_arxiv,MMDS10_arxiv}, which 
address algorithmic and statistical aspects of large-scale data analysis 
more generally.
Not surprisingly, the widespread interest in these algorithms is not simply
due to the strong worst-case bounds they achieve, but also to their 
usefulness in downstream data analysis and scientific computing 
applications.
Thus, it is worth highlighting some of the ``lessons learned'' that are 
applicable more generally than to the particular algorithms and applications 
reviewed here.
\begin{itemize}
\item
\textbf{Objective functions versus certificates.}
TCS is typically concerned with providing bounds on objective functions in 
approximate optimization problems (as in,
\emph{e.g.},~(\ref{eqn:ls-bound-eq1}) and~(\ref{eqn:rel-err})) and makes no 
statement about how close the certificate (\emph{i.e.}, the vector or graph 
achieving that approximate solution) is to a/the exact solution of
the optimization problem (as in, \emph{e.g.}, (\ref{eqn:ls-bound-eq2})).
In machine learning and data analysis, on the other hand, one is often 
interested in statements about the quality of the certificate, largely since 
the certificate is often used more generally for other downstream 
applications like clustering or classification.
\item
\textbf{Identifying structure versus washing out structure.}
TCS is often \emph{not} interested in identifying structure \emph{per se}, 
but instead only in exploiting that structure to provide fast algorithms.
Thus, important structural statements are often buried deep in the analysis
of the algorithm.
Making such structural statements explicit has several benefits:
one can obtain improved bounds if the tools are more powerful than 
originally realized (as when relative-error projection algorithms followed 
additive-error projection algorithms and relative-error sampling algorithms 
simply by performing a more sophisticated analysis); structural properties 
can be of independent interest in downstream data applications; and it can 
make it easer to couple to more traditional numerical methods.
\item
\textbf{Side effects of computational decisions.}
There are often side effects of computational decisions that are at least
as important for the success of novel methods as is the original nominal 
reason for the introduction of the new methods.
For example, randomness was originally used as a resource inside the 
algorithm to speed up the running time of algorithms on worst-case 
input.
On the other hand, using randomness inside the algorithm often leads to 
improved condition number properties, better parallelism properties on 
modern computational architectures, and better implicit regularization
properties, in which case the approximate answer can be even better than the
exact answer for downstream applications.
\item
\textbf{Significance of cultural issues.}
TCS would say that if a randomized algorithm succeeds with constant 
probability, say with probability at least $90\%$, then it can be boosted 
to hold with probability at least $1-\delta$, where the dependence on 
$\delta$ scales as $O(\log(1/\delta))$, using standard 
methods~\cite{MotwaniRaghavan95}.
Some areas would simply say that such an algorithm succeeds with 
``overwhelming probability'' or fails with ``negligible probability.''
Still other areas like NLA and scientific computing are more willing to 
embrace randomness if the constants are folded into the algorithm such that 
the algorithm fails with probability less than, say,~$10^{-17}$.
Perhaps surprisingly, getting beyond such seemingly-minor cultural 
differences has been the main bottleneck to technology transfer such as that 
reviewed~here.
\item
\textbf{Coupling with domain experience.}
Since new methods almost always perform more poorly than well-established 
methods on traditional metrics, a lot can be gained by coupling with domain
expertise and traditional machinery.
For example, by coupling with traditional iterative methods, minor variants 
of the original randomized algorithms for the LS problem can have their 
$\epsilon$ dependence improved from roughly $O(1/\epsilon)$ to 
$O(\log(1/\epsilon))$.
Similarly, since factors of $2$ matter for geneticists, by using the 
leverage scores as a ranking function rather than as an importance sampling 
distribution, greedily keeping, say, $100$ SNPs and then filtering to $50$ 
according to a genetic criterion, one often does very well in those 
applications.
\end{itemize}

\section{Conclusion}
\label{sxn:conclusion}

Randomization has had a long history in scientific 
applications~\cite{HamHan64,Met53}.
For example, originally developed to evaluate phase space integrals in 
liquid-state statistical mechanics, Markov chain Monte Carlo techniques 
are now widely-used in applications as diverse as option valuation in 
finance, drug design in computational chemistry, and Bayesian inference
in statistics.
Similarly, originally developed to describe the energy levels of systems
arising in nuclear physics, random matrix theory has found applications 
in areas as diverse as signal processing, finance, multivariate 
statistics,  and number theory. 
Randomized methods have been popular in these and other scientific 
applications for several reasons:
the weakness of the assumptions underlying the method permits its broad 
applicability; 
the simplicity of these assumptions has permitted a rich body of 
theoretical work that has fruitfully fed back into applications; 
due to the intuitive connection between the method and hypothesized noise 
properties in the data; and
since randomization permits the approximate solution of otherwise
impossible-to-solve problems.

Within the last few decades, randomization has also proven to be useful in 
a very different way---as a powerful resource in TCS for establishing 
worst-case bounds for a wide range of computational problems.
That is, in the same way that space and time are valuable resources 
available to be used judiciously by algorithms, it has been discovered 
that exploiting randomness as an algorithmic resource \emph{inside the 
algorithm} can lead to better algorithms.
Here, ``better'' typically means faster in worst-case theory when 
compared, \emph{e.g.}, to deterministic algorithms for the same problem;
but it could also mean simpler---which is of typically interest since 
simpler algorithms tend to be more amenable to worst-case theoretical 
analysis.
Applications of this paradigm have included algorithms for number 
theoretic problems such as primality testing, algorithms for data 
structure problems such as sorting and order statistics, as well as 
algorithms for a wide range of optimization and graph theoretic problems
such as linear programming, minimum spanning trees, shortest paths, and 
minimum cuts.

Perhaps since its original promise was oversold, and perhaps due to the 
greater-than-expected difficulty in developing high-quality 
numerically-stable software for scientific computing applications, 
randomization \emph{inside the algorithm} for common matrix problems was 
mostly ``banished'' from scientific computing and NLA in the 1950s. 
Thus, it is refreshing that within just the last few years, novel 
algorithmic perspectives from TCS have worked their way back to NLA, 
scientific computing, and scientific data analysis.
These developments have been driven by large-scale data analysis 
applications, which place very different demands on matrices than 
traditional scientific computing applications.
As with other applications of randomization, though, the ideas underlying 
these developments are simple, powerful, and broadly-applicable.

Several obvious future directions seem particularly promising application 
areas for this randomized matrix algorithm paradigm.
\begin{itemize}
\item
\textbf{Other traditional NLA problems and large-scale optimization.}
Although least squares approximation and low-rank matrix approximation are 
fundamental problems that underlie a wide range of problems, there are many 
other problems of interest in NLA---computing polar decompositions, 
eigenvalue decompositions, Cholesky decompositions, etc.
In addition, large-scale numerical optimization code often uses these 
primitives many times during the course of a single computation.
Thus, for example, some of the fast numerical implementations for very 
overdetermined least squares problems that were described in
Section~\ref{sxn:least-squares:faster-pr} can in principle be used to 
accelerate interior-point methods for convex optimization and linear 
programming.
Working through the practice in realistic computational settings remains
an ongoing challenge.
\item
\textbf{Parallel and distributed computational environments.}
In many applications, communication is more expensive than computation. 
This is true both for computations involving a single machine---recall 
recent developments in multicore computing---as well as for computations 
run across many machines---such as in large distributed data centers.
In some cases, such as with Gaussian-based random projections, computations
can be easily parallelized; and numerical implementations for both the least 
squares approximation problem and the low-rank approximation problem have 
already exploited this.
Taking advantage of modern computer architectures and systems requirements
more generally is a substantial challenge.
\item
\textbf{Sparse graphs, sparse matrices, and sparse projections.}
Sparsity is a ubiquitous property of data, and one which is a challenge 
since vanilla applications of randomized algorithms tend to densify the 
input data.
In some cases, sparsity in the input is structured and can be exploited by
the randomized algorithm; while in other cases it is less structured but it 
can be respected with existing projection methods.
More generally, sparse projection matrices are of interest---such 
projection matrices make it easy to perform incremental updates in data 
streaming environments, they can make it easier to perform matrix-vector 
products quickly, etc.
Problems of recovering sparse signals have been approached by researchers 
in theoretical computer science, applied mathematics, and digital signal 
processing; and in many cases the approaches are somewhat orthogonal to that
of the work reviewed here.
\item
\textbf{Laplacian matrices and large informatics graphs.}
Laplacians are fundamental matrices associated with a graph, and they permit
many of the randomized matrix algorithms we have been discussing to be 
applied to graph-based data.
In some cases, the goal might be to sparsify an input graph; but more 
typically graphs arising in data applications are sparse and irregular.
In the case of large social and information networks, for example, it is 
known that, while often there exists good small-scale clustering structure, 
there typically does not exist good large-scale clustering structure.
Part of this has to do with the heavy-tailed properties in these graphs, 
which imply that although there may exist a small number of most important 
nodes in the graph, these nodes do not capture most of the information in 
the data.
This presents numerous fundamental challenges for the algorithms reviewed 
here, and these challenges have just begun to be addressed.
\item
\textbf{Randomized algorithms and implicit regularization.}
In many cases, randomized algorithms or their output are more robust than 
their deterministic variants.
For example, algorithms may empirically be less sensitive to pivot rule 
decisions; and their output may empirically be ``nicer'' and more 
``regular''---in the sense of statistical regularization.
Existing theory (reviewed here) makes precise a sense in which randomized matrix algorithms 
are not much worse than the corresponding deterministic algorithms; but
quantifying a sense in which the output of randomized matrix algorithms
is even ``better'' than the output of the corresponding deterministic 
algorithms is clearly of interest if one is interested in very large-scale 
applications.
\end{itemize}

\noindent
In closing, it seems worth reminding the reader that researchers often look 
with great expectation toward randomness, as if a na\"{i}ve application of 
randomness will somehow solve all of one's problems.
By now, it should be clear that such hopes are rarely realized in practice.
It should also be clear, though, that a careful application of 
randomness---especially when it is coupled closely with domain 
expertise---provides a powerful framework to address a range of matrix-based
problems in modern massive data set analysis.

\vspace{0.25in}
\noindent
\textbf{Acknowledgments.}
I would like to thank the numerous colleagues and collaborators with whom 
these results have been discussed in preliminary form---in particular, Petros 
Drineas, with whom my contribution to the work reviewed here was made;
Michael Jordan and Deepak Agarwal, who first pointed out the connection 
between the importance sampling probabilities used in the relative-error 
matrix approximation algorithms and the concept of statistical leverage;
Sayan Mukherjee, who generously provided unpublished results on the 
usefulness of randomized matrix algorithms in his genetics~applications; 
and Ameet Talwalkar and an anonymous reviewer for providing numerous 
valuable comments.

\noindent

\begin{small}

\end{small}

%
%

\end{document}